\documentclass[a4paper,11pt]{article}
\usepackage[pub]{zyanres}

\DeclareSymbolFont{matha}{OML}{txmi}{m}{it}
\DeclareMathSymbol{v}{\mathord}{matha}{118}

\def\heq{\overset{\mathcal{H}}{=}}
\def\fheq{\overset{\mathcal{H}^+}{=}}
\def\pheq{\overset{\mathcal{H}^-}{=}}
\def\nheq{\overset{\mathcal{N}}{=}}

\def\ceq{\overset{\mathcal{C}(v)}{=}}
\def\cceq{\overset{\mathcal{C}}{=}}

\DeclareSymbolFont{matha}{OML}{txmi}{m}{it}
\DeclareMathSymbol{v}{\mathord}{matha}{118}

\def\aeq{\overset{\mathcal{A}}{=}}
\def\teq{\overset{\mathcal T}{=}}

\usepackage[normalem]{ulem}

\numberwithin{equation}{section}

\title{Properties of Dynamical Black Hole Entropy}
\author{Manus R.~Visser\footnote{mv551@cam.ac.uk} and Zihan~Yan\footnote{zy286@cam.ac.uk}\\~\\~\\
\textit{DAMTP, Centre for Mathematical Sciences, University of Cambridge,\\Wilberforce Road, Cambridge, U.K. CB3 0WA}}
\date{\today}

\begin{document}
    \maketitle
    \begin{abstract}
We study the first law for non-stationary perturbations of a stationary black hole whose event horizon is a Killing horizon, that relates the first-order change in the mass and angular momentum to the change in the entropy of an arbitrary horizon cross-section. Recently, Hollands, Wald and Zhang \cite{Hollands:2024vbe} have shown that the dynamical black hole entropy that satisfies this first law, for general relativity, is $S_{\text{dyn}}=(1-v\partial_v)S_{\text{BH}}$, where $v$ is the affine parameter of the null horizon generators and $S_{\text{BH}}$ is the Bekenstein-Hawking entropy, and for general diffeomorphism covariant theories of gravity $S_{\text{dyn}}=(1-v\partial_v)S_{\text{Wall}}$, where $S_{\text{Wall}}$ is the Wall entropy.  They obtained the first law by applying the Noether charge method to non-stationary perturbations and arbitrary cross-sections. In this formalism, the dynamical black hole entropy is defined as an ``improved'' Noether charge, which is unambiguous to first order in the perturbation. In the present article we provide a pedagogical derivation of the   physical process version of the non-stationary first law for general relativity by integrating the linearised Raychaudhuri equation between two arbitrary horizon cross-sections. Moreover, we generalise the  derivation of the first law in \cite{Hollands:2024vbe} to non-minimally coupled matter fields  that are smooth on the horizon, using  boost weight arguments rather than Killing field arguments, and we relax some of the gauge conditions on the perturbations by  allowing for non-zero variations of the horizon Killing field and surface gravity.  Finally, for $f(\text{Riemann})$ theories of gravity we show explicitly using Gaussian null coordinates that the improved Noether charge is $S_{\text{dyn}}=(1-v\partial_v)S_{\text{Wall}}$, which is a non-trivial check of \cite{Hollands:2024vbe}.    
    \end{abstract}

\newpage
    \tableofcontents

    \newpage

    \section{Introduction}

\subsection{Motivation}
The laws of black hole thermodynamics   belong to the most remarkable discoveries of modern physics, since they combine  general relativity, quantum field theory and thermodynamics into one overarching framework. Moreover, they provide a low-energy window into quantum gravity. The first law of black hole mechanics  relates the variation of the mass $M$ and angular momentum~$J$ of the black hole to the variation  of the black hole entropy $S$ \cite{Bekenstein:1973ur,Bardeen:1973gs}
\begin{equation}
\label{firstlaw1}
    \delta M - \Omega_{\mathcal H} \delta J = T \delta S\,, 
\end{equation}
with $\Omega_{\mathcal H}$   being the angular velocity of the horizon. This variational relation holds for   linear perturbations of  stationary, axisymmetric  black holes, whose event horizon is a Killing horizon, and we assumed the black hole is a solution to the vacuum gravitational field equations, and the perturbation satisfies the linearised field equations.
The black hole temperature is universally given by the Hawking temperature~\cite{Hawking:1975vcx}
\begin{equation}
\label{hawkingtemp}
    T =\frac{\kappa}{2\pi}\,,
\end{equation}
where $\kappa$ is the surface gravity of the Killing horizon (and we have set $\hbar = c=k_B=1$).
The mass, angular momentum, and black hole entropy, on the other hand,  depend  on the  gravitational theory under consideration. For instance, in general relativity the mass and angular momentum  are given by the ADM formulae \cite{Arnowitt:1962hi,Regge:1974zd}, defined at spatial infinity, but in higher curvature gravity there are   suitable generalisations of these definitions, see, e.g., \cite{Jacobson:1993xs,Kastor:2011qp,Kastikainen:2019iaz} for the ADM mass in Lovelock gravity. In this article, however, we focus on the entropy of black holes. For general relativity, the black hole entropy $S$ satisfies the   Bekenstein-Hawking formula \cite{Bekenstein:1973ur,Hawking:1975vcx}
\begin{equation}
\label{bekensteinhaw}
    S_{\text{BH}} = \frac{A}{4 G}\,,
\end{equation}
where  $A$ is the event horizon area and $G$ is Newton's constant. On the other hand,  for an arbitrary diffeomorphism covariant theories of gravity,  $S$ is given by the Wald  entropy,\footnote{Wald entropy  seems to be standard terminology  for \eqref{iyerwald} in the literature (see, e.g., \cite{Wall2015}), even though this particular expression only appeared later in the   work of Iyer and Wald \cite{Iyer:1994ys}. The term Iyer-Wald entropy is reserved for their dynamical black hole entropy proposal \eqref{iyerwald2}, which they retracted in a note added to the journal version.} which for a Lagrangian  $L$ consisting of contractions of the inverse metric and Riemann tensor, henceforth called $f(\text{Riemann})$ theories of gravity, takes the form \cite{Wald:1993nt,Iyer:1994ys} 
\begin{equation}
\label{iyerwald}
    S_{\text{Wald}} =- 8\pi \int_{\mathcal C(v)} \dd{A}  \pdv{L}{R_{u v u v}}\,.
\end{equation}
Here $v$ is the (future-directed) affine   parameter along the  outgoing  null geodesics of the future horizon, $u$ affinely parameterises the ingoing null geodesics of the horizon and is also future-directed,   $\mathcal C(v)$ is a horizon cross-section at affine null time $v$,   $\dd A$ is a ``volume'' (area) element on~$\mathcal C(v)$,   and $ \partial L/\partial R_{u v u v}$ is the functional derivative of the Lagrangian with 
respect to the Riemann tensor component $R_{u v u v}$, with the metric and connection held fixed. 

Furthermore, the Bekenstein-Hawking entropy \eqref{bekensteinhaw} satisfies a   second law, i.e.,  $\partial_v S_{\text{BH}}\ge 0$:  according to the area theorem  \cite{Hawking:1971tu}  the surface area of the future event horizon is  never decreasing in time in any dynamical process, if the null energy condition holds and weak cosmic censorship is valid. For arbitrary diffeomorphism covariant theories of gravity there is no equivalent second law for black hole entropy in a general context. Iyer and Wald  \cite{Iyer:1994ys}  proposed a formula for dynamical black hole entropy for such general   theories of gravity  by evaluating the Wald entropy with only the boost-invariant tensor components of the dynamical fields (metric and matter) and their derivatives as input. This is known as the Iyer-Wald entropy
\begin{equation}
\label{iyerwald2}
    S_\text{Iyer-Wald}[\phi] = S_\text{Wald}[\mathfrak{I} \phi]\,,
\end{equation}
  where $\phi = (g, \varphi)$ is the collective notation for the metric $g$ and matter fields $\varphi$, and $\mathfrak{I}$ is a projection operator that keeps only boost-invariant  components (with respect to the    Killing vector that generates local boosts near the horizon). As a result, the Iyer-Wald entropy contains only zero boost weight terms.\footnote{We introduce the notion ``boost weight'' of a tensor component in Section \ref{sec:gnc}.} However, $S_\text{Iyer-Wald}$ does not seem to satisfy a second law, and   it is not field redefinition invariant, as was noted in a revised version of~\cite{Iyer:1994ys}. 
  
  Recently, Wall \cite{Wall2015} derived  a second law  for higher curvature gravity in a perturbative context, that holds   for linear non-stationary  perturbations to a Killing horizon.    His dynamical  entropy proposal is constant  to linear order for vacuum perturbations of the metric field, i.e., $ \partial_v \delta S_{\text{Wall}}=0$. Moreover, it is non-decreasing, $ \partial_v \delta S_{\text{Wall}}\ge 0$, for perturbations that are sourced by an external stress-energy tensor, satisfying the null energy condition $\delta T_{vv} \ge 0$, and that  ends as a stationary state, i.e., $\partial_v \delta S_{\text{Wall}}|_{v=+\infty}=0 $.  
For $f(\text{Riemann})$ theories of gravity the Wall entropy has an additional term compared to the Wald  entropy \eqref{iyerwald}, that depends on the extrinsic curvatures of the horizon in the $v$- and $u$-directions, $K_{ij}$ and $\bar K_{ij}$, respectively, and for more complicated theories   there will be more terms.  In particular, for    $f(\text{Riemann})$   gravity  the Wall entropy takes the form\footnote{The    extrinsic curvature term in \eqref{wallentropy} has a different sign compared to equation (14) in \cite{Wall2015}, which is due to the fact that in our conventions the tangent $\partial_u$ to the ingoing null geodesics at the future horizon is future-directed, whereas in \cite{Wall2015} (and \cite{Wall:2024lbd}) it is past-directed. 
} \cite{Wall2015} 
  \begin{equation}
   \label{wallentropy}
        S_{\text{Wall}} = - 8 \pi  \int_{\mathcal{C}(v)} \dd{A} \left( \pdv{L}{R_{u v u v}} - 4 \pdv{L}{R_{u i u j}}{R_{v k v l}} \bar K_{ij} K_{kl} \right).
   \end{equation}
The Wall entropy reduces to the Wald  entropy on a    Killing horizon, since $K_{ij}=0$ on the future horizon and $\bar K_{ij}=0$ on the past horizon. Remarkably, the Wall entropy matches to linear order in perturbation theory with   the   holographic entanglement entropy computed by Dong \cite{Dong:2013qoa} for higher curvature gravity (see also \cite{Camps:2013zua} for a derivation of holographic entanglement entropy in quadratic gravity). Since the Dong and Wall entropies are derived in different contexts, and because it is not well understood whether their agreement is a coincidence or whether it holds more generally,     we will refer to the dynamical black hole entropy proposal~\eqref{wallentropy} as Wall entropy.
More recently, the Wall entropy was proven to be gauge invariant to linear order in the perturbation, and was  extended for effective field theories to second order in the perturbation   \cite{Hollands:2022fkn} (see also \cite{Davies:2022xdq}) and  in a non-perturbative context~\cite{Davies:2023qaa}.   Further, $ S_{\text{Wall}}$ has   been generalised to any diffeomorphism covariant theory of gravity non-minimally  coupled to scalar fields,  gauge fields \cite{Biswas:2022grc} and (non-gauged) vector fields~\cite{Wall:2024lbd}. And, a covariant entropy current  for the Wall entropy in general  theories of gravity was obtained  in  \cite{Hollands:2024vbe} (see also  \cite{Bhattacharyya:2021jhr,Hollands:2022fkn}).   At the bifurcation surface both $S_{\text{Iyer-Wald}}$ and $S_{\text{Wall}}$ reduce to $S_{\text{Wald}}$, and for general relativity all the higher curvature black hole  entropy definitions are equal to~$S_{\text{BH}}$.

Now,  let us discuss the regime of validity of these entropy functionals, especially for the black hole entropies that obey the first law \eqref{firstlaw1}. The second law for $S_{\text{BH}}$ and the linearised second law for $S_{\text{Wall}}$ hold for non-stationary situations and for any horizon cross-section, as $ \partial_v S_{\text{BH}} \ge 0$ and $\partial_v \delta S_{\text{Wall}}\ge0$ are valid  for any time $v$ on the horizon,  but this is not the case for the first law. There are  two major limitations of the standard treatments of the first law   of black hole thermodynamics: 
\begin{itemize}
  \item[(i)] the first law often does not hold for non-stationary perturbations of a stationary black hole, and, if it does,    
  \item[(ii)]  the black hole entropy   cannot be evaluated at an arbitrary  cross-section of the event horizon of the perturbed non-stationary black hole. 
\end{itemize}
We start with the first limitation (i). The Bekenstein-Hawking  entropy and Wald entropy both are valid for arbitrary   cross-sections  of Killing horizons, since they are constant in Killing time for stationary black holes. That is, they obey   the   first law  \eqref{firstlaw1} for  stationary perturbations of the Killing horizon. There are two versions of such a stationary first law \cite{Wald:1995yp}. First, the \emph{stationary comparison version} of the first law compares     two different, but infinitesimally close, stationary black hole geometries. Originally,  Bekenstein \cite{Bekenstein:1973ur}   derived  this first law by varying the parameters   describing the Kerr-Newman black hole. Further, Bardeen, Carter and Hawking \cite{Bardeen:1973gs} (see also \cite{Carter:2009nex})  extended it,   using the Komar integral method,       to   asymptotically flat, stationary, axisymmetric black holes in general relativity surrounded by fluid matter.    And, Wald \cite{Wald:1993nt} derived the stationary comparison first law using the Noether charge method, showing that the black hole entropy of a Killing horizon is given by the Noether charge associated to the horizon Killing field integrated over a horizon cross-section.

  Second, for the \emph{physical process version} of the first law we consider a stationary black hole and slightly perturb it by a  flux of ingoing matter. The standard derivation \cite{Wald:1995yp} assumes that the perturbation starts in a stationary state (at the bifurcation surface) and settles down to a stationary solution at future infinity. The physical process first law was originally derived for vacuum black hole solutions in general relativity \cite{Hawking:1972hy}, and was later extended  to charged, stationary black holes  by  Gao and Wald \cite{Gao:2001ut} (see also \cite{Rignon-Bret:2023lyn}). In fact, it also holds for Rindler horizons \cite{Jacobson:1999mi}, and more generally, for any causal horizon~\cite{Jacobson:2003wv} and   bifurcate Killing horizon~\cite{Amsel:2007mh}. 

  All these derivations of the first law have in common that they assume  the perturbation of the metric is stationary, i.e., $\delta (\mathcal L_\xi   g_{ab}) =0$, where $\mathcal L_\xi$ is the Lie derivative with respect to the horizon generating Killing field.  However,  non-stationary perturbations of a stationary black hole also present a   well-defined setup for the   first law.  This  is precisely the setup in which Wall~\cite{Wall2015} derived the linearised second law for higher curvature gravity (see  \cite{Sarkar:2013swa,Bhattacharjee:2015yaa,Mishra:2017sqs, Bhattacharyya:2021jhr} for a similar setup). To recap, in thermodynamic language, the Bekenstein-Hawking entropy and Wald entropy are valid for   black holes in thermal equilibrium (with their Hawking radiation), but they do not describe the entropy of near-equilibrium black holes. 
  
This brings us to the second limitation (ii). Iyer and Wald \cite{Iyer:1994ys} generalised the first law using the Noether charge method to non-stationary perturbations, and showed that the Wald entropy \eqref{iyerwald} does satisfy   a non-stationary first law.
The caveat is, however, that the Wald entropy must be   evaluated at the bifurcation surface in that case. But the entropy of non-stationary black holes evolves in principle over time, so the entropy at the bifurcation surface is generically not the same as    the entropy of other horizon cross-sections.   It turns out that the Wald entropy on arbitrary cross-sections of a non-stationary black holes suffers from so-called JKM ambiguities, named after  Jacobson, Kang and Myers (JKM) \cite{Jacobson:1993vj}. Hence, the Wald entropy cannot be the dynamical black hole entropy for arbitrary horizon slices. On the other hand, the Wall entropy does apply to non-stationary geometries, but it does not hold for arbitrary horizon cross-sections, since in the derivation of the physical process first law for Wall entropy it is assumed that the initial and final state of the perturbation are at the bifurcation surface and at future infinity, respectively \cite{Wall2015}.

 Recently, Hollands, Wald and Zhang \cite{Hollands:2024vbe} have shown how to overcome these two  limitations. They derived a first law for non-stationary perturbations of a stationary, axisymmetric black hole, whose event horizon is a Killing horizon, and obtained an unambiguous dynamical entropy functional  satisfying the first law for arbitrary  cross-sections of the perturbed event horizon. Their formula for the dynamical black hole entropy, reviewed below,  differs from the Bekenstein-Hawking entropy and Wall entropy by a dynamical correction term. They derived the  first law by applying the Noether charge method to non-stationary variations and arbitrary horizon cross-sections.  In this formalism, the dynamical black hole entropy that satisfies the non-stationary first law is  defined as an ``improved'' Noether charge, which we  review below. Further, their derivation of the first law holds for any diffeomorphism covariant Lagrangian for which the metric is the only dynamical field. We also note that  the dynamical black hole entropy  for general relativity was previously documented by Rignon-Bret  in  \cite{Rignon-Bret:2023fjq} (citing the upcoming work~\cite{Hollands:2024vbe}), and he also proposed a different dynamical entropy for null surfaces that vanishes for every cross-section of a light cone in Minkowski spacetime. And, interestingly, in~\cite{Chandrasekaran:2018aop,Ciambelli:2023mir} a boost charge was found  that satisfies the same formula as the dynamical black hole entropy. 
 
 In this article we generalise and improve upon their work \cite{Hollands:2024vbe} in a number of ways, and we give a more pedagogical proof of the first law, that is:  
 
 \begin{itemize}
 \item[a)]
 We explain how the non-stationary physical process first law for general relativity simply follows  from the Raychaudhuri equation. This derivation does not make use of the covariant phase space formalism, and is hence considerably less technical. In going beyond general relativity, for arbitrary diffeomorphism covariant theories, we do introduce the covariant phase space formalism, as does \cite{Hollands:2024vbe}, but our analysis is based mostly on a Gaussian null coordinate (GNC) system, and uses the associated boost weight arguments,  which is arguably more powerful and less technical than the  tetrad approach in  \cite{Hollands:2024vbe}. The drawback of working in a particular GNC system, however, is that it is not manifestly covariant. For general relativity and $f(R)$  we do give a fully covariant proof of the dynamical first law using the covariant phase space formalism. 
 
 \item[b)] Our gauge conditions for the perturbations are less restrictive, i.e., we do not fix the horizon Killing vector field ($\delta \xi^a \neq 0$) nor the surface gravity ($\delta \kappa \neq 0 $). Especially the latter condition is important as it allows for nonzero variations of the black hole temperature. In thermodynamics changing an
equilibrium state to a nonequilibrium state in general also alters the temperature. There
is no a priori reason why the temperature should be fixed in such a process, which is precisely what   \cite{Hollands:2024vbe} does, hence from this point of view they excludes a physical class of metric perturbations. However, for near-equilibrium variations the change in temperature should \emph{not} enter into the first law, i.e. $dE = T dS$  should still hold.  We show explicitly that $\delta \kappa$ indeed drops out in the dynamical first law   for any diffeomorphism invariant theory of gravity, so that the first law reads $\delta M = T \delta S_{\text{dyn}} $ (and does not contain an additional term $S_{\text{dyn}} \delta T$), which is an important consistency check.
 
 \item[c)] We derive the non-stationary first law for arbitrary theories of gravity non-minimally coupled to any  bosonic  matter field (whereas in \cite{Hollands:2024vbe} the metric is the only dynamical field)  that are smooth on the entire horizon. Notably, we derive the dynamical black hole entropy in this case without imposing  additional restrictions on the perturbations of the matter fields.
 One caveat though is that for minimally coupled gauge fields whose pullback is   smooth on the  horizon, there are no additional charge terms (of the form $\Phi \delta Q$)  in the black hole first law,  since the electric potential $\Phi \equiv -\xi \cdot A|_{\mathcal H} $ vanishes on the horizon  if the gauge field is stationary and smooth on the horizon.  We plan to treat these  charge terms  in the dynamical first law   in a forthcoming paper \cite{VisserYan}.

 \item[d)] We evaluate the improved Noether charge explicitly for a generic  $f$(Riemann) theory using Gaussian null coordinates and show that it equals the dynamical entropy formula  in \cite{Hollands:2024vbe}  (whereas in \cite{Hollands:2024vbe}  they only do this for  the quadratic gravity  Lagrangian $\mathbf L=\bm \epsilon R_{ab}R^{ab}$).  This lengthy calculation makes extensive use of the boost weight analysis, and seems less accessible in the tetrad approach of \cite{Hollands:2024vbe}. 
 \end{itemize}


Finally, we comment that the  dynamical entropy for arbitrary higher curvature   theories that satisfies the black hole first law does not necessarily hold  in a fully non-perturbative regime outside the regime of validity of effective field theory,  in which case black hole entropy  might not even be well defined or it might be ambiguous.  Throughout this paper  we will consider only   perturbations off a stationary background to first order in perturbation theory (whereas in \cite{Hollands:2024vbe} they also consider second-order perturbations). Interestingly, this geometric setup  is  a sweet spot where the black hole entropy and energy can   be  defined in an unambiguous manner using   the background  Killing time, while the resulting expression for dynamical entropy  is nontrivial and different from the known entropy functionals for black holes discussed above (even for general relativity!). Further, in this setup  the black hole temperature $T$ in the first law \eqref{firstlaw1} is still given by the Hawking temperature \eqref{hawkingtemp}, since it is evaluated on the background Killing horizon. 

\subsection{Dynamical Black Hole Entropy} 
 Let us now review the  results in \cite{Hollands:2024vbe} for the  non-stationary first law and the dynamical black hole entropy that satisfies this first law. 
 The non-stationary first law for arbitrary horizon cross-sections still takes the form \eqref{firstlaw1}, where the mass and angular momentum terms are unchanged, but the   entropy $S $ is no longer equal to the Bekenstein-Hawking entropy for general relativity, or to the Wall entropy for higher curvature gravity. In fact, the dynamical black hole entropy differs from these entropy functionals on arbitrary horizon slices by a dynamical correction term.      For general relativity,   the   dynamical black hole entropy   was defined in \cite{Hollands:2024vbe} as 
\begin{equation}
\label{dynamicalintro}
    S_{\text{dyn}} =\left ( 1- v \dv{v}\right) S_{\text{BH}} \,,
\end{equation}
where $v$ is the affine null parameter along the future horizon, ranging from $v=0$ at $\mathcal B$ to $v=\infty$ at future infinity. The derivative term is the dynamical correction term which   vanishes on a Killing horizon, since the area is constant along the horizon, and it is zero on the bifurcation surface~$\mathcal B$. Note that the product $v \dv*{v}$ is   invariant under the gauge transformation that rescales the affine parameter along each horizon generator, $v \to f(x^i) v,$ where $x^i$ are the codimension-2, spatial coordinates on the horizon.  It is also noteworthy that $ S_{\text{dyn}} $ is smaller than $S_{\text{BH}}$, if the derivative of the horizon area $\dd A / \dd  v $ with respect to affine time is positive, which follows from the area theorem. This suggests that the entropy \eqref{dynamicalintro} is associated to an area of a  surface inside the event horizon. Indeed, as shown in \cite{Hollands:2024vbe}, the dynamical black hole entropy is equal to the Bekenstein-Hawking entropy of the apparent horizon to linear order in   perturbation theory. We also provide a pedagogical proof of this claim in the present paper. 

Hollands, Wald and Zhang showed that the dynamical black hole entropy   satisfies both a non-stationary comparison version  and a physical process version of the first law. The comparison first law \eqref{firstlaw1}  compares two vacuum black hole geometries and  holds for arbitrary cross-sections, hence the entropy is the same at all horizon cross-sections, i.e. $S_{\text{dyn}} [\mathcal C(v_1)]= S_{\text{dyn}}[\mathcal C(v_2)]$. In particular, the dynamical entropy at a cross-section $\mathcal C(v)$ is equal to the Bekenstein-Hawking entropy at the bifurcation surface, i.e. $S_{\text{dyn}}[\mathcal C(v)] = S_{\text{BH}}[\mathcal B]$, since  $S_{\text{dyn}}$ reduces to $S_{\text{BH}}$ at $\mathcal B$. Thus, the dynamical black hole entropy is constant to first order for   perturbations that only change the metric.

 This will change when the perturbation changes the stress-energy tensor, i.e., $\delta T_{ab} \neq 0$, as is the case for the physical process first law, which   reads
\begin{equation}
\label{physicalintroo}
    \Delta \delta M - \Omega_{\mathcal H} \Delta \delta J = T \Delta \delta S_{\text{dyn}}\,.
\end{equation}
Here $\Delta$ stands for the difference between two horizon cuts $\mathcal C(v_1)$ and $\mathcal C(v_2)$, whereas $\delta$ denotes a linear perturbation around the stationary background. $\Delta \delta M$ and $\Delta \delta J $ stand for the change in mass and angular momentum of the matter passing through the horizon.
An immediate consequence of the physical process first law is that the dynamical black hole entropy \eqref{dynamicalintro} obeys the second law to first order in perturbation theory \cite{Hollands:2024vbe}. This can be seen as follows. For  linearised perturbations that are sourced by external matter, described by a   stress-energy tensor $\delta T_{ab}$, the variation of the mass and angular momentum of the matter are related to the matter Killing energy flux $\Delta \delta E$ through the horizon between affine times $v_1$ and $v_2$
\begin{equation}
\label{matterkillingee}
   \Delta \delta M - \Omega_{\mathcal H} \Delta \delta J  =  \int_{v_1}^{v_2} \dd{v} \int_{\mathcal C(v)} \dd{A}  \kappa \,v\, \delta T_{vv}=\Delta \delta E\,.
\end{equation}
Combining this with the first law \eqref{firstlaw1} yields
\begin{equation}
    T \Delta \delta S_{\text{dyn}} =  \int_{v_1}^{v_2} \dd{v} \int_{\mathcal C(v)} \dd{A} \kappa \,v \, \delta T_{vv}\,.
\end{equation}
If the stress-energy variation satisfies the null energy condition, $\delta T_{vv} \ge 0$ then  it follows  immediately from  \eqref{firstlaw1} that the dynamical entropy is non-decreasing: 
\begin{equation}
    \Delta \delta S_{\text{dyn}}\ge0.
\end{equation}
This signals an important difference between the dynamical black hole entropy and Bekenstein-Hawking entropy, as noted in \cite{Hollands:2024vbe}. On the one hand, $S_{\text{BH}}$ (the horizon area) already changes before matter crosses the horizon due to the teleological definition  of the event horizon.  On the other hand, the linearised second law implies that $S_{\text{dyn}}$ only gets modified when matter crosses between $v_1$ and~$v_2. $ If   matter crosses the horizon at a later stage, for $v_3 > v_2$, then the dynamical entropy remains the same between $v_1$ to $v_2.$ This is  because the change in the dynamical correction term $- v \dv*{S_{\text{BH}}}{v}$ precisely cancels  against the change in  $S_{\text{BH}}$ such that the dynamical entropy is unchanged. 
Thus, dynamical black hole entropy only changes locally in affine time.

\begin{table}[t]
    \centering
    \begin{tabular}{l|c|c|c}
    \hline
         Dynamical Black Hole Entropy Proposals &  $S_\text{Iyer-Wald}$ & $S_\text{Wall}$ & $S_\text{dyn}$\\
         \hline
         CFL between $\mathcal B$ and $\mathcal S_{\infty}$ & yes & yes & yes\\
         CFL between $\mathcal C(v)$ and $\mathcal S_{\infty}$ & no & yes  & yes\\
         PPFL between $\mathcal B$ and $i^+$ & yes & yes & yes\\
         PPFL between $\mathcal C(v_1)$ and $\mathcal C(v_2)$ & no & no & yes\\
         Linearised Second Law & no & yes & yes\\
         \hline
    \end{tabular}
    \caption{Proposals for dynamical black hole entropy. Here we compare the Iyer-Wald entropy~\eqref{iyerwald2}, Wall entropy \eqref{wallentropy} and dynamical black hole entropy \eqref{dynwall} and state whether they satisfy the first laws and linearised second law. The perturbations are first-order and non-stationary by default. CFL stands for comparison version of first law and PPFL means physical process version of the first law. We have also used the notation  $\mathcal B$ for the bifurcation surface, $\mathcal S_{\infty}$ for spatial infinity, $i^+$ for future  timelike infinity, $\mathcal C(v)$ for the horizon cross-section at affine  time~$v$.  The CFL has not been proven for the Wall entropy, but can be deduced from the fact that $S_\text{Wall}$ is constant on the horizon for vacuum perturbations, and at $\mathcal B$ equals the Wald entropy. In fact, at $\mathcal B$ all proposals agree and coincide with the Wald entropy: $S_\text{Iyer-Wald}\overset{\mathcal B}=S_\text{Wall}\overset{\mathcal B}=S_\text{dyn}\overset{\mathcal B}= S_{\text{Wald}}$.  For general relativity, $S_\text{Iyer-Wald}$ and $S_\text{Wall}$ are equal to the Bekenstein-Hawking entropy \eqref{bekensteinhaw}, whereas $S_\text{dyn}$ is given by   \eqref{dynamicalintro}. }
    \label{tb:diff-dyn-ent}
\end{table}

Furthermore, the dynamical black hole entropy was generalised in \cite{Hollands:2024vbe} to higher curvature gravity, for which it was shown that
\begin{equation}
\label{dynwall}
    S_{\text{dyn}} =\left ( 1- v \dv{v}\right) S_{\text{Wall}}   \,.
\end{equation}
We provide a non-trivial check of  this formula by an explicit calculation    using Gaussian null coordinates for $f(\text{Riemann})$ theories, for which the Wall entropy is given by \eqref{wallentropy}. More generally,  the defining relation of the Wall entropy is  that its second $v$-derivative is related to the $vv$-component of the gravitational field equations as \cite{Wall:2024lbd}
  \begin{equation}
        \partial_v^2 \delta S_{\text{Wall}} = - 2 \pi \int_{\mathcal{C}(v)} \dd{A} \delta E_{vv}\,.
    \end{equation}
The physical process first law follows straightforwardly from this defining equation (see also \cite{Kar:2024dqk} for a similar  discussion) by multiplying it on both sides with $-\kappa v$, and then integrating over the affine parameter from $v_1$ to~$v_2$  
\begin{equation}
\label{intfwall1}
   \kappa \int_{v_1}^{v_2} \dd{v} (- v \partial_v^2 \delta S_{\text{Wall}}) = 2 \pi  \int_{v_1}^{v_2} \dd{v} \int_{\mathcal C(v)} \dd{A} \kappa  v \delta T_{vv} \,,
\end{equation}
where we used the linearised gravitational field equation $\delta E_{vv} = \delta T_{vv}$. The right-hand side of this equation is $2\pi$ times the matter Killing energy flux \eqref{matterkillingee}, and we may rewrite the left-hand side using
\begin{equation}
\label{intfwall3}
    - v \partial_v^2 \delta S_{\text{Wall}} = \partial_v \left ( (1 - v \partial_v) \delta S_{\text{Wall}}\right)= \partial_v \delta S_{\text{dyn}}\,.
\end{equation}
Thus, the physical process first law in higher curvature gravity follows from combining   \eqref{matterkillingee}, \eqref{intfwall1} and \eqref{intfwall3}, and identifying $T= \kappa / 2\pi$, i.e., 
\begin{equation}
\label{physicalintroo2}
\Delta  \delta M - \Omega_{\mathcal H} \Delta \delta J= T \Delta \delta \left ( S_{\text{Wall}} -    v \dv{v} S_{\text{Wall}} \right)\,, 
\end{equation}
   The linearised second law continues to hold for   dynamical entropy in higher curvature gravity, since combining \eqref{intfwall1} and \eqref{intfwall3} yields 
\begin{equation}
        \partial_v  \delta S_{\text{dyn}} =  2 \pi \int_{\mathcal{C}(v)} \dd{A} v \delta T_{vv}.
    \end{equation}
If the perturbation is sourced by a stress-energy tensor that satisfies the null energy condition, $\delta T_{vv}\ge 0$,  then $S_{\text{dyn}} $ satisfies a  linearised second law  
\begin{equation}
     \partial_v  \delta S_{\text{dyn}} \ge 0.
\end{equation}
Finally, in  Table \ref{tb:diff-dyn-ent} we compare three different proposals for dynamical black hole entropy and indicate whether they obey the comparison first law, physical process first law and linearised second law.

\subsection{Methodology: Raychaudhuriology and Noetherology}
Hollands, Wald and Zhang \cite{Hollands:2024vbe} established the non-stationary first law for arbitrary horizon cross-sections by using the Noether charge method. Below we review this derivation and explain in more detail how we generalise and improve upon this derivation in the present work. However, in this article we also present a different, pedagogical derivation of the physical process version of the first law for general relativity, that was not given in \cite{Hollands:2024vbe}. That is, we derive the physical process first law 
by integrating the linearised Raychaudhuri equation between two arbitrary horizon cross-sections. For perturbations off a stationary black hole that satisfy the linearised Einstein equation,  the Raychaudhuri equation reads to first order
\begin{equation}
    \dv{\delta \theta_v}{v} =   -8 \pi G \delta T_{vv}   \,,
\end{equation}
where $\theta_v = (1/\dd{A}) \partial_v \dd{A}$ is the outgoing null expansion of the future horizon. Since the matter Killing energy variation is given by \eqref{matterkillingee}, we multiply on both sides by $\kappa v$ and integrate over the horizon between two cross-sections $\mathcal C(v_1)$ and $\mathcal C(v_2)$, 
\begin{equation}
    -\kappa \int_{v_1}^{v_2} \dd{v}\int_{\mathcal C(v)} \dd{A} v \frac{\dd \delta \theta_v}{\dd v} =   8\pi G \int_{v_1}^{v_2} \dd{v}\int_{\mathcal C(v)} \dd{A} \kappa v \delta T_{vv} \,.
\end{equation}
Then, the right side    is equal to the matter Killing energy flux $\Delta \delta E$ and the left side can be computed by integrating by parts, which gives rise to a boundary term and the integral of $\delta \theta_v$ along the horizon, which  is equal to the horizon area change. Usually the boundary term   is set to zero, because it is assumed that the perturbed black hole starts and ends in  a stationary state, at $v=0$ and $v=+\infty$, respectively. However, when we integrate between two intermediate cross-sections at $v_1$ and $v_2$ the boundary term is nonzero and proportional to 
\begin{equation}
\label{boundarytermdyn}
  - \int_{\mathcal C(v)} \dd{A} (v \delta \theta_v)\Big |_{v_1}^{v_2} = -\Delta \delta \int_{\mathcal C(v)} \dd{A} v \theta_v = - \Delta \delta \left( v \frac{\dd }{\dd v} A\right) \,.
\end{equation}
Note in the second equality we are allowed to pull the $\delta$ through, since $\theta_v =0$ on the background Killing horizon. \eqref{boundarytermdyn} is precisely the dynamical correction term to the Bekenstein-Hawking entropy. As we will show in more detail in the paper, after combining all the different terms, one obtains the physical process first law \eqref{physicalintroo}, where the dynamical black hole entropy obeys~\eqref{dynamicalintro}. 
 Thus,  we see that the dynamical black hole entropy in general relativity follows in a straightforward way  from Raychaudhuriology. 

Now, in \cite{Hollands:2024vbe} Hollands, Wald and Zhang derived the physical process and comparison version of the first law  for general diffeomorphism covariant Lagrangians that only depend on the metric  by applying the
 Noether charge method  \cite{Wald:1993nt,Iyer:1994ys}  to non-stationary perturbations. We generalise their  derivation to the case where the metric is non-minimally coupled to arbitrary bosonic matter fields  that are smooth on the horizon. Moreover, we allow for non-vanishing variations of the horizon generating Killing field and surface gravity, i.e., $\delta \xi^a \neq 0$ and $\delta \kappa \neq 0 $, and show that the first law is independent of these variations. Below, we summarise the derivation of the non-stationary comparison first law. The main steps of the derivation are the same as in \cite{Hollands:2024vbe}, such as \eqref{exactnessonhorizon} and \eqref{imprnoethch1}, but some of the technicalities are different due to our less restrictive gauge conditions on the perturbation, especially \eqref{variationofQonhorizon}.

We employ  the fundamental variational identity \eqref{eq:q-theta-c} for  general perturbations  away from a stationary background with Killing field~$\xi^a$,
    \begin{equation}
        \dd{(\delta_\phi \mathbf Q_\xi - \xi \cdot  \bm \Theta(\phi, \delta \phi))} = 0, \label{eq:d-Qimprov-zero}
    \end{equation} 
    where we assumed the background   equations of motion  for the dynamical fields, collectively denoted as $\phi$, and the linearised constraint equations are satisfied.
    This identity holds    in particular  also  for non-stationary perturbations, for which $\delta (\mathcal L_\xi g_{ab})\neq 0 $. Here, $\mathbf Q_\xi$ is the Noether charge codimension-2 form with respect to $\xi$,   $\bm \Theta(\phi,\delta\phi)$ is the symplectic potential codimension-1 form,   $\delta_\phi$ denotes a field variation that does not act on the vector field $\xi$, so that $\delta_\phi \mathbf Q_\xi = \delta \mathbf Q_\xi -   \mathbf Q_{\delta \xi}$, and $\delta$ acts both on the dynamical fields and on the vector field. 
    
    Next, we integrate the variational identify over a spacelike hypersurface with  a single asymptotic boundary and a compact interior boundary at the horizon. Because of Stokes' theorem, the boundary integral at a codimension-2 sphere at infinity $\mathcal S_{\infty}$  is equal to the boundary integral at a   cross-section $\mathcal C$  of the horizon
    \begin{equation}
        \int_{\mathcal S_{\infty}}  (\delta_\phi \mathbf Q_\xi - \xi \cdot  \bm \Theta(\phi, \delta \phi))= \int_{\mathcal C} (\delta_\phi \mathbf Q_\xi - \xi \cdot  \bm \Theta(\phi, \delta \phi))\,. \label{eq:stokes1}
    \end{equation}
    At asymptotic infinity we identify 
      \begin{equation}
      \label{dmomegajnoether}
            \int_{\mathcal S_\infty} (\delta_\phi \mathbf Q_\xi - \xi \cdot  \bm \Theta(\phi, \delta \phi)) = \delta M - \Omega_{\mathcal H} \delta J\,,
    \end{equation}
where $M$ and $J$ are the mass and angular momentum of the black hole. The mass is well defined if   a codimension-1 form $\mathbf B_\infty$ exists at a timelike codimension-1 hypersurface whose radial coordinate $r$ tends to infinity, such that
\begin{equation}
\label{Batinfinty}
    \bm \Theta \overset{r\to \infty}{=} \delta \mathbf B_\infty\,.
\end{equation}
We assume the Killing field  is normalised at infinity as $\xi^a = (\partial_t)^a + \Omega_{\mathcal H} (\partial_{\vartheta})^a$, and we keep the time translation    and   rotational Killing fields fixed at asymptotic infinity, $\delta (\partial_t)^a=\delta (\partial_{\vartheta})^a=0.$
Then one may define the   mass and angular momentum as \cite{Wald:1993nt}
 \begin{align}
     M &=  \int_{\mathcal S_\infty}  (\mathbf Q_{\partial_t}- \partial_t \cdot \mathbf B_\infty ) \,,\\
     J &=  - \int_{\mathcal S_\infty} \mathbf Q_{\partial_{\vartheta}}\,.
 \end{align}
     Note that $\partial_{\vartheta} \cdot \mathbf B_\infty$ does not appear in the formula for $J$, because $\partial_{\vartheta}$ is tangent to $S_\infty$. For asymptotically flat solutions to general relativity, Iyer and Wald      \cite{Iyer:1994ys}  showed     that a $\mathbf B_\infty$ exists that satisfies \eqref{Batinfinty}, and they recovered the correct  ADM mass  and angular momentum.

Further, for the interior boundary integral,  Wald  \cite{Wald:1993nt}    evaluated  it at the bifurcation surface~$\mathcal B$, where $\xi^a =0$, hence the term $\xi \cdot  \bm \Theta$ vanishes. He   identified the black hole entropy with the Noether charge integrated over the bifurcation surface
\begin{equation}
\label{noetherwaldb}
      \int_{\mathcal B}  \delta_\phi \mathbf Q_\xi = \frac{\kappa}{2\pi} \delta S_{\text{Wald}}\,.
\end{equation}
  Subsequently, Iyer and Wald \cite{Iyer:1994ys}  showed  that the Noether charge integral yields the Wald entropy formula \eqref{iyerwald} for arbitrary diffeomorphism covariant theories of gravity. The first law of black hole mechanics for such a general theory of gravity then follows from inserting    \eqref{dmomegajnoether} and \eqref{noetherwaldb} into the variational identity \eqref{eq:stokes1}. 

    Alternatively, for arbitrary horizon cross-sections, the term $\xi \cdot \mathbf \Theta$ does not vanish in general. Therefore,   in that case Hollands, Wald and Zhang \cite{Hollands:2024vbe} proposed that the full  interior integral  should be proportional to the variation of the dynamical  black hole entropy 
    \begin{equation}
        \int_{\mathcal C } (\delta_\phi \mathbf Q_\xi - \xi \cdot  \bm \Theta(\phi, \delta \phi)) = \frac{\kappa}{2 \pi} \delta S_\text{dyn}\,,
    \end{equation}
    where $\kappa$ is the  surface gravity of the background Killing horizon. This identification leads to the desired  non-stationary comparison first law \eqref{firstlaw1} for arbitrary horizon cross-sections in arbitrary diffeomorphism covariant theories. Crucially, the dynamical black hole entropy is only well defined if   there exists a codimension-1 form $\mathbf B_{\mathcal H^+}$ such that \cite{Hollands:2024vbe}
    \begin{equation}
    \label{exactnessonhorizon}
        \bm \Theta  \fheq \delta \mathbf B_{\mathcal H^+}\,, \quad \text{and} \quad \mathbf B_{\mathcal H^+}\fheq 0, 
    \end{equation} 
    where the second equality holds on the background Killing horizon. The two properties \eqref{exactnessonhorizon} of $\mathbf B_{\mathcal H^+}$  together   imply that $\xi \cdot \bm \Theta \cceq  \delta (\xi \cdot \mathbf B_{\mathcal H^+}) \cceq \kappa \delta (\xi \cdot \mathbf B_{\mathcal H^+} / \kappa_3).$    In \cite{Hollands:2024vbe} it was proven that such a $\mathbf B_{\mathcal H^+}$ form  exists using Killing field arguments for general  diffeomorphism covariant Lagrangians for which the metric is the only dynamical field. Alternatively, in this paper we construct   a $\mathbf B_{\mathcal H^+}$ form    assuming  a   fixed Gaussian null coordinates  system near the   horizon,  and using the associated boost weight arguments. Our  proof   of \eqref{exactnessonhorizon} holds for any diffeomorphism covariant Lagrangians that depends on the metric and arbitrary non-minimally coupled bosonic matter fields that are smooth on the horizon (see Section \ref{sec:exactness}), and thus is more general than   in \cite{Hollands:2024vbe}.
    
    Further, in order for $S_\text{dyn}$ to be well defined, we require 
   \begin{equation}
   \label{variationofQonhorizon}
       \delta_\phi  \mathbf Q_\xi  \overset{\mathcal C}{=} \kappa \delta  (\mathbf Q_\xi / \kappa_3)\,.
   \end{equation}
 Here $\kappa_3  $ is the surface gravity  defined as $\kappa_3^2\fheq - \frac{1}{2}(\nabla^a \xi^b )  ( \nabla_{[a} \xi_{b]})   $ on the perturbed  horizon. For a Killing horizon different definitions of surface gravity all coincide, but for a perturbed Killing horizon  they differ from each other, hence we need to specify which surface gravity we refer to   (see Section \ref{ssec:gauge-cond}). We prove \eqref{variationofQonhorizon}   in Section \ref{sec:structural} using  fixed GNC   for general theories of gravity. The condition \eqref{variationofQonhorizon} does not appear in \cite{Hollands:2024vbe} since they keep the surface gravity fixed. 
    
    Finally, it follows from \eqref{exactnessonhorizon} and \eqref{variationofQonhorizon}  that  the dynamical black hole entropy can  be defined as the \emph{improved Noether charge} $\tilde{\mathbf Q}_\xi$   \cite{Wald:1999wa,Harlow:2019yfa,Freidel:2020svx,Freidel:2020ayo, Shi:2020csw,Chandrasekaran:2020wwn,Freidel:2021cjp}, i.e.,   \cite{Hollands:2024vbe}
    \begin{equation}
    \label{imprnoethch1}
        S_{\text{dyn}} = \int_{\mathcal C} \frac{2\pi}{\kappa_3} \tilde{\mathbf {Q}}_\xi= \int_{\mathcal C}  \frac{2\pi}{\kappa_3}(\mathbf {Q_\xi} - \xi \cdot \mathbf B_{\mathcal H^+})\,.
    \end{equation}
Note that at the bifurcation surface, where $\xi^a =0$,  and   on a Killing horizon, where     $\mathbf B_{\mathcal H^+} \fheq 0$, the dynamical black hole entropy reduces to    Wald's definition of black hole entropy as Noether charge. But for arbitrary cross-sections and non-stationary black holes, the term $\xi \cdot \mathbf B_{\mathcal H^+}$ gives a  dynamical correction to Wald's definition. In \cite{Hollands:2024vbe} it was shown that the improved Noether charge is related to the Wall entropy by \eqref{dynwall} using a covariant definition of the Wall entropy current in general theories of gravity.    Instead, in this paper   we compute the improved Noether charge explicitly for some examples using GNC, and derive  the dynamical entropy  formula~\eqref{dynamicalintro} for general relativity and its generalisation \eqref{dynwall} for    $f(\text{Riemann})$ theories (see Section~\ref{sec:examples}). Moreover, since the improved Noether  charge is invariant under the JKM ambiguities to leading order in the perturbation (see  \cite{Hollands:2024vbe} for a proof, and also Section \ref{sec:jkm-inv}),  Noetherology thus yields a unique dynamical black hole entropy for first-order perturbations off a stationary background. 

\subsection{Plan of the Paper}
The paper is organised as follows.  In Section \ref{sec:geometricsetup} we introduce our geometric setup in more detail, describe our gauge conditions for the perturbation,  and review Gaussian null coordinates  in   affine parameterisation.  Further, in Section \ref{sec:dynrayc} we derive the physical process first law for non-stationary perturbations and arbitrary horizon cross-sections from the Raychaudhuri equation for null geodesic congruences. In Section \ref{sec:covphas} we explain how a comparison version and physical process version of the first law can be derived for arbitrary diffeomorphism covariant theories of gravity using the Noether charge formalism.  In the final Section \ref{sec:examples} we explicitly compute the dynamical  entropy for three examples: general relativity, $f(R)$ gravity and  $f(\text{Riemann})$ theories.  Finally,   Appendix \ref{app:surf-grav} contains technical details about the   definitions of surface gravity on a  null hypersurface, and in Appendix \ref{appGNC} we compute the connection coefficients and covariant derivatives on the horizon in Gaussian null coordinates.

Our conventions mainly follow those in Wald's textbook \cite{Wald:1984rg}. We assume a mostly positive signature metric, $D$ is the number of spacetime dimensions,  Latin indices $a,b,c,\dots$ denote abstract spacetime indices, $i,j,k, \dots$ denote codimension-2 spatial indices, $(v,u,x^i)$ label the Gaussian null coordinates near the future event horizon, and we use  boldface notation for differential forms. The orientation of the volume form is chosen to be $\bm \epsilon = \dd{u} \wedge \dd{v} \wedge \bm \epsilon_{\mathcal C}$ near the horizon, where $\bm \epsilon_{\mathcal C}$ is the spatial codimension-2 volume (area) form, see equation \eqref{eq:orientation}. We set $\hbar = c=1$ in the entire paper, but keep Newton's constant $G$ explicit.

    \section{Geometric Setup}
\label{sec:geometricsetup}

    In this section we introduce the geometric setup of the paper (Section \ref{sec2.1}), we impose gauge conditions on the perturbations at the horizon (Section \ref{ssec:gauge-cond}), and we review   the   Gaussian null coordinates   based on an affine parameterisation of the null geodesics on the future event horizon (Section \ref{sec:gnc}).

    \subsection{Stationary Black Hole Background Geometry}
\label{sec2.1}
    
   Consider a  stationary black hole  background geometry  $(\mathcal{M},  g)$ in $D$ spacetime dimensions. We take the event horizon of the stationary black hole to be   a bifurcate Killing horizon $\mathcal H$, and   label the future horizon by $\mathcal H^+$, the past horizon by $\mathcal H^-$ and the codimension-2 bifurcation surface by $\mathcal B.$ The Killing field that is normal to the Killing horizon is denoted by $\xi^a$. We assume   $\xi^a$ is a Killing symmetry of all the background dynamical fields, including the    metric $g$ and matter fields $\varphi$, 
    \begin{equation}
        \mathcal L_\xi g = 0, \quad \mathcal L_\xi \varphi = 0.
    \end{equation}
    
    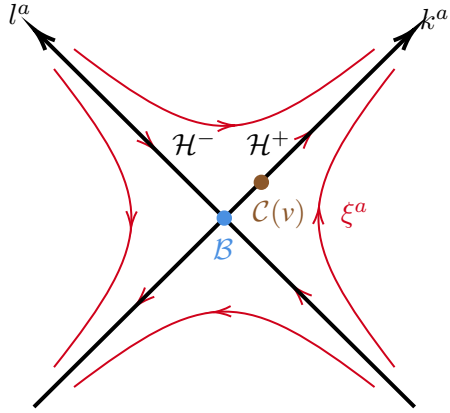
\begin{figure}[t]
        \centering

\tikzset{every picture/.style={line width=0.75pt}} 

\begin{tikzpicture}[x=0.75pt,y=0.75pt,yscale=-1,xscale=1]

\draw [color={rgb, 255:red, 208; green, 2; blue, 27 }  ,draw opacity=1 ]   (240,70) .. controls (188.85,135.77) and (189.82,154.23) .. (240,220) ;
\draw [shift={(202.28,139.31)}, rotate = 92.42] [color={rgb, 255:red, 208; green, 2; blue, 27 }  ,draw opacity=1 ][line width=0.75]    (8.74,-2.63) .. controls (5.56,-1.12) and (2.65,-0.24) .. (0,0) .. controls (2.65,0.24) and (5.56,1.12) .. (8.74,2.63)   ;
\draw [color={rgb, 255:red, 208; green, 2; blue, 27 }  ,draw opacity=1 ]   (111.41,188.59) -- (198.59,101.41) ;
\draw [shift={(200,100)}, rotate = 135] [color={rgb, 255:red, 208; green, 2; blue, 27 }  ,draw opacity=1 ][line width=0.75]    (10.93,-3.29) .. controls (6.95,-1.4) and (3.31,-0.3) .. (0,0) .. controls (3.31,0.3) and (6.95,1.4) .. (10.93,3.29)   ;
\draw [shift={(110,190)}, rotate = 315] [color={rgb, 255:red, 208; green, 2; blue, 27 }  ,draw opacity=1 ][line width=0.75]    (10.93,-3.29) .. controls (6.95,-1.4) and (3.31,-0.3) .. (0,0) .. controls (3.31,0.3) and (6.95,1.4) .. (10.93,3.29)   ;
\draw [color={rgb, 255:red, 208; green, 2; blue, 27 }  ,draw opacity=1 ]   (121.41,111.41) -- (188.59,178.59) ;
\draw [shift={(188.59,178.59)}, rotate = 45] [color={rgb, 255:red, 208; green, 2; blue, 27 }  ,draw opacity=1 ][line width=0.75]    (10.93,-3.29) .. controls (6.95,-1.4) and (3.31,-0.3) .. (0,0) .. controls (3.31,0.3) and (6.95,1.4) .. (10.93,3.29)   ;
\draw [shift={(121.41,111.41)}, rotate = 225] [color={rgb, 255:red, 208; green, 2; blue, 27 }  ,draw opacity=1 ][line width=0.75]    (10.93,-3.29) .. controls (6.95,-1.4) and (3.31,-0.3) .. (0,0) .. controls (3.31,0.3) and (6.95,1.4) .. (10.93,3.29)   ;
\draw [line width=1.5]    (60,240) -- (247.88,52.12) ;
\draw [shift={(250,50)}, rotate = 135] [color={rgb, 255:red, 0; green, 0; blue, 0 }  ][line width=1.5]    (11.37,-3.42) .. controls (7.23,-1.45) and (3.44,-0.31) .. (0,0) .. controls (3.44,0.31) and (7.23,1.45) .. (11.37,3.42)   ;
\draw [line width=1.5]    (62.12,52.12) -- (250,240) ;
\draw [shift={(60,50)}, rotate = 45] [color={rgb, 255:red, 0; green, 0; blue, 0 }  ][line width=1.5]    (11.37,-3.42) .. controls (7.23,-1.45) and (3.44,-0.31) .. (0,0) .. controls (3.44,0.31) and (7.23,1.45) .. (11.37,3.42)   ;
\draw [color={rgb, 255:red, 74; green, 144; blue, 226 }  ,draw opacity=1 ]   (155,145) ;
\draw [shift={(155,145)}, rotate = 0] [color={rgb, 255:red, 74; green, 144; blue, 226 }  ,draw opacity=1 ][fill={rgb, 255:red, 74; green, 144; blue, 226 }  ,fill opacity=1 ][line width=0.75]      (0, 0) circle [x radius= 3.35, y radius= 3.35]   ;
\draw [color={rgb, 255:red, 208; green, 2; blue, 27 }  ,draw opacity=1 ]   (70,70) .. controls (121.52,136.13) and (121.52,154.6) .. (70,220) ;
\draw [shift={(108.39,150.33)}, rotate = 272.1] [color={rgb, 255:red, 208; green, 2; blue, 27 }  ,draw opacity=1 ][line width=0.75]    (8.74,-2.63) .. controls (5.56,-1.12) and (2.65,-0.24) .. (0,0) .. controls (2.65,0.24) and (5.56,1.12) .. (8.74,2.63)   ;
\draw [color={rgb, 255:red, 208; green, 2; blue, 27 }  ,draw opacity=1 ]   (80,230) .. controls (146.13,179.21) and (164.6,179.58) .. (230,230) ;
\draw [shift={(149.58,192.34)}, rotate = 357.03] [color={rgb, 255:red, 208; green, 2; blue, 27 }  ,draw opacity=1 ][line width=0.75]    (8.74,-2.63) .. controls (5.56,-1.12) and (2.65,-0.24) .. (0,0) .. controls (2.65,0.24) and (5.56,1.12) .. (8.74,2.63)   ;
\draw [color={rgb, 255:red, 208; green, 2; blue, 27 }  ,draw opacity=1 ]   (80,60) .. controls (146.13,111.52) and (164.6,111.52) .. (230,60) ;
\draw [shift={(160.33,98.39)}, rotate = 177.9] [color={rgb, 255:red, 208; green, 2; blue, 27 }  ,draw opacity=1 ][line width=0.75]    (8.74,-2.63) .. controls (5.56,-1.12) and (2.65,-0.24) .. (0,0) .. controls (2.65,0.24) and (5.56,1.12) .. (8.74,2.63)   ;
\draw    (80,140) ;
\draw [color={rgb, 255:red, 139; green, 87; blue, 42 }  ,draw opacity=1 ]   (173.5,127) ;
\draw [shift={(173.5,127)}, rotate = 0] [color={rgb, 255:red, 139; green, 87; blue, 42 }  ,draw opacity=1 ][fill={rgb, 255:red, 139; green, 87; blue, 42 }  ,fill opacity=1 ][line width=0.75]      (0, 0) circle [x radius= 3.35, y radius= 3.35]   ;

\draw (260,45) node    {$k^a$};
\draw (53,42.5) node    {$l^a$};
\draw (177,107) node    {$\mathcal{H}^+$};
\draw (140,107) node    {$\mathcal{H}^-$};
\draw (154.5,160.5) node  [color={rgb, 255:red, 74; green, 144; blue, 226 }  ,opacity=1 ]  {$\mathcal{B}$};
\draw (220,143.5) node  [color={rgb, 255:red, 208; green, 2; blue, 27 }  ,opacity=1 ]  {$\xi ^{a}  $};
\draw (182.5,143) node  [color={rgb, 255:red, 139; green, 87; blue, 42 }  ,opacity=1 ]  {$\mathcal{C}( v)$};

\end{tikzpicture}

        \caption{Bifurcate Killing horizon $\mathcal H$. The horizon is comprised of two null surfaces $\mathcal H^+$ at $u=0$, the future horizon, and $\mathcal H^-$ at $v=0$, the past horizon, that intersect at the  bifurcation surface~$\mathcal B.$ The horizon generating Killing field is denoted by $\xi^a$, and   $k^a=(\partial_v)^a$ and $l^a=(\partial_u)^a$ are the (future-directed)  tangents to the   affinely parameterised   geodesics of   $\mathcal H^+$ and $\mathcal H^-$, respectively, where $v$ and $u$ are  the   affine parameters. $\mathcal C(v)$ labels a cross-section of $\mathcal H^+$ at affine  time~$v$. } 
        \label{fig:bkg}
    \end{figure}
    
\noindent  Note that for matter fields with gauge symmetry, we choose a gauge such that the Killing equation above holds.\footnote{See Section \ref{ssec:cps} for a further justification of this gauge choice.} 
Below we describe some kinematic properties of bifurcate Killing horizons, for which  we will derive a dynamical non-stationary first law in the following sections.   Similar descriptions can be found for instance in \cite{Carter:2009nex,Carter1987,Compere:2006my,Poisson:2009pwt}. We   are mainly interested   in the part of the horizon $\mathcal H^+$ that lies to the future of $\mathcal B$, with cross-sections   labelled by $\mathcal C$, since we will be deriving  a dynamical black hole entropy associated to (perturbations of) these cross-sections.

We wish to decompose the background metric on $\mathcal H^+$  using the outgoing and ingoing null vector fields on $\mathcal H^+$, $k^a$ and~$l^a$, respectively. Together ($k^a,l^a$) form a  null zweibein basis, that is Lie transported by $k^a$, for a two-dimensional  subspace of the tangent space of  $\mathcal H^+$. To construct the tangent null vector fields we employ an affine parameterisation for the null geodesic generators of the horizon, instead of a Killing parameterisation.   The affine parameterisation is more convenient to describe the perturbation that we will introduce momentarily, since we want to keep the zweibein  ($k^a, l^a$) fixed when we perturb the geometry,  while at the same time we want to impose the gauge condition $\delta \xi^a \neq 0 $ for the Killing field. In terms of the Killing parameterisation, the condition $\delta \xi^a \neq 0$ is incompatible with a fixed zweibein basis, if the Killing field is part of this basis. In terms of the affine parameterisation, on the other hand, this is not a problem, because, as we will explain, fixing $k^a$ does not fix $\xi^a$ if  their proportionality constant (i.e., the surface gravity)   on the horizon is allowed to vary.

    The double null decomposition of the metric on $\mathcal H^+$  is carried out as follows: 

     \begin{enumerate}
        \item Let the future horizon $\mathcal{H}^+$ be located at the null hypersurface $\tilde u (x)=0$, for some smooth function $\tilde u (x^a)$.  We define $k_a$ as the normal to $\mathcal H^+$ that satisfies the affinely parameterised geodesic equation
          \begin{equation}
          \label{defk}
            k_a \overset{\mathcal H^+}{\propto}  -  \partial_a \tilde u,  \quad k^b \nabla_b k^a \fheq 0.
        \end{equation}
         The minus sign is chosen in the first expression so that $k^a$ is future directed when $\tilde u$ increases toward the future. 
         From  the first expression   it follows that $k^a$ is hypersurface orthogonal on $\mathcal H^+$ and that the irrotationality condition $k_{[a} \nabla_b k_{c]}\fheq 0$ holds. Since $\mathcal H^+$ is a null hypersurface its normal $k_a$  is null
        \begin{equation} 
        k^ak_a \fheq 0.
        \end{equation}
         Because the normal is orthogonal to itself, it is also tangent to the null generators of~$\mathcal H^+.$ Moreover, the second equality in \eqref{defk} implies $k^a$ is an affinely parameterised tangent, satisfying    
        \begin{equation}
           k^a   \fheq (\partial_v)^a,
        \end{equation}
        where $v$ is the affine   parameter along the null geodesics of $\mathcal H^+$.
         Without loss of generality, we choose $v=0$ at the bifurcate surface $\mathcal{B}$, so that it is positive to the future of $\mathcal B$. This does not fix the affine parameter uniquely, since there is still a scaling freedom $v \to av $, where  $a$ can differ from generator to generator but is constant on each generator of $\mathcal H^+.$  
        \item  To isolate the codimension-2 part of the metric that is transverse to $k^a$, we need to   introduce an auxiliary null vector field $l^a$, that is defined on $\mathcal H^+$ via 
        \begin{equation} \label{defl}
             g_{ab} l^a l^b \fheq 0, \quad l^a k_a \fheq -1 .
        \end{equation}
        The normalisation minus one in the second condition is chosen so that if $k^a$ is tangent to the (future-directed) outgoing null geodesics of $\mathcal H^+$, then $l^a$ is tangent to the (future-directed) ingoing null geodesics of $\mathcal H^+$, hence it 
        is tangent to the past horizon $\mathcal{H}^-$ at  $\mathcal{B}$. Note these conditions \eqref{defl} do  not specify $l^a$ uniquely, because they   are invariant under the transformation 
        $l^a \to {l^a}'=l^a + \frac{1}{2} c_ic^i k^a + c^i m^a_i$, where $c^i$ are arbitrary coefficients (with $i=1,\dots, D-2$) and $m^a_i$ are  spacelike vectors  on $\mathcal H^+$ that are orthogonal to   $k^a$ and~$l^a$, $k_am^a_i = 0=  l_a m^a_i$, and satisfy $g_{ab}m^a_i m^b_{j} = \delta_{ij}$.\footnote{Moreover, it can be shown that the geometries quantities appearing in the Raychaudhuri equation (see equation \eqref{raychau} below), i.e. the outgoing null expansion $\theta_v$ and the square of the shear and rotation, $\sigma^{ab} \sigma_{ab}$ and $\omega^{ab}\omega_{ab}$, also remain invariant under the transformation.}  
        We will shortly describe our choice for $l^a$ that is designed to be compatible with the Killing field $\xi^a$.  
    
        \item Once the choice of $l^a$ is settled, we extend it off the future horizon by solving the affine null geodesic equation 
        \begin{equation} 
        l^b \nabla_b l^a = 0, \label{eq:affinegeodesicl}
        \end{equation}
        which will give the integral curves whose tangent vector is $l^a$. This guarantees that $l^a$ is tangent to the past horizon on the whole of $\mathcal{H}^-$, which is hence located at $v = 0$. It also implies that $l^a$ is null everywhere, $g_{ab} l^a l^b = 0 $, since $l^b \nabla_b (l^a l_a)=0$. 
        
   Since $l^a$ is null and tangent to $\mathcal H^-$, it is also normal to it. Further, we denote the affine null  distance away from the horizon by $u$ and we identify 
    \begin{equation}
     l^a = (\partial_u)^a.
    \end{equation}
    Notice that $u$ and $\tilde u$ may differ away from the horizon, but they agree at the horizon where they  both vanish.  
   
    \item Similarly, we can extend $k^a$ off the future horizon. This is done by keeping the    parameter $v$  fixed along the null geodesics generated by $l^a$, and demanding that $k^a = \left(\partial_v\right)^a$ everywhere.  In other words, we extend $k^a$ such that it commutes with $l^a$, i.e., $[k,l]^a=0$, which is equivalent to the requirement that the Lie derivative of $l^a$ along $k^b$ vanishes, i.e.,  $\mathcal L_k l^a = 0.$
     This means the extension of $k^a$ off the horizon should satisfy
    \begin{equation} \label{klcommute}
        l^b \nabla_b k^a = k^b \nabla_b l^a. 
    \end{equation}
   This  reflects the fact that different choices of $l^a$ on $\mathcal H^+$ determine how we extend $k^a$ off the horizon. 
    The commutativity of $k^a$ and $l^a$ also means the   parameters $(u,v)$ can  act as the null  coordinates near the horizon, whose origin $u=v=0$ is located at the bifurcate surface~$\mathcal{B}$.    Further,  away from the horizon the parameter $v$ is in general not affine, i.e., $k^b \nabla_b k^a \neq 0$,  and  $k^a$ is not   null, $k_a k^b \neq 0.$
   We also note the extension of $k^a$ off the horizon implies the normalisation   $l^a k_a = -1$ holds everywhere, since 
     \begin{equation}
        l^b \nabla_b (l^a k_a) = l^a l^b \nabla_b k_a = l^a k^b \nabla_b l_a = \frac{1}{2} k^b \nabla_b (l^a l_a) = 0\,,
    \end{equation}
    where in  the first equality we used the affine geodesic equation \eqref{eq:affinegeodesicl}, in the second equality we employed  \eqref{klcommute}, and the last equality follows from the fact that $l^a$ is null everywhere.    
    \end{enumerate}
    
 \noindent  From the above construction it follows that the metric can be decomposed on $\mathcal H^+$  as 
    \begin{equation}
    \label{doublenull}
        g_{ab} \fheq  - k_a l_b - l_a k_b + \gamma_{ab}\,,
    \end{equation}
    where $\gamma_{ab}=\gamma_{(ab)}$ is the intrinsic  codimension-2 spatial metric of each cross-section of the future horizon, that is purely transverse, i.e., orthogonal to $k^a$ and $l^a$ on $\mathcal H^+$, $\gamma_{ab}k^a \fheq 0 \fheq \gamma_{ab}l^a$. This double null decomposition does not extend  away from the horizon, as the vector field $k^a$ is not necessarily null off the horizon.

    As we want to use this decomposition to study the bifurcate Killing horizon, below we   demonstrate how the choice of ($k^a,l^a$) can be  made compatible with the horizon generating Killing field $\xi^a$.  We first review how this works on the horizon, and then we construct a $\xi^a$-compatible zweibein off the horizon.  Since $\xi^a$ is normal to $\mathcal H$, it is  tangent to the null geodesic generators of $\mathcal H. $ Along the future horizon it is thus   proportional to   $k^a$,  whereas along the past horizon it is proportional to minus $l^a$, because $\xi^a$ is past directed on $\mathcal H^-$ whereas $l^a$ is future directed. In fact, the precise relation between $\xi^a$ and $k^a$  on $\mathcal H^+$ immediately follows from the   non-affine geodesic equation  obeyed by $\xi^a$ on the future horizon
    \begin{equation}\label{kappa}\xi^b \nabla_b \xi^a \fheq \kappa \xi^a\,,
    \end{equation}   where $\kappa>0$ is the surface gravity of the bifurcate Killing horizon.\footnote{On the past horizon we define the surface gravity via  $\xi^b \nabla_a \xi^a \pheq  - \kappa \xi^a$,  such  that $\kappa$ is also positive on $\mathcal H^-.$} In other words, $    \kappa$ measures the failure of the Killing parameter $\tau$, satisfying $\xi^a \nabla_a \tau =1$, to be affine.  The surface gravity is constant along each null generator of an arbitrary Killing horizon $\mathcal H$,  i.e. $\xi^a \nabla_a \kappa = 0$ on $\mathcal H$, which essentially follows from the Killing equation $\nabla_{(a} \xi_{b)}=0$ and   the fact that $\kappa$ is completely determined in terms of $\xi^a$ by \eqref{kappa}. Moreover, bifurcate Killing horizons have the additional property that  $\kappa$   does not vary from generator to generator (see \cite{Kay:1988mu} for a proof). Thus, $\kappa$ is constant on any bifurcate Killing horizon, which is the zeroth law of black hole mechanics. 
    
    Now, comparing \eqref{kappa} with the affine geodesic equation \eqref{defk} for $k^a$, one can show that    the Killing field and the affinely parameterised tangent 
to $\mathcal H^+$ are related on the horizon by 
      \begin{equation}\label{xiandk}
        \xi^a = (\partial_\tau)^a \fheq e^{  \kappa \tau} k^a =e^{  \kappa \tau}( \partial_v)^a,
    \end{equation}
    where $v$ is the affine   parameter of the null generators of $\mathcal H^+$, satisfying $ k^a \nabla_a v =1.$    This implies  to the future of $\mathcal B$  on $\mathcal H^+$  the relation between   $v$ and the Killing parameter $\tau$ is  given by\footnote{    This relation between $v$ and $\tau$ is only valid to the future of the bifurcation surface, since $\tau$ ranges from  $ -\infty$ at $\mathcal B$ to $+\infty$ at future infinity. To the past of $\mathcal B$ on $\mathcal H^+$  the vector field $k^a$ is future directed while $\xi^a$ is past directed, hence the relation becomes $v = - \frac{1}{\kappa} \exp (\kappa \tau)$,  where the Killing parameter $\tau$ covers another patch of $\mathcal H^+$.}
    \begin{equation}
        v = \frac{1}{\kappa} \exp(\kappa \tau). \label{eq:v-tau-horizon}
    \end{equation}
        
  Next, we wish to express   the Killing field in terms of $k^a$ and $l^a$ away from the horizon, which  depends on the specific choice of $k^a$ and $l^a$. We choose $k^a$ and $l^a$ off the horizon such that the Killing field everywhere takes the form  
     \begin{equation}
     \label{horizonkilling}
        \xi^a = \kappa (v k^a - u l^a).
     \end{equation}
     This  ensures that $\xi^a$ acts like a local Lorentz boost near the   horizon and it respects the fact that $\xi^a = 0$ at   $\mathcal B$, where $v=u=0$. We now show that such a choice of  zweibein ($k^a, l^a$) exists.

To extend the zweibein off the horizon in a $\xi^a$-compatible manner, we   construct  another affinely parameterised null vector field  $\beta^a$ that satisfies
     \begin{equation}
     \label{beta}
          \beta^a \beta_a = 0,\quad   \xi^a \beta_a = -1, \quad \beta^a \nabla_a \beta^b =0\,,\quad [\xi, \beta]^a = 0\,,
     \end{equation}  
     throughout   spacetime, because $\xi^a$ is defined everywhere in the background. Notice that   $\beta^a$ must be singular  at the bifurcation surface, $\beta^a = O(1/v)$ at $v=0$, whereas $\xi^a$ vanishes there, $\xi^a = O(v)$ at $v=0$, but this will not be a problem for our discussion below. Now $(\xi^a, \beta^a)$ form a   null zweibein basis for the tangent space of $\mathcal H$. Further, by denoting $\rho$ the affine null distance   away from the horizon $\mathcal H$ along the geodesics to which  $\beta^a$ is tangent, we can see that  
\begin{equation}
    \beta^a = (\partial_\rho)^a\,, 
\end{equation}
where $\rho = 0$ at the Killing horizon $\mathcal H$.
A $\xi^a$-compatible choice of $(k^a, l^a)$ can be made by relating the affine parameters $(u,v)$ for the null geodesics generators of $\mathcal H$ to the   parameters $(\rho, \tau)$, in the region to the future of $\mathcal B$, as follows\footnote{The inverse relation is: $ \tau = \frac{1}{\kappa} \log (\kappa v),\, \rho = \kappa u v$.}
\begin{equation}
\label{coordtransf}
  v = \frac{1}{\kappa} \exp(\kappa \tau), \quad u = \rho \exp (- \kappa \tau)  .
\end{equation}
The reasons for this choice for $(u,v)$  are threefold: 1) we want $u=0$ to label the future horizon; 2) the relation  \eqref{eq:v-tau-horizon} between $v$ and $\tau$ on $\mathcal H^+$ should hold; 3) we require the Killing symmetry to be manifest as a boost $(u,v) \mapsto (au,  v/a)$, so the geometry and the matter fields should depend on $(u,v)$ in terms of the product $uv$. Since the Killing background should only depend on $\rho$ besides codimension-2 spatial parameters, and not on $\tau$, by matching the codimension-2 data and  by  dimensional analysis it follows that $\rho = \kappa u v$.

Finally, the   zweibein $(k^a,l^a)$ is fully determined by our choice of $(u,v)$. In terms of $(\xi^a, \beta^a)$ they can thus be expressed as 
\begin{equation}
\label{klxb}
    k^a =  \frac{1}{\kappa v} \xi^a +\kappa u \beta^a, \quad l^a = \kappa v \beta^a.
\end{equation}
Notice the powers  of $u$ and $v$ match with the singularity of $\beta^a$ and the zero value of $\xi^a$ at $\mathcal B$,  so that ($k^a, l^a$) are   smooth at $\mathcal B$. 
By combining these two  equations  
we reach at our desired expression \eqref{horizonkilling} for $\xi^a$.

    \subsection{Gauge Conditions for  First-Order Non-Stationary Perturbations}
    \label{ssec:gauge-cond}
    Next, we consider a   linear   perturbation of a stationary black hole  background metric  $g \to g + \epsilon\, \delta g$ and the stationary background matter fields $\varphi \to \varphi + \epsilon \,\delta \varphi$, where $\epsilon\ll 1$ is a small perturbation parameter. We are especially interested in first-order \emph{non-stationary} perturbations, defined as\footnote{Note if the Killing field is allowed to vary, then  it would contribute as: $\delta (\mathcal L_\xi g_{ab}) = \mathcal L_{\xi}\delta g_{ab} + \mathcal L_{\delta \xi} g_{ab}$.}
    \begin{equation}
    \label{pert}
        \delta (\mathcal{L}_\xi  g) \neq 0, \quad  \delta ( \mathcal{L}_\xi \varphi ) \neq 0.
    \end{equation}
    This means that the perturbed geometry is not   a stationary black hole geometry, in particular the true event horizon of the perturbed black hole is not a Killing horizon. 
    When perturbing a geometry   there is a certain gauge freedom in which points are chosen to correspond in the two slightly different geometries.  This gauge freedom consists of infinitesimal diffeomorphisms from the unperturbed differentiable manifold to the perturbed manifold.    Moreover, there is a freedom in how the tensor fields on the manifold, such as vector fields or the metric field,  change  under the variation. By the variation of  a vector field $\chi^a = \chi^\mu e^a_\mu$, written in terms of a tetrad basis, we mean that both the coordinate components $\chi^\mu$ and the tetrad basis $e^a_\mu$ may vary.  One can fix the freedom to transform the tensor fields by imposing gauge conditions on the perturbations.   Whether the gauge conditions can be imposed without loss of generality can be checked, for instance, by showing that they automatically hold for any perturbation in Gaussian null coordinates in affine parametrisation.  In section \ref{sec:gnc} we demonstrate   this is indeed the case for our gauge conditions \eqref{fixedkandl}-\eqref{affinegaugecond} below, whereas   gauge condition \eqref{killinggauge} does not follow from GNC but it can be imposed for non-stationary perturbations --- for stationary perturbations it automatically holds --- because $\delta \xi^a$ does not have a fixed meaning in the perturbed geometry.
    
    In order to simplify the derivation of the black hole first law for the linear non-stationary variations \eqref{pert}, we choose the following gauge conditions:

    \begin{enumerate}
        \item[a)] The event horizon of the perturbed black hole geometry is identified with the bifurcate Killing horizon $\mathcal H$ of the background geometry, i.e.,  $\mathcal H^+$ is still located at $u=0$ and $\mathcal H^-$ at $v=0$ after the perturbation. 
        
        \item[b)] The  affinely parameterised null normals to $\mathcal H^+$ and $\mathcal H^-$  are fixed under the variation, 
        \begin{equation}
        \label{fixedkandl}
            \delta k^a =0, \quad \delta l^a =0,
        \end{equation}
        and $k^a$ remains the   null normal   to $\mathcal H^+$ and $\ell^a$ remains null everywhere under the perturbation,    which  yield    the following conditions on the variation of the  metric:
        \begin{equation}
        \label{fixednullnormal}
            k^a \delta g_{ab} \fheq 0, \quad l^a \delta g_{ab} = 0.
        \end{equation}   Together with \eqref{horizonkilling} this implies that $\xi^a\delta g_{ab}=0$ on the Killing horizon $\mathcal H.$  
        Moreover, we require that  after the perturbation $k^a$ is still affinely parameterised on $\mathcal H^+$ and $\ell^a$ everywhere, i.e.,    
        \begin{equation}
        \label{affinegaugecond}
             \delta (k^a \nabla_a k^b) \fheq 0, \quad \delta (l^a \nabla_a l^b) = 0.    \end{equation} 
             These equations are equivalent to  the   conditions on the variation of the Christoffel connection: $k^a k^c \delta \Gamma^b_{ac}\fheq 0$ and $l^a l^c\delta \Gamma^b_{ac}=0$, respectively.
        \item[c)]  The Killing vector field $\xi^a$   remains null and tangent to the geodesic generators of    the event horizon of the perturbed black hole, 
        \begin{equation}
        \label{killinggauge}
        \delta (g_{ab}\xi^a \xi^b)\heq 0, 
        \quad \eta_a \delta \xi^a  \heq 0 \,.
        \end{equation}where $\eta^a$ is a spacelike vector orthogonal to both $k^a$ and $l^a$, $k_a\eta^a=0=l_a \eta^a$. Together with condition \eqref{fixednullnormal}, $\xi^a \delta g_{ab}\heq 0$, the first equation implies   $\xi_a \delta \xi^a \heq 0$, in particular $k_a \delta \xi^a \fheq 0$ and   $l_a \delta \xi^a \pheq 0.$ Combined with the second equation in \eqref{killinggauge} this means on the future horizon $\delta \xi^a$ is proportional to $k^a$ and on the past horizon to $l^a.$
        
    \end{enumerate}

    \noindent We emphasise  these gauge conditions do not fix the Killing field to be the same in the background and perturbed geometry --- nor do they fix the auxiliary  null vector field $\beta^a$ defined in \eqref{beta} ---  
    \begin{equation}
        \delta \xi^a \neq 0\,. 
    \end{equation} 
 The Killing field is often held fixed in the variation, e.g. in \cite{Wald:1993nt}, however it may vary for certain perturbations.\footnote{Another setup where  a  vector field $\xi^a$, that is not necessarily Killing,  may change due to a perturbation, is when it      depends on the background dynamical fields, i.e. $\xi^a = \xi^a(g,\varphi)$. This is  relevant,  for instance, for studying asymptotic Killing symmetries \cite{Barnich:2001jy,Barnich:2010eb,Adami:2020ugu}, symplectic symmetries, \cite{Compere:2015knw} and corner symmetries \cite{Donnelly:2016auv,Freidel:2021cjp}.} For instance, if the horizon  Killing field of a stationary, axisymmetric black hole   is normalised as $\xi^a = (\partial_t)^a + \Omega_{\mathcal H} (\partial_\vartheta )^a$ and the time translation    and   rotational Killing fields  are kept the same, $\delta (\partial_t)^a=0$ and $\delta (\partial_\vartheta)^a=0$, then for perturbations that change the angular horizon velocity, $\delta \Omega_{\mathcal H}\neq0$, the Killing field $\xi^a$ varies. 
    Moreover, when  the horizon Killing field is normalised so as not to have unit surface gravity, $\kappa \neq 1$, then by equation~\eqref{horizonkilling}, $\xi^a = \kappa (v k^a - u l^a)$,   the horizon Killing field     varies if     the surface gravity changes due to the perturbation, since $(k^a,l^a)$ and $(v,u)$ are fixed by assumption b), i.e.,
    \begin{equation}
        \delta \kappa \neq 0 \,.
    \end{equation}
Perturbations that change the surface gravity were already considered in the original work on black hole mechanics by Bardeen, Carter and Hawking \cite{Bardeen:1973gs}, and also for instance in~\cite{Carter:2009nex,Gao:2003ys,Compere:2006my}.   Crucially, they showed \cite{Bardeen:1973gs} that   the surface gravity variation  drops out in the first law of black hole mechanics, $\delta M =\frac{\kappa}{8\pi G} \delta A$, so that it can be interpreted as a proper fundamental equation in thermodynamics, $dE = TdS$.    As we will see, in our    non-stationary first law of black holes the variation of the surface gravity  is also absent, which forms  an important consistency check of the derivation.

Interestingly, the variation of the surface gravity is not uniquely defined for  non-stationary  variations of a Killing horizon, as there are   different definitions of the surface gravity associated to a   vector field, which all agree on  Killing horizons, but they       disagree  from each other on  the  event horizon of a perturbed stationary black hole. This is relevant for our setup since it means we should  be careful about which definition of (the variation) of the surface gravity we are using in the derivation of the first law. 
    The  three   definitions of ``surface gravity''  $\kappa_1, \kappa_2,$ and $\kappa_3$ that we consider are~\cite{Jacobson:1993pf} (see also \cite{Belin:2022xmt})
     \begin{equation}
        \nabla_a (   \xi_b \xi^b)=  -2 \kappa_1 \xi_a,
    \end{equation}
    \begin{equation}
        \xi^b \nabla_b \xi^a = \kappa_2 \xi^a,
    \end{equation}
    \begin{equation}
    \label{k3here}
  (\nabla^a \xi^b )  ( \nabla_{[a} \xi_{b]})  = -2 \kappa_3^2  \,.
    \end{equation}
These surface gravities are usually defined on  a Killing horizon $\mathcal H$, for which they are all the same: $\kappa_1=\kappa_2=\kappa_3= \kappa$ (see, e.g., Sec. 12.5 in  \cite{Wald:1984rg} for a proof). It is maybe less well known that these   quantities are also well defined for any null hypersurface $\mathcal N$ for which $\xi_a$ is the  normal (but not necessarily a Killing field). In particular, this means the surface gravities are well defined for non-stationary perturbations of Killing horizons to which     the horizon Killing field remains normal. The first quantity $\kappa_1$ is well defined because the normal to a null hypersurface is null on the surface, $\xi_b\xi^b = 0$ on $\mathcal N$, so $\nabla_a(\xi_b\xi^b)$ must be   normal to $\mathcal N$, and is hence     proportional to $\xi^a.$ The second definition is the geodesic equation for $\xi^a$  in non-affine parameterisation, which holds because $\xi^a$ is tangent to the null generators of $\mathcal N$, which are geodesics since $\mathcal N$ is a null hypersurface. And the third definition is covariant, hence  $\kappa_3$   is also well defined.  We anticipate already   that the third definition $\kappa_3$ is   the one that is relevant for the Noether charge, associated with~$\xi^a$, evaluated on a dynamical black hole horizon (see Section \ref{sec:examples}).  

We emphasise, however, that these surface gravities are   not constant on a generic null hypersurface. As stated above, on a Killing horizon  the surface gravity is constant along each null generator, i.e. $\mathcal L_\xi \kappa = 0$. Remarkably, this property extends to the case of     conformal Killing horizons \cite{DyerHonig}, to which a conformal Killing vector is tangent,  but only for the quantity $\kappa_1$,  not for $\kappa_2$ and~$\kappa_3$.\footnote{The definition $\kappa_3$ is also conformally invariant \cite{Jacobson:1993pf} and it does not vary from generator to generator on  bifurcate conformal Killing horizons (see     \cite{Jacobson:2018ahi} for a proof).}  Thus, it is a special property of (conformal) Killing horizons that   they obey a ``zeroth law''.\footnote{A zeroth law has also been proven for isolated horizons in \cite{Ashtekar:1999yj} (see also \cite{Ashtekar:2004cn}), where the second definition of the surface gravity $\kappa_2$ was being used. } 

Furthermore, on any null surface $\mathcal N$ the surface gravities are not entirely independent, but satisfy the additional relation 
 (see equation (5) \cite{Jacobson:1993pf} in and  (D.13) in \cite{Belin:2022xmt})  
\begin{equation}
\label{k3relation1}
    \kappa_3 \nheq \frac{1}{2}(\kappa_1 + \kappa_2)\,.
\end{equation}
As we show in Appendix     \ref{app:surf-grav} this relation follows from the fact that   $\xi_{[a}\nabla_{b}\xi_{c]}=0$ at $\mathcal N$, which holds by Frobenius's theorem since $\xi_a$ is orthogonal to the   hypersurface $\mathcal N$. We also derive the following expressions for the surface gravities in the Appendix \ref{app:surf-grav}
\begin{equation}
\label{surfgrav1}
         \kappa_1 \nheq  l^a \nabla_a (  k_b \xi^b), \quad   \kappa_2 \nheq- k^a \nabla_a (  l_b \xi^b)\, ,  \quad  \kappa_3  \nheq l_{[a} k_{b]} \nabla^a \xi^b\,.
    \end{equation}
These expressions hold on any null hypersurface $\mathcal N$, where $k^a$ is the affinely parameterised null normal to $\mathcal N$, and $l^a$ is an auxiliary null vector field that commutes with $k^a$, satisfying $k_a l^a =-1$ on $\mathcal N$ and $\mathcal L_k l^a =0$. 
Crucially the equations in \eqref{surfgrav1} are also valid if $\xi^a$ is not a Killing field. Therefore, we can use them to compute the variations of the   surface gravities for (non-stationary) perturbations of Killing horizons.
 From our gauge conditions in assumption b) we show in Appendix  \ref{app:surf-grav} that the variations are given by 
\begin{equation}
\label{varkappa}
        \delta \kappa_1 \fheq   l^a \nabla_a (k_b \delta \xi^b), \quad \delta \kappa_2 \fheq- k^a \nabla_a(l_b \delta \xi^b)\, , \quad \delta \kappa_3 \fheq  l_{[a} k_{b]} \nabla^a \delta \xi^b\,.
    \end{equation}
We note that for stationary perturbations all the variations of the surface gravity are the same, i.e. $\delta \kappa_1 = \delta \kappa_2  = \delta \kappa_3 = \delta \kappa$, but for non-stationary variations they do not agree. Thus, we  see that for our gauge conditions the variations of the surface gravities depend only on the variation of the background Killing field.

   \subsection{Gaussian Null Coordinates in Affine Parameterisation} \label{sec:gnc}

\begin{figure}[t]
        \centering

        \tikzset{every picture/.style={line width=0.75pt}} 

        \begin{tikzpicture}[x=0.75pt,y=0.75pt,yscale=-1,xscale=1]
        
        \draw [color={rgb, 255:red, 208; green, 2; blue, 27 }  ,draw opacity=0.45 ]   (37.5,70) -- (142.29,128.59) ;
        \draw  [color={rgb, 255:red, 155; green, 155; blue, 155 }  ,draw opacity=1 ] (145.5,37) -- (303,37) -- (235.5,277.5) -- (78,277.5) -- cycle ;
        \draw [color={rgb, 255:red, 74; green, 144; blue, 226 }  ,draw opacity=0.45 ]   (251.71,37) -- (184.21,277.5) ;
        \draw [color={rgb, 255:red, 74; green, 144; blue, 226 }  ,draw opacity=0.45 ]   (279.66,37) -- (212.16,277.5) ;
        \draw [color={rgb, 255:red, 74; green, 144; blue, 226 }  ,draw opacity=0.45 ]   (224.25,37) -- (156.75,277.5) ;
        \draw [color={rgb, 255:red, 74; green, 144; blue, 226 }  ,draw opacity=0.45 ]   (196.51,37) -- (129.01,277.5) ;
        \draw [color={rgb, 255:red, 74; green, 144; blue, 226 }  ,draw opacity=0.45 ]   (167.86,37) -- (100.36,277.5) ;
        \draw [color={rgb, 255:red, 74; green, 144; blue, 226 }  ,draw opacity=1 ][line width=1.5]    (142.29,128.59) -- (158.96,67.93) ;
        \draw [shift={(159.75,65.04)}, rotate = 105.37] [color={rgb, 255:red, 74; green, 144; blue, 226 }  ,draw opacity=1 ][line width=1.5]    (14.21,-4.28) .. controls (9.04,-1.82) and (4.3,-0.39) .. (0,0) .. controls (4.3,0.39) and (9.04,1.82) .. (14.21,4.28)   ;
        \draw [color={rgb, 255:red, 208; green, 2; blue, 27 }  ,draw opacity=1 ][line width=1.5]    (142.29,128.59) -- (87.62,97.97) ;
        \draw [shift={(85,96.5)}, rotate = 29.26] [color={rgb, 255:red, 208; green, 2; blue, 27 }  ,draw opacity=1 ][line width=1.5]    (14.21,-4.28) .. controls (9.04,-1.82) and (4.3,-0.39) .. (0,0) .. controls (4.3,0.39) and (9.04,1.82) .. (14.21,4.28)   ;
        \draw [color={rgb, 255:red, 245; green, 166; blue, 35 }  ,draw opacity=0.45 ][line width=0.75]    (117.5,136) .. controls (162.5,111) and (224,155) .. (278,125.5) ;
        \draw [color={rgb, 255:red, 245; green, 166; blue, 35 }  ,draw opacity=1 ][line width=1.5]    (142.29,128.59) -- (168.52,125.82) ;
        \draw [shift={(171.5,125.5)}, rotate = 173.96] [color={rgb, 255:red, 245; green, 166; blue, 35 }  ,draw opacity=1 ][line width=1.5]    (9.95,-2.99) .. controls (6.32,-1.27) and (3.01,-0.27) .. (0,0) .. controls (3.01,0.27) and (6.32,1.27) .. (9.95,2.99)   ;
        \draw    (142.29,128.59) ;
        \draw [shift={(142.29,128.59)}, rotate = 0] [color={rgb, 255:red, 0; green, 0; blue, 0 }  ][fill={rgb, 255:red, 0; green, 0; blue, 0 }  ][line width=0.75]      (0, 0) circle [x radius= 3.35, y radius= 3.35]   ;
        \draw    (37.5,70) ;
        \draw [shift={(37.5,70)}, rotate = 0] [color={rgb, 255:red, 0; green, 0; blue, 0 }  ][fill={rgb, 255:red, 0; green, 0; blue, 0 }  ][line width=0.75]      (0, 0) circle [x radius= 3.35, y radius= 3.35]   ;
        
        \draw (178.08,60.67) node  [color={rgb, 255:red, 74; green, 144; blue, 226 }  ,opacity=1 ]  {$k^{a}$};
        \draw (186.89,117.07) node  [color={rgb, 255:red, 245; green, 166; blue, 35 }  ,opacity=1 ]  {$m^a_i$};
        \draw (147,147.7) node    {$\left( v,0,x^{i}\right)$};
        \draw (85.49,110.5) node  [color={rgb, 255:red, 208; green, 2; blue, 27 }  ,opacity=1 ]  {$l^{a}$};
        \draw (297.39,124.57) node  [color={rgb, 255:red, 245; green, 166; blue, 35 }  ,opacity=1 ]  {$\mathcal{C}( v)$};
        \draw (287.58,49.67) node  [color={rgb, 255:red, 155; green, 155; blue, 155 }  ,opacity=1 ]  {$\mathcal{H}^+$};
        \draw (73,57.7) node    {$\left( v,u,x^{i}\right)$};

\end{tikzpicture}

        \caption{Gaussian null coordinates $(v,u,x^i)$. Here, $\mathcal H^+$ labels the future event horizon ($u=0$) of the black hole, $\mathcal C(v)$ is a time slice of $\mathcal H^+$ at the null time $v,$ $k^a$ is the affinely parameterised (future-directed) tangent to the null geodesic generators of $\mathcal H^+$,  $l^a$ is the affinely parameterised (future-directed) null vector field transverse to $k^a$, and $m^a_i$ are $D-2$ spacelike vector fields that are orthogonal to $k^a$ and $l^a$.}
        \label{fig:gnc}
    \end{figure}
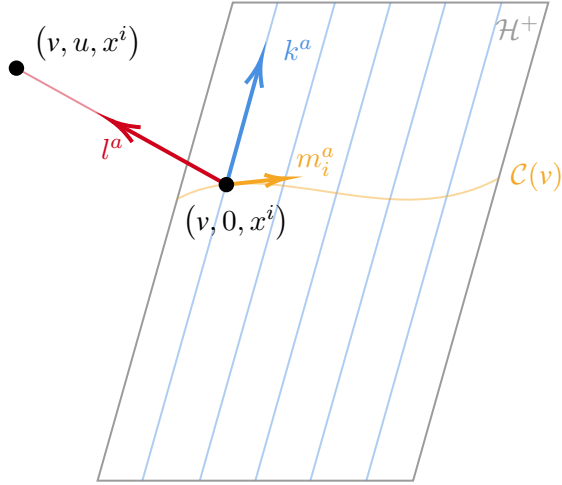

    To simplify our calculations of the dynamical black hole entropy for $f$(Riemann) theories of gravity (see Section \ref{sec:fofriemann}),   here  we introduce   \emph{Gaussian null coordinates} (GNC) labelled by $(v,u,x^i)$ near the future horizon $\mathcal H^+$. We use the affine   parameter $v$ of the null geodesic generators of $\mathcal H^+$ as one of the coordinates, and mark the location of $\mathcal H^+$ as $u=0$. These coordinates  are widely used in previous work on the second law of black hole mechanics in higher curvature gravity~\cite{Wall2015,Bhattacharyya:2021jhr,Hollands:2022fkn,Wall:2024lbd}. They can be obtained using our construction  of the zweibein $(k^a,l^a)$ and the transverse metric $\gamma_{ab}$, as follows. We   label  a point on the horizon by the   affine parameter $v$ and codimension-two (with respect to the full spacetime) spatial coordinates $x^i$ with $i=1,\cdots,D-2$. We can choose such $x^i$ after projecting out the directions labeled by $k^a$ and $l^a$ using the projection operator $\gamma^a_b \fheq \delta^a_b + k^a l_b + l^a k_b$ defined on the horizon. Then, following the geodesics generated by $l^a$ away from the horizon, we   label points with affine parameter $u$ away from the point $(v,x^i)$ on the horizon as $(v,u,x^i)$. This construction of the GNC system is illustrated  in Figure \ref{fig:gnc}.
    
  In GNC gauge, the line element takes the form
    \begin{equation}
    \label{gnc}
        \dd{s^2} = - 2 \dd{u}\dd{v} - u^2 \alpha(v, u, x^i) \dd{v^2} - 2 u \omega_i(v,u,x^i) \dd{v}\dd{x^i} + \gamma_{ij}(v,u,x^i) \dd{x^i} \dd{x^j},
    \end{equation}  
    and the inverse metric is given by
    \begin{equation}
    \label{inversegnc}
        g^{ab} = \mqty(u^2 (\alpha + \omega_i \omega^i) & -1 & u \omega^i\\-1 & 0 & 0\\u \omega^i & 0 & \gamma^{ij})\,,
    \end{equation}
    where $\gamma^{ij}$ is the inverse of $\gamma_{ij}$, and $\omega^i = \gamma^{ij} \omega_j$. 
Note the metric components $g_{uu}$ and $g_{ui}$ vanish everywhere in GNC gauge. The $u^2$ dependence (instead of linear   dependence) of $g_{vv}$   is a result of the affine geodesic equation  on the horizon
    \begin{equation}
        0 \fheq k^b \nabla_b k^a = \conn{a}{v}{v} = \frac{1}{2} g^{ab} \left( 2\partial_v g_{bv} - \partial_b g_{vv}\right)\,, 
    \end{equation}
  which implies
    \begin{equation}
        \partial_u g_{vv} \fheq 0, 
    \end{equation}
    hence $g_{vv} \propto u^2.$
    In Appendix \ref{appGNC} we compute the Christoffel connection of the metric in  GNC and give a prescription for how to calculate covariant derivatives on the horizon. There we  also review   non-affine GNC based on the Killing parameterisation of the null generators of~$\mathcal H^+$, in order to contrast it with the affine GNC used in the main body of the article.

    We verify that the metric indeed satisfies the double null decomposition on the horizon, $ g_{ab} \fheq - 2 k_{(a} l_{b)} + \gamma_{ab}$. The basis vector fields are defined as
     \begin{equation}
        k^a = \left(\pdv{v}\right)^a, \quad l^a = \left(\pdv{u}\right)^a, \quad m_i^a = \left(\pdv{x^i}\right)^a
    \end{equation}
    and, off the horizon,  the  dual covectors are given by 
    \begin{align}
        k_a & = - (\dd{u})_a - u^2 \alpha (\dd{v})_a - u \omega_i (\dd{x^i})_a \label{dualvectorsgnc1}\\
        l_a & = - (\dd{v})_a      \label{dualvectorsgnc}\\
        m^i_a & = (\dd{x^i})_a - u \omega^i (\dd{v})_a \label{dualvectorsgnc3}
    \end{align}
    where we have raised the codimension-2 index $i$ using the inverse transverse metric $\gamma^{ij}$.  The transverse metric can be expressed as
    \begin{equation}
        \gamma_{ab} = \gamma_{ij} (\dd x^i)_a (\dd x^j)_b \fheq \gamma_{ij} m^i_a m^j_b\,. 
        \label{tranversemetricgnc}
    \end{equation} 
    Thus, using \eqref{dualvectorsgnc1}-\eqref{tranversemetricgnc}, one can check that on the horizon the double null decomposition \eqref{doublenull} of the metric is satisfied. In the next section we explain how to extend this double null decomposition on an apparent horizon that is located at a distance $\epsilon$ from the event horizon.

Notice in    \eqref{gnc}   the metric functions in GNC  have an arbitrary dependence on $u,v $ and $x^i$ for a generic geometry. However, for a stationary geometry the functions $\alpha, \omega_i$ and $\gamma_{ij}$ are further constrained. Namely, the isometry  generated by the   Killing field $\xi^a = \kappa (v k^a - u l^a)$  constrains the form of $\alpha, \omega_i, \gamma_{ij}$, such that they only depend on $u,v$ through the product $\kappa u v$~\cite{Bhattacharyya:2021jhr,Hollands:2022fkn}. 
   For a  dynamical perturbation  that does  not respect the Killing symmetry,  the metric functions will       have arbitrary  dependence  on $u$ and $v$  after the perturbation. 
    
    In this article we work with a fixed GNC system. It follows from our gauge condition \eqref{affinegaugecond} that the affine parameters $v$ and $u$ of the null generators of $\mathcal H^+$ and $\mathcal H^-$, respectively, are fixed under the perturbation, and we also keep the spatial coordinates $x^i$ fixed. Hence, 
    a non-stationary   perturbation of a stationary background    changes only the metric functions   
    \begin{equation}
        \begin{split}
            \alpha(\kappa u v, x^i) & \to \alpha(\kappa u v, x^i) + \delta \alpha(v,u,x^i),\\
            \omega_i(\kappa u v, x^i) & \to \omega_i(\kappa u v, x^i) + \delta \omega_i(v,u,x^i),\\
            \gamma_{ij}(\kappa u v, x^i) & \to \gamma_{ij}(\kappa u v, x^i) + \delta \gamma_{ij}(v,u,x^i).
        \end{split}
    \end{equation}
   We now confirm that the   fixed GNC system near $\mathcal H^+$ is compatible with our other gauge conditions for the perturbations  
    in the previous section. The gauge condition \eqref{fixedkandl} is equivalent to keeping $(v,u)$ fixed. The second condition \eqref{fixednullnormal} implies   \ 
    \begin{equation}
        \delta g_{va} \fheq 0,   \quad \delta g_{ua} = 0,
    \end{equation}
    which  holds for   fixed GNC, because $\delta g_{vv}=-u^2 \delta \alpha, \delta g_{vi}=-2 u \delta w_i$ and $\delta g_{uv}=\delta g_{uu}=\delta g_{ui}=0.$   Thirdly, the condition \eqref{affinegaugecond} that $k^a$ and $l^a$ remain  affinely parameterised after the perturbation  is equivalent to 
       \begin{equation}
    \delta \Gamma^a_{vv} \fheq 0, \quad \delta \Gamma_{uu}^a =0.
    \end{equation}
    The second equation is automatic for a fixed GNC system, since $\Gamma_{uu}^u = \Gamma_{uu}^v=\Gamma_{uu}^i=0.$ Writing out the first equation gives
     \begin{equation}
        0 \fheq     \delta \conn{a}{v}{v} = \frac{1}{2} g^{ab} \left( 2\partial_v \delta  g_{bv} - \partial_b \delta g_{vv}\right). 
    \end{equation}
    This is satisfied for  fixed GNC, because of $\delta g_{uv}=0$ and 
    \begin{equation}
        \partial_a \delta g_{vv} = -2 u (\partial_a u )  \delta \alpha - u^2 \partial_a (\delta \alpha) \fheq 0\,, \quad \partial_v \delta g_{vi}=- 2 u \partial_v \delta \omega^i \fheq 0\,.
    \end{equation}
    Next,  we turn to the gauge conditions in \eqref{killinggauge}, which imply that $\delta \xi^a$ on the future horizon is proportional to $k^a$ and on the past horizon to $l^a.$ The proportionality factor can be determined from \eqref{varkappa}, to wit
    \begin{equation}
\delta \xi^a \fheq \left( \int_0^v \delta \kappa_2 \dd{v'} \right) k^a, \qquad \delta \xi^a \pheq  - \left (\int_0^u \delta \kappa_1 \dd{u'} \right)  l^a\,.
\end{equation}
This is consistent with varying the form \eqref{horizonkilling} of the horizon Killing field in   GNC gauge,  from one stationary geometry to another stationary one, since $\delta \kappa$ is constant in that case and $(k^a, l^a)$ and $(v,u)$ are kept fixed.

Further,  above we considered a fixed GNC system, but there is   a gauge freedom in the choice of GNC, corresponding to the   rescaling of the affine parameter along the horizon generators. That is, the form of the metric \eqref{gnc} is invariant under the coordinate transformation $v \to a(x^i) v$ and a simultaneous redefinition   of  the coordinate $u$.   This gauge freedom is thoroughly studied in~\cite{Hollands:2022fkn} (see also \cite{Bhattacharyya:2021jhr}), where the Wall entropy is proven to be gauge invariant  to first order in the perturbation around a stationary black hole.

Finally,  we introduce the so-called boost weight \cite{Wall2015,Bhattacharyya:2021jhr,Hollands:2022fkn,Wall:2024lbd}, which is a  useful notion to keep track of linear perturbations around a stationary background. 
   In affine GNC $(v,u,x^i)$,  a quantity $q$ is said to have   \emph{boost weight} or \emph{Killing weight}  $w$ if it transforms as $q \to a^w q$ under a rigid rescaling $u \to a u$, $v \to v/a$ of the $(u,v)$ coordinates. For example, we can define the boost weight $w$ of any component of a covariant tensor field $T_{a_1\dots a_n}$ (with all indices lowered using the metric) as
    \begin{equation}
        w = \text{number downstairs   $v$-indices} - \text{number of downstairs $u$-indices}
    \end{equation}
    where we note the codimension-2 spatial indices do not contribute to the weight, e.g., $T_{vv}$ has weight 2, $X_{vijk}$ has weight $1$, $Y_{uuui}$ has weight $-3$, et cetera. As for raised indices, each upstairs $u$, $v$ contribute $+1$, $-1$ to the boost weight, respectively. For instance, $X^{uiuj}$ has weight $2$, $Y^v{}_v$ has weight $0$, et cetera.

    Now consider a weight $w$ component $T_{(w)}$ of a tensor field $T$ that is Lie derived by the horizon Killing field $\xi^a$, i.e., $\mathcal{L}_\xi T = 0$. In~GNC the corresponding $w$ component   of the Lie derivative of $T$  with respect to  $\xi= \kappa (v \partial_v - u \partial_u)$ is \cite{Bhattacharyya:2021jhr}
    \begin{equation}
        \left( \mathcal{L}_\xi T \right)_{(w)} = \kappa \left( v \partial_v - u \partial_u + w \right) T_{(w)}\,. \label{eq:lie-deriv}
    \end{equation} 
    Then, on the future horizon $(u=0)$, it can be shown that  the solution to the stationarity condition, $(\mathcal L_\xi T)_{(w)}=0$, for the tensor component $T_{(w)}$  is 
 \cite{Wall:2024lbd}
 \begin{equation}
        T_{(w)} (v,0,x^i) = C_{-w}(x^i) v^{-w},
    \end{equation}
    where $C_{-w}(x^i)$ is a function of the codimension-2 coordinates $x^i$.
Note the tensor components blow up  at the bifurcation surface $u=v=0$ for positive boost weight, $w>0$, except if $C_{-w}(x^i)=0$.  For \emph{physically observable} tensor fields, we assume that it is regular everywhere on the future horizon, in particular at the bifurcation surface. Hence, we find that the function $C_{-w}(x^i)$ must vanish in order to prevent singular backreaction on the geometry. We conclude that in GNC, near a future   Killing horizon $\mathcal H^+$ with a bifurcation surface,  \emph{the positive boost weight components of any stationary physical tensor vanish on the Killing horizon $\mathcal H^+$}.

An example of tensor fields that could be  allowed to be singular is a gauge field $A$, which is not physically observable. There are cases where $A$ is singular at the bifurcation surface while the physically observable field strength tensor $F = \dd A$ is regular everywhere on the horizon.  For example, $A=\frac{1}{\kappa v}\dd{v}$ satisfies $\mathcal L_\xi A = 0$ and is irregular at $v=0$, but has a regular field strength $F=0.$\footnote{We thank an anonymous referee for this point.} The singularity in $A$ does not impose any problem on gauge-invariant theories, because only $F$ appears in the dynamics.  However, in the rest of paper we assume that all matter tensor fields are smooth on the entire horizon, since     gauge fields that blow up at the bifurcation surface should introduce charge terms in the dynamical first law (which we will consider elsewhere).

      Hence,   whenever we encounter a stationary tensor component  with positive boost weight  that is not being varied, it should be treated as zero on the horizon  (in the presence of gauge fields that are not smooth on the horizon, this only applies to gauge-invariant tensors). Another corollary of the claim  above is that a perturbed positive boost weight tensor component that is stationary in the background   is at least first order in the perturbation. This means that a product of   perturbed     positive boost weight   tensor components, that are stationary in the background, is at least second order in the perturbation, and can thus be neglected at first order. The boost weight thus gives  a nice accounting of first-order perturbations around a stationary background, which is useful for deriving the non-stationary first law of black holes. We will employ these  boost weights arguments especially in Sections~\ref{sec:exactness}~and~\ref{sec:fofriemann}, as they   greatly simplify the computation of the dynamical black hole entropy  for higher curvature gravity to linear order in perturbation theory.

    \section{Dynamical Black Hole Entropy from Raychaudhuri Equation}
\label{sec:dynrayc}
    
    In this section we derive  a   ``physical process version'' of the first law  for non-stationary perturbations of a stationary black hole (Section \ref{sec:ray}). Our derivation is based on the Raychaudhuri equation and holds for  black holes in general relativity. Furthermore, we show that the dynamical black hole entropy   satisfying the first law is equal to the Bekenstein-Hawking entropy of the apparent horizon to first order in the perturbation (Section \ref{apparent}). This was proven prior to us in Appendix A of \cite{Hollands:2024vbe}, but we also give a proof of this claim using GNC for pedagogical reasons.

    \subsection{The Physical Process First Law for   Non-Stationary Perturbations}
\label{sec:ray}

Consider a  stationary black hole solution to the vacuum Einstein equation and  perturb it by throwing in a small amount of matter, described by  the variation of  the energy-momentum tensor~$\delta T_{ab}$ \cite{Hawking:1972hy}. We assume the black hole is not destroyed by this infinitesimal physical process, so that, after the matter has fallen into the black hole, there still exists an event horizon. 
 The physical process version of the first law relates the change in black hole entropy to the change in  the mass and angular momentum of the matter passing  through the horizon. In the standard  treatment  \cite{Wald:1995yp,Gao:2001ut,Poisson:2009pwt}  of the physical process first law   it is assumed  that the black hole starts and ends in a stationary state. That is,  the black hole horizon initially coincides with the bifurcation surface of the   Killing horizon, corresponding to Killing time $\tau = - \infty$, and after the perturbation it   settles down  to a Killing horizon again   at future infinity, $\tau =+ \infty$. 
 In this section, we will relax these assumptions  by considering a non-stationary initial and final state for the black hole at two   arbitrary times $\tau_1$ and $\tau_2$.   As a consequence, we  show that   more general boundary conditions yield  a  dynamical correction term to the Bekenstein-Hawking entropy.

  We want to point out that a similar geometric setup was considered in \cite{Mishra:2017sqs},  but the extra term in the physical process first law was interpreted as  the membrane energy  associated with the horizon fluid, instead of as a new contribution to the entropy.  In addition, this   term was also observed by Sorkin in \cite{Sorkin:1995ej}, but he   viewed it as ``unwanted'' and argued it to be zero by a suitable identification of the unperturbed and perturbed horizon.

   \begin{figure}[t]
        \centering

\tikzset{every picture/.style={line width=0.75pt}} 

\begin{tikzpicture}[x=0.75pt,y=0.75pt,yscale=-1,xscale=1]

\draw [line width=0.75]    (70,260) -- (280,50) ;
\draw [line width=0.75]    (70,50) -- (280,260) ;
\draw [color={rgb, 255:red, 0; green, 0; blue, 0 }  ,draw opacity=1 ]   (175,155) ;
\draw [shift={(175,155)}, rotate = 0] [color={rgb, 255:red, 0; green, 0; blue, 0 }  ,draw opacity=1 ][fill={rgb, 255:red, 0; green, 0; blue, 0 }  ,fill opacity=1 ][line width=0.75]      (0, 0) circle [x radius= 3.35, y radius= 3.35]   ;
\draw [color={rgb, 255:red, 139; green, 87; blue, 42 }  ,draw opacity=1 ]   (190,140) ;
\draw [shift={(190,140)}, rotate = 0] [color={rgb, 255:red, 139; green, 87; blue, 42 }  ,draw opacity=1 ][fill={rgb, 255:red, 139; green, 87; blue, 42 }  ,fill opacity=1 ][line width=0.75]      (0, 0) circle [x radius= 3.35, y radius= 3.35]   ;
\draw [color={rgb, 255:red, 139; green, 87; blue, 42 }  ,draw opacity=1 ]   (230,100) ;
\draw [shift={(230,100)}, rotate = 0] [color={rgb, 255:red, 139; green, 87; blue, 42 }  ,draw opacity=1 ][fill={rgb, 255:red, 139; green, 87; blue, 42 }  ,fill opacity=1 ][line width=0.75]      (0, 0) circle [x radius= 3.35, y radius= 3.35]   ;
\draw [color={rgb, 255:red, 144; green, 19; blue, 254 }  ,draw opacity=1 ]   (218.94,131.06) -- (204.6,116.72)(221.06,128.94) -- (206.72,114.6) ;
\draw [shift={(200,110)}, rotate = 45] [color={rgb, 255:red, 144; green, 19; blue, 254 }  ,draw opacity=1 ][line width=0.75]    (10.93,-3.29) .. controls (6.95,-1.4) and (3.31,-0.3) .. (0,0) .. controls (3.31,0.3) and (6.95,1.4) .. (10.93,3.29)   ;
\draw [color={rgb, 255:red, 208; green, 2; blue, 27 }  ,draw opacity=1 ][line width=1.5]    (100,70) .. controls (166.13,121.52) and (184.6,121.52) .. (250,70) ;
\draw [shift={(182.24,108.17)}, rotate = 175.81] [color={rgb, 255:red, 208; green, 2; blue, 27 }  ,draw opacity=1 ][line width=1.5]    (11.37,-3.42) .. controls (7.23,-1.45) and (3.44,-0.31) .. (0,0) .. controls (3.44,0.31) and (7.23,1.45) .. (11.37,3.42)   ;
\draw [color={rgb, 255:red, 208; green, 2; blue, 27 }  ,draw opacity=1 ][line width=2.25]    (240,90) -- (267.17,62.83) ;
\draw [shift={(270,60)}, rotate = 135] [color={rgb, 255:red, 208; green, 2; blue, 27 }  ,draw opacity=1 ][line width=2.25]    (10.49,-3.16) .. controls (6.67,-1.34) and (3.17,-0.29) .. (0,0) .. controls (3.17,0.29) and (6.67,1.34) .. (10.49,3.16)   ;
\draw [color={rgb, 255:red, 74; green, 144; blue, 226 }  ,draw opacity=1 ][line width=1.5]    (240,90) -- (257.88,72.12) ;
\draw [shift={(260,70)}, rotate = 135] [color={rgb, 255:red, 74; green, 144; blue, 226 }  ,draw opacity=1 ][line width=1.5]    (8.53,-2.57) .. controls (5.42,-1.09) and (2.58,-0.23) .. (0,0) .. controls (2.58,0.23) and (5.42,1.09) .. (8.53,2.57)   ;

\draw (295,41) node    {$\mathcal{H}^{+}$};
\draw (175,170) node  [color={rgb, 255:red, 0; green, 0; blue, 0 }  ,opacity=1 ]  {$\mathcal{B}$};
\draw (205.75,152.5) node  [color={rgb, 255:red, 139; green, 87; blue, 42 }  ,opacity=1 ]  {$\mathcal{C}( v_{1})$};
\draw (250.25,108.5) node  [color={rgb, 255:red, 139; green, 87; blue, 42 }  ,opacity=1 ]  {$\mathcal{C}( v_{2})$};
\draw (264.25,85.75) node  [color={rgb, 255:red, 74; green, 144; blue, 226 }  ,opacity=1 ]  {$k^{a}$};
\draw (243.5,133.25) node  [color={rgb, 255:red, 144; green, 19; blue, 254 }  ,opacity=1 ]  {$\delta T_{\textcolor[rgb]{0.56,0.07,1}{ab}}$
};
\draw (256.29,55.75) node  [color={rgb, 255:red, 208; green, 2; blue, 27 }  ,opacity=1 ]  {$\xi ^{a}$};

\end{tikzpicture}

        \caption{Physical process first law for arbitrary cross-sections of a bifurcate Killing horizon. A small flux of matter, described by the stress-energy tensor variation $\delta T_{ab},$   crosses the future horizon $\mathcal H^+$ between two generic cross-sections $\mathcal C(v_1)$ and $\mathcal C(v_2)$. The first law   relates the matter Killing energy flux, relative to the horizon generating Killing field $\xi^a$, through the horizon to the   entropy change between $\mathcal C(v_1)$ and $\mathcal C(v_2)$ due to the perturbation. Usually it is assumed that $v_1 =0$ and $v_2 = \infty$, but we keep the affine times arbitrary.}
        \label{fig:ppfl}
    \end{figure}
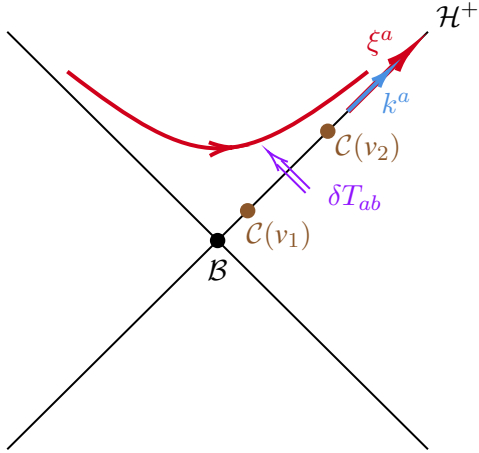

To begin with, we define the matter Killing energy flux, relative to the horizon Killing field~$\xi^a,$ through the future horizon between two arbitrary cross-sections $\mathcal C(v_1)$ and $\mathcal C(v_2)$, where $v_1$ and $v_2$ are the affine ``times'',
\begin{equation}
\label{matterKillingflux}
    \Delta \delta E = \int_{v_1}^{v_2} \dd{v} \int_{\mathcal C(v)} \dd{A} \delta T_{ab} \xi^a k^b\,,
\end{equation}
where $\dd{A}= \sqrt{\gamma} \dd[D-2]{x}$ is the area element of a cross-section $\mathcal C(v)$ of the horizon at time $v$. Here the large $\Delta$ stands for the difference between two horizon  cross-sections, whereas the small $\delta$ denotes the perturbation. Note the perturbation acts only on $T_{ab}$, and not on $\dd{A}$ or $\xi^a$, since the stress-energy tensor vanishes in the unperturbed black hole background.  Usually the range of integration for $v$ is chosen  from $0$ (bifurcation surface) to $\infty $ (future infinity), but here we consider two arbitrary affine times. 
Assuming the horizon Killing field is normalised as $\xi^a = (\partial_t)^a + \Omega_{\mathcal H} (\partial_\vartheta)^a$, where $\Omega_{\mathcal H}$ is the   angular velocity of the horizon, the  matter Killing energy flux is related to the change in the mass and angular momentum of the matter that is passing the horizon  by \cite{Gao:2001ut,Poisson:2009pwt}
\begin{equation}
\label{energyflux}
   \int_{v_1}^{v_2} \dd{v} \int_{\mathcal C(v)} \dd{A} \delta T_{ab} \xi^a k^b=\Delta  \delta  M - \Omega_{\mathcal H} \Delta\delta J\,.
\end{equation}
Next, we recall the outgoing null expansion $\theta_v = \nabla_a k^a$ of the future horizon is equal to the rate of change of an area element $\dd{A}$  along the affine null parameter $v$: 
\begin{equation}
\label{vexpansion}
    \theta_v \dd{A} =  \dv{v} (\dd{A}) \,,
\end{equation}
where $\dd{A}= \sqrt{\gamma} \dd[D-2]{x}$ is the area element of a cross-section of the horizon.  The Raychaudhuri equation for the congruence of null geodesics of $\mathcal H^+$ is
\begin{equation}
\label{raychau}
    \dv{\theta_v}{v} = - \frac{1}{D-2} \theta_v^2 - \sigma^{ab} \sigma_{ab} + \omega^{ab}\omega_{ab} - R_{ab} k^a k^b \,.
\end{equation}
Here,   $\sigma_{ab}$ and $\omega_{ab}$ are the shear and rotation tensors, respectively,  of the horizon generators with respect to the affine parameter $v$. Since $k^a$ is orthogonal to the hypersurface (the future horizon), the rotation tensor vanishes identically on the horizon:  $\omega_{ab}=0$. Further, the expansion and shear are quantities of first order in the perturbation,  $\theta_v, \sigma_{ab}\sim \mathcal O(\epsilon)$, hence the quadratic terms $ \theta_v^2$ and $\sigma^{ab} \sigma_{ab}$ in the Raychaudhuri equation may be neglected for the purpose of deriving the physical process first law. 

Now, the physical process first law follows from  varying the Raychaudhuri equation and assuming the linearised  Einstein equation  holds: $\delta R_{ab}k^a k^b =8 \pi G \delta T_{ab}k^a k^b$. Since on the horizon we have $\xi^a \fheq \kappa v k^a$, we multiply the varied Raychaudhuri equation on both sides by $\kappa v$ and integrate over the horizon between the affine times $v_1$ and $v_2$. Then, recalling     the affine parameter $v$ is fixed under the perturbation,  \eqref{affinegaugecond}, we find to first order in the perturbation
\begin{equation}
\label{integratedray}
\kappa \int_{v_1}^{v_2} \dd{v} \int_{\mathcal C(v)} \dd{A} v \dv{\delta \theta_v}{v} = - 8\pi G \int_{v_1}^{v_2} \dd{v} \int_{\mathcal C(v)} \dd{A} \delta T_{ab} \xi^a k^b \,.
\end{equation}
Next, we integrate the left side of this equation by parts
\begin{equation}
\label{intbyparts}
     \int_{\mathcal C(v)} \dd{A} \int_{v_1}^{v_2} \dd{v}  v \dv{\delta \theta_v}{v} 
=   \int_{\mathcal C(v)} \dd{A} (v \delta \theta_v) \Big|_{v_1}^{v_2} -  \int_{\mathcal C(v)} \dd{A} \int_{v_1}^{v_2} \dd{v} \delta \theta_v\,.
\end{equation}
Note, on the right-hand side, we may pull the variation to the front of the integrals, since the expansion $\theta_v$ vanishes    on the future Killing horizon of the unperturbed black hole.  Moreover, it follows from \eqref{vexpansion} that the second term on the right side is minus  the horizon area  change.  The boundary term on the right side vanishes in the standard derivation \cite{Wald:1995yp} of the physical process first law, where the range for $v$ is taken between $0$ and $\infty$, because  the lower limit vanishes at the bifurcation surface   $v_1=0$, and  the upper limit is also zero because $\theta_v$ vanishes faster than $1/v$ as   $v_2 \to 
\infty $ for  a stationary final state with a finite horizon area. However,   this term does not vanish for two arbitrary affine times, and, crucially, gives  a nontrivial dynamical correction to the Bekenstein-Hawking entropy. Hence, we find
\begin{equation}
     \int_{\mathcal C(v)} \dd{A} \int_{v_1}^{v_2} \dd{v}  v \dv{\delta \theta_v}{v} 
= - \Delta \delta \left ( \int_{\mathcal C(v)} \dd{A} (1-v \theta_v)  \right) .
\end{equation}
Finally, we insert  this equation and the matter Killing energy flux \eqref{energyflux} into the integrated Raychaudhuri equation \eqref{integratedray}. 
This yields   the physical process first law for arbitrary cross-sections of the horizon and non-stationary perturbations
\begin{equation}
\label{physicalprocessfirstlawfinal}
   \frac{\kappa}{2\pi}\Delta\delta S_{\text{dyn}   }  = \Delta \delta M - \Omega_{\mathcal H} \Delta \delta J \,,
\end{equation}
where $S_{\text{dyn}   }$ is the dynamical black hole entropy of the cross-section $\mathcal C$ of the horizon
\begin{equation}
\label{dynentr2}
S_{\text{dyn}   } [\mathcal C]= \frac{1}{4G}  \int_{\mathcal C(v)} \dd{A} \left ( 1- v \theta_v  \right)\,.
\end{equation}
This is equivalent to the formula \eqref{dynamicalintro} for $S_{\text{dyn}   }$ in the introduction, because of \eqref{vexpansion}. A few comments are in order about the dynamical black hole entropy. First, the product $v \theta_v$ is gauge invariant under the scaling transformation $v \to f(x^i) v,$ hence the entropy does not depend to first order in the perturbation on the choice of affine parameter. Further, the entropy does not depend on the auxiliary null vector field $l^a$ and hence also does not depend on the ambiguities in its definition. Second, at the bifurcation surface the dynamical black hole entropy reduces to the standard Bekenstein-Hawking entropy, since $v=0$ at $\mathcal B$. Third, for stationary variations we have $\delta \theta_v = 0$, because the expansion in the $v$-direction vanishes on a Killing horizon, hence we also recover the Bekenstein-Hawking entropy in that case. In the most general case, for arbitrary cross-sections and a non-stationary variation, the dynamical black hole entropy receives a correction term compared to the Bekenstein-Hawking entropy due to the nonzero expansion. If the expansion is positive, which follows from the null energy condition, then the entropy is smaller than the Bekenstein-Hawking entropy, suggesting that it is associated to a surface inside the event horizon (see our discussion in Section \ref{apparent}).

As an aside, we mention that the dynamical black hole entropy  can also be expressed in terms of the Killing parameterisation of the null geodesics of the unperturbed Killing horizon.
From the relation between the Killing time parameter $\tau$ and the affine parameter $v$, $\kappa v = \exp(\kappa \tau)$, it follows that 
\begin{equation}
    S_{\text{dyn}   } [\mathcal C]= \frac{1}{4G}  \int_{\mathcal C(\tau)} \dd{A} \left ( 1- \frac{1}{\kappa} \theta_\tau \right)\,,
\end{equation}
where $\theta_\tau$ is the expansion along the Killing parameter. Note that   scaling   the Killing field $\xi \to c \xi$ also rescales the surface gravity $\kappa \to c \kappa$, hence the product $\theta_\tau/ \kappa$ is reparameterisation invariant. 

An important implication of the physical process first law is that the dynamical black hole entropy satisfies a ``linearised'' second law. This is because, assuming the null energy condition $\delta T_{ab} k^a k^b \ge 0 $, the matter Killing energy flux \eqref{matterKillingflux} is non-negative, hence the dynamical black hole entropy is non-decreasing  to first order in the perturbation
\begin{equation}
\label{2ndlaw}
  \Delta   \delta S_{\text{dyn}   } \ge 0 .
\end{equation}
Thus, for first-order perturbations sourced by external matter that satisfies the null energy condition, the dynamical black hole entropy obeys the classical second law of black hole thermodynamics. 

There is, however, a fundamental difference between the  Bekenstein-Hawking entropy, satisfying the area theorem, and    dynamical black hole entropy, satisfying \eqref{2ndlaw}. On the one hand,  the dynamical black hole entropy \eqref{dynentr2} changes only when matter crosses the horizon.  This is because the physical process first law is valid for any two cross-sections, in particular also for times $v_1$ and $v_2$ that are very close to each other, hence the entropy change occurs locally. If there is no energy flux through the horizon between times $v_1$ and $v_2$, then according to the first law the entropy does not change. 
On the other hand,  Bekenstein-Hawking entropy already changes in anticipation of matter   crossing the horizon. This follows from the fact that the black hole event horizon is defined in a teleological way as the 
causal boundary of the past of future null infinity. At future null infinity the horizon generators must have zero expansion, i.e. they are parallel.
  However, according to the Raychaudhuri equation \eqref{raychau}, the expansion   decreases if positive matter Killing energy crosses the future horizon. Hence,   before   matter even crosses the horizon, the   generators must   have positive expansion in order for the     expansion to be zero   at future infinity.   In other words, the horizon area already increases before matter is thrown into the black hole. 

To  illustrate this point, we consider   a stress-energy tensor that is proportional to a delta function on the horizon: $8 \pi G T_{vv} = c\,\delta(v - v_0)$,  where $c$ is a small constant and $v_0$ is the affine time when the matter source travels through the horizon. Then by the linearised Raychaudhuri equation \eqref{raychau} the  derivative of the expansion is, to first order in the perturbation, proportional to a delta function: \begin{equation}
    \partial_v \theta_v = - c\,\delta(v - v_0)\,.
\end{equation}
The solution to this differential equation for the expansion is a step function 
\begin{equation}
\label{expdelta}
     \theta_v  = c-c\,H (v-v_0)=\begin{cases}
         c  , & v < v_0\,,\\
       0, & v > v_0\,,
    \end{cases}
\end{equation}
where $H$ is the Heaviside step function. 
Note that   we imposed the teleological boundary condition: $\theta_v =0$ at future infinity.
The expansion itself, on the other hand, is related to the derivative of the horizon area by \eqref{vexpansion}, which hence   has the shape of a kink, i.e. it increases until it reaches a constant at the time $v_0$,
\begin{equation}
\label{areadelta}
    A = \begin{cases}
        A_0 + c(v-v_0), & v < v_0\,,\\
        A_0, & v > v_0\,.
    \end{cases}
\end{equation}
Further, the dynamical entropy \eqref{dynentr2} of a black hole can be computed as follows 
\begin{equation}
    S_\text{dyn} =\frac{1}{4G}(1- v \partial_v) A= \begin{cases}
       \frac{1}{4G}( A_0 - c v_0), & v < v_0\\
       \frac{1}{4G} A_0, & v > v_0.
    \end{cases}
\end{equation}
Thus, we  conclude   the profile of the Bekenstein-Hawking entropy (the horizon area)   is a kink as a function of affine time, whereas the dynamical black hole entropy behaves more like a step function, i.e. it  changes only when the matter source crosses the horizon,  in agreement with the discussion above. 

    \subsection{Relation to  Bekenstein-Hawking Entropy of the Apparent Horizon}
    \label{apparent}
    An apparent horizon is defined on a Cauchy surface as the boundary of an outer trapped region, which consists   of   surfaces whose outgoing null expansion is negative  \cite{Hawking:1973uf}. The apparent horizon is foliated by    future  marginally outer trapped surfaces, which means that its   outgoing null expansion vanishes. The location of an apparent horizon is highly  ambiguous as it  depends on the choice of Cauchy slice. Nevertheless, for some fixed foliation of the spacetime, apparent horizons can provide a more local notion of the boundary of a black hole than the event horizon, as the expansion of a null geodesic congruence is defined locally. Thus, it is interesting to investigate the relation between the area entropy of an apparent horizon and the dynamical black hole entropy, as the latter also only changes locally in affine time. This relationship was previously established in  \cite{Hollands:2024vbe}.

    Another reason to study the relationship between the entropy of the  event horizon~$\mathcal H^+$ and that of the apparent horizon $\mathcal A$, is that  the apparent horizon must lie on or inside the event horizon.  In a stationary background  the apparent horizon of a Cauchy slice coincides with the cross-section of the Killing horizon at that slice.   When non-stationary perturbations are switched on, however,  the apparent horizon lies within the black hole,   because   the expansion~$\theta_v$ is non-negative along the outgoing null geodesics of the event horizon --- assuming the null energy condition and weak cosmic censorship  --- whereas the outgoing null  expansion $\tilde \theta_{\tilde k}$ of the apparent horizon vanishes (along the null normal $\tilde k$ to $\mathcal A$). Since the dynamical black hole entropy \eqref{dynentr2} is smaller than the area entropy of the event horizon, if $\theta_v \ge 0,$   a natural question  is whether the dynamical entropy is equal to the area entropy associated to an apparent horizon. In this section we show   this is indeed the case for linear perturbations around a stationary black hole.

    We foliate the spacetime using   the GNC system near the future horizon $\mathcal H^+$ of the black hole.
    Suppose the GNC system is adapted such that    $\mathcal H^+$ is still at $u=0$ after the perturbation (assumption a) in Section \ref{ssec:gauge-cond}). We denote the location of $\mathcal A$ as $u = \mathcal U(v,x^i)\ge 0 $, where we notice that it is in general not a null hypersurface, and its $u$-position   depends on $v$ and $x^i$, so it may look like a wiggly surface in  GNC.  We fix the   spatial foliation of $\mathcal A$ by demanding that   every constant $v$ surface within $\mathcal A$ is future marginally outer trapped. Any other foliation of $\mathcal A$ by $v \to v + f(x^i)$ in principle could be studied using different schemes for extending $k^a$ (hence the $v$-coordinate) off the event horizon.  Below we assume that the affine null distance $\mathcal U(v,x^i)$ to the event horizon, and its spacetime derivatives, are of first order in the perturbation, i.e. their   magnitude is comparable to the perturbation parameter $\epsilon$. 
We will work to first order in $\epsilon$, hence we ignore quantities quadratic in $\mathcal U$ and its derivatives.

    \begin{figure}[btp]
        \centering

\tikzset{every picture/.style={line width=0.75pt}} 

\begin{tikzpicture}[x=0.75pt,y=0.75pt,yscale=-1,xscale=1]

\draw [color={rgb, 255:red, 189; green, 16; blue, 224 }  ,draw opacity=1 ][line width=1.5]    (148.73,161.91) .. controls (194.07,154.04) and (224.07,190.6) .. (259.64,170.21) .. controls (295.2,149.81) and (321.88,184.53) .. (358.47,170.59) ;
\draw [color={rgb, 255:red, 208; green, 2; blue, 27 }  ,draw opacity=1 ]   (149.2,161.91) .. controls (220.51,102.98) and (318.47,106.7) .. (343.27,22.33) ;
\draw [color={rgb, 255:red, 208; green, 2; blue, 27 }  ,draw opacity=1 ]   (358.93,170.59) .. controls (422.18,128.41) and (391.64,62.03) .. (468.52,50.87) ;
\draw [color={rgb, 255:red, 208; green, 2; blue, 27 }  ,draw opacity=1 ]   (343.27,22.33) .. controls (382.96,75.68) and (425.12,33.5) .. (468.52,50.87) ;
\draw [color={rgb, 255:red, 208; green, 2; blue, 27 }  ,draw opacity=1 ]   (47.5,241.5) .. controls (112.5,217.5) and (119,185) .. (148.73,161.91) ;
\draw [color={rgb, 255:red, 208; green, 2; blue, 27 }  ,draw opacity=0.25 ]   (215.08,249.38) .. controls (280.8,188.58) and (290.88,218.36) .. (358.47,170.59) ;
\draw [color={rgb, 255:red, 208; green, 2; blue, 27 }  ,draw opacity=1 ]   (148.5,294) .. controls (177,265.5) and (195.5,269.5) .. (215.08,249.38) ;
\draw [color={rgb, 255:red, 208; green, 2; blue, 27 }  ,draw opacity=1 ]   (148.5,294) .. controls (121.5,274) and (113,291.5) .. (91.5,280.5) .. controls (70,269.5) and (68.54,257.13) .. (47.5,241.5) ;
\draw [color={rgb, 255:red, 74; green, 144; blue, 226 }  ,draw opacity=1 ]   (79.29,68.24) -- (272.74,328.78) ;
\draw [color={rgb, 255:red, 74; green, 144; blue, 226 }  ,draw opacity=1 ]   (272.74,328.78) -- (475.5,328.78) ;
\draw [color={rgb, 255:red, 74; green, 144; blue, 226 }  ,draw opacity=1 ]   (475.5,328.78) -- (381.64,201.68) ;
\draw [color={rgb, 255:red, 74; green, 144; blue, 226 }  ,draw opacity=1 ]   (282.04,68.24) -- (79.29,68.24) ;
\draw [color={rgb, 255:red, 74; green, 144; blue, 226 }  ,draw opacity=1 ]   (294.29,84.52) -- (282.04,68.24) ;
\draw [color={rgb, 255:red, 74; green, 144; blue, 226 }  ,draw opacity=0.25 ]   (294.29,84.52) -- (358.47,170.59) ;
\draw [color={rgb, 255:red, 74; green, 144; blue, 226 }  ,draw opacity=1 ]   (358.93,170.59) -- (381.64,201.68) ;
\draw [color={rgb, 255:red, 74; green, 144; blue, 226 }  ,draw opacity=1 ]   (247.2,137.52) -- (268.14,166.18) ;
\draw [shift={(245.43,135.09)}, rotate = 53.85] [fill={rgb, 255:red, 74; green, 144; blue, 226 }  ,fill opacity=1 ][line width=0.08]  [draw opacity=0] (8.93,-4.29) -- (0,0) -- (8.93,4.29) -- cycle    ;
\draw [color={rgb, 255:red, 65; green, 117; blue, 5 }  ,draw opacity=1 ]   (291.97,140.21) -- (268.14,166.18) ;
\draw [shift={(294,138)}, rotate = 132.55] [fill={rgb, 255:red, 65; green, 117; blue, 5 }  ,fill opacity=1 ][line width=0.08]  [draw opacity=0] (8.93,-4.29) -- (0,0) -- (8.93,4.29) -- cycle    ;
\draw [color={rgb, 255:red, 74; green, 144; blue, 226 }  ,draw opacity=1 ]   (94.11,102.71) -- (115.64,132.18) ;
\draw [shift={(92.93,101.09)}, rotate = 53.85] [color={rgb, 255:red, 74; green, 144; blue, 226 }  ,draw opacity=1 ][line width=0.75]    (10.93,-3.29) .. controls (6.95,-1.4) and (3.31,-0.3) .. (0,0) .. controls (3.31,0.3) and (6.95,1.4) .. (10.93,3.29)   ;
\draw [color={rgb, 255:red, 0; green, 0; blue, 0 }  ,draw opacity=1 ]   (425.36,185.14) -- (373.64,221.18) ;
\draw [shift={(427,184)}, rotate = 145.14] [color={rgb, 255:red, 0; green, 0; blue, 0 }  ,draw opacity=1 ][line width=0.75]    (10.93,-3.29) .. controls (6.95,-1.4) and (3.31,-0.3) .. (0,0) .. controls (3.31,0.3) and (6.95,1.4) .. (10.93,3.29)   ;
\draw [line width=1.5]    (192.88,221.17) -- (395.64,221.17) ;

\draw (314.13,233.5) node    {$\mathcal{C}( v_{0}) \subset \mathcal{H}^{+}$};
\draw (412.36,29.7) node  [color={rgb, 255:red, 208; green, 2; blue, 27 }  ,opacity=1 ]  {$u=\mathcal{U}\left( v,x^{i}\right)$};
\draw (123.49,81.69) node  [color={rgb, 255:red, 74; green, 144; blue, 226 }  ,opacity=1 ]  {$v=v_{0}$};
\draw (166.23,221.7) node    {$u=0$};
\draw (94.59,163.7) node  [color={rgb, 255:red, 189; green, 16; blue, 224 }  ,opacity=1 ]  {$u=\mathcal{U}\left( v_{0} ,x^{i}\right)$};
\draw (196.91,151.69) node  [color={rgb, 255:red, 189; green, 16; blue, 224 }  ,opacity=1 ]  {$\mathcal{T}( v_{0})$};
\draw (239.49,131) node  [color={rgb, 255:red, 74; green, 144; blue, 226 }  ,opacity=1 ]  {$l^{a}$};
\draw (307.61,134.2) node  [color={rgb, 255:red, 65; green, 117; blue, 5 }  ,opacity=1 ]  {$\tilde{k}^{a}$};
\draw (341.9,63.2) node  [color={rgb, 255:red, 208; green, 2; blue, 27 }  ,opacity=1 ]  {$\mathcal{A}$};
\draw (84.72,93.2) node  [color={rgb, 255:red, 74; green, 144; blue, 226 }  ,opacity=1 ]  {$u$};
\draw (438.22,186.7) node    {$v$};

\end{tikzpicture}

        \caption{Apparent horizon $\mathcal A$ at $u=\mathcal U(v,x^i)$, based on a spacetime foliation by affine Gaussian null coordinates. A constant-$v$ section  of the apparent horizon at $v=v_0$ is by construction a future marginally outer trapped surface $\mathcal T(v_0)$ with null normals $l^a$ and $\tilde k^a$. The black hole event horizon $\mathcal H^+$ is located at $u=0$, and $\mathcal C(v_0)$ is a cross-section at $v=v_0$. In a stationary background the apparent and black hole horizon coincide, but for a small non-stationary perturbation the apparent horizon lies slightly inside the black hole event horizon.}
        \label{fig:mar-tra}
    \end{figure}
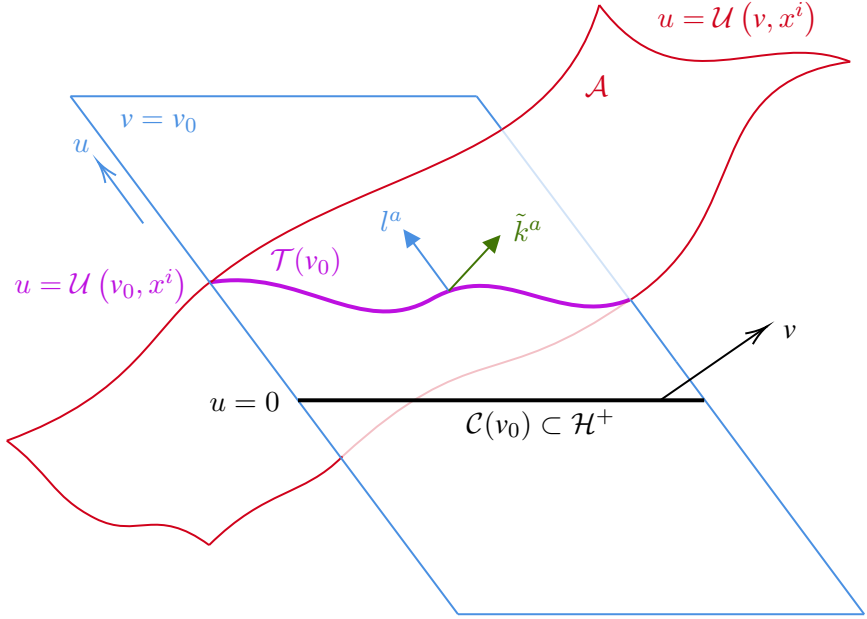

First, we use GNC to  define the   null normals to any future marginally outer trapped surface   $\mathcal T(v_0) $ at  $v=v_0$, which is  a constant  $v$-slice  of the apparent horizon $\mathcal A = \{ u = \mathcal U(v,x^i)\}$. One of the null normals to $\mathcal T$ is  $l_a=-(\dd{v})_a$, \eqref{dualvectorsgnc}, since $\mathcal T$ is by construction  a constant-$v$ surface. The other null normal can be found by the condition that 
    $\dd{u} - (\partial_i \mathcal U) \dd{x^i} \teq 0$, so it reads 
    \begin{equation}
        \tilde k_a \teq - (\dd{u})_a + \partial_i \mathcal U (\dd{x^i})_a\,,
    \end{equation}
    where we have chosen the sign of $\tilde k$ such that it is future directed. Using the inverse metric \eqref{inversegnc} in GNC, we find 
    \begin{equation}
        \tilde k^a = g^{ab} \tilde k_b \teq \left(\pdv{v}\right)^a + (D^i \mathcal U - \omega^i \mathcal U ) \left(\pdv{x^i}\right)^a + \mathcal O(\epsilon^2) \teq k^a + (D^i \mathcal U - \omega^i \mathcal U ) m_i^a + \mathcal O(\epsilon^2)\,,
    \end{equation}
    where $D^i = \gamma^{ij} D_j$, and $D_i$ is the codimension-2 intrinsic covariant derivative.

    Before decomposing the metric on $\mathcal T(v)$ in terms of the null normals $l^a$ and $\tilde k^a$, we need to check the following consistency conditions up to first order:
    \begin{enumerate}
        \item $\tilde k^a$ should  be null at first order, 
        \begin{equation}
            \tilde k^a \tilde k_a = k^a k_a + 2 (D_i \mathcal U - \mathcal U \omega_i) k^a m^i_a + (D^i \mathcal U - \omega^i \mathcal U )(D_j \mathcal U - \omega_j \mathcal U) m_i^a m^j_a = \mathcal O(\epsilon^2)\,,
        \end{equation}
        which follows from the relevant metric components in GNC.
        \item $\tilde k^a$ should obey the same condition as $k^a$ when contracted with $l_a$:
        \begin{equation}
            \tilde k^a l_a = k^a l_a + (D^i \mathcal U - \omega^i \mathcal U ) m_i^a l_a = - 1\,,
        \end{equation}
        because $g_{ua} = 0$ everywhere.
    \end{enumerate}
    Using the null normals $(\tilde k, l)$, we carry out the following metric decomposition on $\mathcal T$,
    \begin{equation}
        g_{ab} \teq - 2 \tilde k_{(a} l_{b)} +\gamma_{ab}\,.
    \end{equation}
    Note this equation is only valid   to first order in $\epsilon.$
   We employ this double null decomposition to study the expansion of the apparent horizon slice $\mathcal T(v)$  along the   null normal $\tilde k$
    \begin{equation}\label{intstepexp}
        \tilde \theta_{\tilde k} = \gamma^a_b \nabla_a \tilde k^b \aeq \gamma^a_b \nabla_a k^b + \gamma^a_b \nabla_a((D^i \mathcal U - \omega^i \mathcal U ) m_i^b)\,.
    \end{equation}
    To identify the second term as a codimension-2 total derivative, we consider 
    \begin{equation}
        m_i^b = (\gamma^b_c - \tilde k^b l_c - l^b \tilde k_c) m_i^c = \gamma^b_c m_i^c - (D_i \mathcal U - 2 \omega_i \mathcal U) l^b \,,
    \end{equation}
    where we used $m^j_a m_i^a = \delta^j_i$. From this it follows that the second term in \eqref{intstepexp} can be written as 
    \begin{equation}
        \gamma^a_b \nabla_a((D^i \mathcal U - \omega^i \mathcal U ) m_i^b) \teq \gamma_b^a \nabla_a (\gamma^b_c ((D^i \mathcal U - \omega^i \mathcal U ) m_i^c)) + \mathcal O(\epsilon^2) = D_i (D^i \mathcal U - \omega^i \mathcal U ) + \mathcal O(\epsilon^2)\,,
    \end{equation}
    where we identified the projected covariant derivative as the codimension-2 intrinsic covariant derivative $D_i$.
    
    Focusing on the first term in \eqref{intstepexp}, we find 
    \begin{equation}
        \gamma^a_b \nabla_a k^b = \frac{1}{2} \gamma^a_b g^{bc} ( \partial_a g_{cv} + \partial_v g_{ac} - \partial_c g_{av}) \teq \frac{1}{2} \gamma^{ac} ( \partial_a g_{cv} + \partial_v g_{ac} - \partial_c g_{av}) \teq \frac{1}{2} \gamma^{ac} \partial_v \gamma_{ac} \teq \theta_v\,,
    \end{equation}
    where we used the property that $\gamma^a_b$ is a projection operator at first order. In the last equality, we have identified the `expansion' of $\mathcal T$ in the $v$-direction as  
    \begin{equation}
         \frac{1}{2} \gamma^{ac} \partial_v \gamma_{ac} = \frac{1}{\sqrt{\gamma}} \partial_v \sqrt{\gamma} = \theta_v\,.
    \end{equation}
We emphasise  this $\theta_v$ is an expansion along a \emph{non-normal} $k$ of $\mathcal T$, hence  it may not vanish on~$\mathcal T$, whereas  for the event horizon $\theta_v$ is the outgoing null expansion, because $k$ is  the null normal  of~$\mathcal H^+$.

    Therefore,  to first order in the perturbation, the expansion along $\tilde k$ can be expressed as
    \begin{equation}
            \delta \tilde \theta_{\tilde k} \teq \delta \theta_v(v, \mathcal U, x^i) + D_i (D^i \mathcal U - \omega^i \mathcal U )\,,           
    \end{equation}
 Note that upon integration on a compact horizon, the second term on the right   would vanish.
 
Now, we Taylor expand the first-order variation of the $k$-expansion  $\theta_v(v,\mathcal U,x^i)$  around the location of the event horizon $(u=0)$,\footnote{We can Taylor expand this around the event horizon to first order, because $\theta_v = \gamma_{ab}\nabla^a k^b$ is a  covariant scalar function on spacetime to first order in the perturbation.} and impose the defining condition for an apparent horizon that the $\tilde k$-expansion $\tilde \theta_{\tilde k}$ must vanish\footnote{We thank Bob Wald for pointing out an error in this equation in a previous version of the paper.}
    \begin{equation}
    \label{apphorizoncond}
        0 \teq \delta \tilde \theta_{\tilde k}(v,\mathcal U,x^i) = \delta \theta_v(v,0,x^i) + \mathcal U(v,x^i) \partial_u \theta_v(v,0,x^i)+ D_i (D^i \mathcal U - \omega^i \mathcal U ). 
    \end{equation}
   This equation is exact to first order in the perturbation parameter $\epsilon$.  We want to solve it  for $\delta \theta_v (v,0,x^i)$, since that   appears in the variation of the dynamical black hole entropy \eqref{dynentr2}.   In order to do so, we make use of two facts about the expansion. First, we have that, to zeroth-order in the perturbation,  
    \begin{equation}
    \label{partialderexp}
        \partial_u \theta_v = \partial_u \left( \frac{1}{\dd{A}} \partial_v \dd{A} \right) = - \frac{1}{(\dd{A})^2} (\partial_u \dd{A} )(\partial_v \dd{A}) + \frac{1}{\dd{A}} \partial_u \partial_v \dd{A} = \partial_v \left ( \frac{1}{\dd{A}} \partial_u  \dd{A}\right) = \partial_v \theta_u,
    \end{equation}
    where $\theta_u= \gamma_{ab}\nabla^a l^b$ is the ingoing null expansion in the $u$-direction. 
    Second, a boost weight analysis (see Section \ref{sec:gnc}) on the event horizon suggests that, at zeroth order,  the ingoing null expansion is proportional to $v$
    \begin{equation}
    \label{boostweightexp}
        \theta_u(v,0,x^i) = - F(x^i) v\,,
    \end{equation}
    for some  function $F(x^i) > 0$. The expansion $\theta_u$ on the event horizon is negative because      null geodesics are focusing in the $u$-direction. Using \eqref{partialderexp} and \eqref{boostweightexp}, we   can  obtain the solution to the apparent horizon condition \eqref{apphorizoncond} for the  expansion $\theta_v$ of the perturbed event horizon  
   \begin{equation}
        \delta \theta_v (v,0,x^i) = \mathcal{U}(v,x^i) F(x^i) -D_i (D^i \mathcal U - \omega^i \mathcal U )\,. \label{eq:theta-u-c}
    \end{equation}        
   Next, we Taylor expand   the area element $\dd{A}$ of  the apparent horizon at the affine time $v$, around the cross-section of the event horizon at $v$, which to leading order in the perturbation gives 
    \begin{equation}
    \label{deltagammau}
        \delta \dd{A}(v,\mathcal U,x^i) = \delta \dd{A}(v,0,x^i) + \mathcal U(v,x^i) \partial_u \dd{A}(v,0,x^i).
    \end{equation}
On the event horizon it follows from \eqref{boostweightexp} that
    \begin{equation}
        \partial_u \dd{A} = \theta_u \dd{A}  = - v F(x^i) \dd{A}.
    \end{equation}
   By inserting this and   \eqref{eq:theta-u-c} into \eqref{deltagammau},  we obtain
    \begin{equation}
    \begin{split}
        \delta \dd{A}(v,\mathcal U,x^i) &= \delta \dd{A}(v,0,x^i) - v \dd{A}(v,0,x^i) \delta \theta_v(v,0,x^i) - v \dd{A}(v,0,x^i)  D_i (D^i \mathcal U - \omega^i \mathcal U )\\
        &= (1 - v \partial_v) \delta \dd{A}(v,0,x^i)- v \dd{A}(v,0,x^i)  D_i (D^i \mathcal U - \omega^i \mathcal U ).
        \end{split}
    \end{equation}
Hence,     integrating this equation over the codimension-2 horizon slice at constant $v$ yields
    \begin{equation}
        A(v,\mathcal U) = (1 - v \partial_v) A(v,0)\,,
    \end{equation}
    where we have taken out the $\delta$, as both sides are exact in the variation.

Thus,     in general relativity  the dynamical black hole entropy captures the area of the apparent horizon, to first order in the perturbation around a stationary black hole, 
    \begin{equation}
    \label{appdynidentify}
        S_\text{dyn} =\frac{1}{4G} (1 - v \partial_v) A[\mathcal C_{\mathcal H^+}(v)] =  \frac{A[\mathcal T(v)]}{4 G}\,,
    \end{equation}
    where $\mathcal C_{\mathcal H^+}(v)$ is the cross-section of the event horizon $\mathcal H^+$ at affine time $v$, and $\mathcal T(v)$ is the cross-section of the apparent horizon $\mathcal A$ at $v$. We have seen that the   dynamical black hole entropy satisfies a linearised second law,  if the perturbation is sourced by stress-energy tensor that obeys the null energy condition. Therefore, the identification \eqref{appdynidentify}  implies that the   Bekenstein-Hawking entropy of the apparent horizon is also  non-decreasing.  This agrees with the     classical second law for the area of  future outer trapping horizons  \cite{Hayward:1993wb}  and dynamical horizons~\cite{Ashtekar:2002ag}, which coincide   in physical setups. Dynamical horizons   are  defined as spacelike hypersurfaces foliated by marginally   trapped surfaces 
    satisfying $\theta_\text{out} =0$ and $\theta_\text{in} <0$. The apparent horizon considered here also satisfies these conditions, so it is an example of a dynamical horizon and hence it should obey the second law.

   \section{Dynamical Black Hole Entropy is Improved Noether Charge}
\label{sec:covphas}

In this section  we derive the dynamical black hole entropy,  valid to leading order  for  non-stationary perturbations of a Killing horizon, for arbitrary diffeomorphism covariant theories of gravity  using the       Noether charge method, also known as covariant phase space formalism,  developed in  \cite{Lee:1990nz,Wald:1993nt, Iyer:1994ys,Wald:1999wa,Hollands:2012sf,Harlow:2019yfa}. We first introduce our notation and briefly review the covariant phase space formalism (Section \ref{ssec:cps}), where our treatment is slightly different from the standard one by Wald and collaborators, since we allow for   variations of both the dynamical fields and the vector fields generating diffeomorphisms (see, e.g,  \cite{Compere:2015knw,Donnelly:2016auv,Freidel:2021cjp} for similar treatments).    Then we derive the non-stationary comparison version (Section \ref{ssec:comparison-1st-law}) and the physical process version (Section \ref{sec:physical2}) of the first law and show that the dynamical black hole entropy that appears in this first law is defined as  an ``improved'' Noether charge.   Furthermore, we analyse several aspects of the dynamical  entropy: we show that it satisfies a linearised second law (Section \ref{sec:linearized2ndlaw}); we discuss its relation to Wall entropy \cite{Wall2015} (Section \ref{sec:wallrelation}); and   we demonstrate its invariance under JKM ambiguities \cite{Jacobson:1993vj}  to leading order in   perturbation theory (Section \ref{sec:jkm-inv}). We end with some technical details: we show using affine Gaussian null coordinates  that the symplectic potential is a total variation when pulled back to the event horizon (Section \ref{sec:exactness}); and we analyse the structure of the   Noether charge and dynamical black hole entropy for arbitrary diffeomorphism covariant theories (Section \ref{sec:structural}).

    \subsection{Covariant Phase Space Formalism} \label{ssec:cps}
We consider a general, classical theory of gravity with arbitrary matter fields in $D$ spacetime dimensions, arising from a diffeomorphism covariant Lagrangian $D$-form  $\mathbf L$, which can always be put in the form \cite{Iyer:1994ys}
    \begin{equation}
    \label{lllagrang}
            \mathbf L = L(g^{ab}, R_{abcd}, \nabla_{e_1} R_{abcd}, \cdots, \nabla_{(e_1 \cdots e_n)}R_{abcd}, \varphi_A, \nabla_{b_1} \varphi_A, \cdots, \nabla_{(b_1\cdots b_m)}\varphi_A) \bm \epsilon.
    \end{equation}
 Here  $\bm \epsilon$ is the volume form,  $g^{ab}$ is the (inverse) spacetime metric, $\nabla$ denotes the covariant derivative with respect to this metric,  $R_{abcd}$ is the Riemann curvature tensor of $g_{ab}$, and    $\varphi_A$ are   arbitrary (bosonic) matter tensor fields with indices $A \equiv a_1\cdots a_s$. We use $\phi \equiv (g_{ab},\varphi_A)$  to collectively denote all   dynamical fields. Further, in the discussion below, we will use the notation 
    \begin{equation}
         \epsilon_{a_1\cdots a_p}  =\epsilon_{a_1\cdots a_p a_{p+1}\cdots a_{D}},
     \end{equation}
     so, for example, $\bm \epsilon_a$ denotes the spacetime volume form with one index displayed and the other indices suppressed.
   In addition, for later convenience, we set the orientation of the volume form   on the horizon to be 
    \begin{equation}
        \bm \epsilon \fheq k \wedge l \wedge \bm \epsilon_{\mathcal C} \label{eq:orientation}
    \end{equation}
    where $k, l$ should be interpreted as 1-forms, and $\bm \epsilon_{\mathcal C}$ is the codimension-2 spatial ``volume'' (area) form of a horizon cross-section.

Under a first-order variation of the dynamical fields, the variation of  $\mathbf L$ can always be expressed as
    \begin{equation}
            \delta \mathbf L = \mathbf E(\phi) \delta \phi + \dd{\bm \Theta(\phi, \delta \phi)} \label{eq:var-L}\,,
    \end{equation}
    where 
    \begin{equation}
        \mathbf E(\phi) \delta \phi = \frac{1}{2} \bm E_{ab} \delta g^{ab} + \bm{\mathcal{E}}^A \delta \varphi_A,
    \end{equation}
   and $\bm E_{ab} = E_{ab} \bm \epsilon$ and $\bm{\mathcal{E}}^A = \mathcal E^A \bm \epsilon$ are the Euler-Lagrange equation of motion forms for $g^{ab}$ and $\varphi_A$, respectively. Moreover, $\bm \Theta(\phi, \delta \phi)$ is the symplectic potential codimension-1 form, which is locally constructed out of $\phi, \delta \phi$ and their derivatives and is linear in $\delta \phi$.   

    Now let $\chi^a$ be an arbitrary smooth vector field, and consider a variation, $\delta \phi = \mathcal L_\chi \phi$,  induced by a diffeomorphism generated by     $\chi^a$. Then,   diffeomorphism covariance of  $\mathbf L$ implies that, under this variation, \eqref{eq:var-L} becomes 
    \begin{equation}
    \label{diffeolagra}
            \mathcal{L}_\chi \mathbf L = \dd{(\chi \cdot \mathbf L)}= \mathbf E(\phi) \mathcal L_\chi \phi + \dd{\bm \Theta(\phi, \mathcal{L}_\chi \phi)}\,.
    \end{equation}
   The second equality follows from the Cartan-Killing equation 
    \begin{equation}
    \label{cartaneq}
            \mathcal{L}_\chi  \bm \Lambda= \dd{(\chi \cdot \bm \Lambda)} + \chi \cdot \dd{\bm \Lambda},
    \end{equation}
    where $\bm \Lambda$ is some differential form, and $\chi \cdot \bm \Lambda$ means contraction of $\chi$ with the first index of $\bm \Lambda$, and from the fact that the Lagrangian form $\mathbf L$ is a top form, hence $\dd {\mathbf L} =0$. Equation \eqref{diffeolagra} implies there exists a \emph{Noether current} codimension-1 form $\mathbf J_\chi$, associated to $\chi^a$, that is closed on shell
    \begin{equation}
    \label{noethercurrent}
        \mathbf J_\chi = \bm \Theta(\phi, \mathcal{L}_\chi \phi) - \chi \cdot  \mathbf L\,.
    \end{equation}
    The Noether current also satisfies an off-shell identity (see the appendix of \cite{Iyer:1995kg} for a proof)
    \begin{equation}
            \dd{(\mathbf J_\chi + \mathbf C_\chi)} = 0\,.
    \end{equation}
    This follows from using the generalised Bianchi identity,
    \begin{equation}
        \nabla^a(E_{ab} - c_{ab}) + \mathcal E^A \nabla_b \varphi_A = 0\,,
    \end{equation}
   to express $\mathbf E(\phi) \mathcal L_\chi \phi = \dd \mathbf C_\chi$, with $\mathbf C_\chi$   the constraint codimension-1 form
    \begin{equation}
        \mathbf C_\chi = \left( - E^{ab} + c^{ab} \right) \chi_b \bm \epsilon_{a},
    \end{equation}
   and
    \begin{equation}
        c^{ab} = \mathcal{E}^{a a_2 \cdots a_s} \varphi\indices{^b_{a_2 \cdots a_s}} + \mathcal{E}^{a_1 a a_3 \cdots a_s} \varphi\indices{_{a_1}^{b}_{a_3 \cdots a_s}} + \cdots + \mathcal{E}^{a_1 \cdots a_{n-1} a} \varphi\indices{_{a_1 \cdots a_{n-1}}^{b}}. \label{eq:cab}
    \end{equation}
   The constraint equations of motion for the metric are $E_{ab}\bm \epsilon^b =0$ and for the matter fields $c_{ab} \bm \epsilon^b =0$. We note that $c^{ab} = 0$, if $\varphi$ is a scalar field.

    Next, by the Poincar\'e lemma, and an explicit construction in \cite{Wald1990}, we can write 
    \begin{equation}
    \label{noethercurrentandcharge}
            \mathbf J_\chi + \mathbf C_\chi = \dd{\mathbf Q_\chi}\,,
    \end{equation}
    where $\mathbf Q_\chi$ is the \emph{Noether charge} codimension-2 form associated  to   $\chi$.

   Furthermore, we consider a variation of the Noether current $\mathbf J_\chi$  that varies both the dynamical fields $\phi$ and the vector field $\chi$
    \begin{equation}
        \begin{split}
        \label{varj}
            \delta \mathbf J_\chi & = \delta_\phi \bm \Theta(\phi, \mathcal{L}_\chi \phi) + \bm \Theta(\phi, \mathcal L_{\delta \chi} \phi) - \delta \chi \cdot \mathbf L - \chi \cdot  \delta \mathbf L\\
            & = \bm \omega(\phi,\delta \phi, \mathcal{L}_\chi \phi) + \mathbf{J}_{\delta \chi} + \dd({\chi \cdot  \bm \Theta(\phi, \delta \phi)}) - \chi \cdot \mathbf E(\phi) \delta \phi\,,
        \end{split}
    \end{equation} 
    where 
    \begin{equation}
        \delta_\phi \bm \Theta(\phi, \mathcal L_\chi \phi)= \delta \bm \Theta(\phi, \mathcal L_\chi \phi) - \bm \Theta(\phi, \mathcal L_{\delta \chi} \phi)
    \end{equation}
    is the variation of the symplectic potential with respect to the fields only, and $\bm \omega$ is the \emph{symplectic current} codimension-1 form
    \begin{equation}
            \bm \omega(\phi,\delta_1 \phi, \delta_2 \phi) = \delta_1 \bm \Theta(\phi, \delta_2 \phi) - \delta_2 \bm \Theta(\phi, \delta_1 \phi).
    \end{equation} 
    Inserting the variation of \eqref{noethercurrentandcharge} for   $\delta\mathbf J_\chi  $, and similarly for $\mathbf{J}_{\delta \chi}$, into   \eqref{varj} yields 
    \begin{equation}\label{eq:omega}
            \bm \omega(\phi,\delta \phi, \mathcal{L}_\chi \phi) = \dd{(\delta_\phi \mathbf Q_\chi - \chi \cdot  \bm \Theta(\phi, \delta \phi))} - \delta_\phi \mathbf C_\chi + \chi \cdot  \mathbf E(\phi) \delta \phi\,,
    \end{equation}
    where 
    \begin{equation}
        \delta_\phi \mathbf Q_\chi \equiv \delta \mathbf Q_\chi - \mathbf{Q}_{\delta \chi} \quad \text{and} \quad \delta_\phi \mathbf C_\chi \equiv \delta \mathbf C_\chi - \mathbf{C}_{\delta \chi}
    \end{equation}
    are the field  variations of the Noether charge $\mathbf{Q}_\xi$ and constraint form $\mathbf{C}_\chi$. This is known as the \emph{fundamental   identity} \cite{Hollands:2012sf} of the covariant phase space formalism. We have allowed for variations of the vector field $\chi$, since for our geometric setup they lead to nonzero variations of the surface gravity (see Section \ref{ssec:gauge-cond}). We will show that in the first law for non-stationary perturbations of stationary black holes  the  variation of the horizon Killing field $\xi^a$ cancels out. In particular, we check in Section \ref{sec:structural} that the variation of the dynamical black hole entropy does not depend on~$\delta \xi^a$. In a different context, variations of the vector field  $\chi$ arise when it depends on  the background dynamical  fields, $\chi^a = \chi^a(\phi)$. The covariant phase space formalism for  field-dependent vector fields is, for instance, studied in detail in \cite{Compere:2006my,Compere:2015knw,Donnelly:2016auv,Freidel:2021cjp}.

    When we apply the fundamental    identity on a stationary black hole background with horizon Killing vector field   $\chi^a = \xi^a$, then the symplectic current evaluated on the Lie derivative of the fields along $\xi^a$ vanishes
    \begin{equation}
         \bm \omega(\phi,\delta \phi, \mathcal{L}_\xi \phi) =0\,,
    \end{equation}
    as it depends linearly on $\mathcal{L}_\xi \phi$, which is zero on a stationary background. 
    
    For fields with gauge symmetries, e.g., a $\mathrm U(1)$ gauge field $A$, this is more subtle as the background gauge fields are stationary up to a pure gauge: $\mathcal L_\xi A = \dd{\lambda_0}$ where $\lambda_0$ is some gauge parameter. However, we can   carry out a gauge transformation $A \to A + \dd{\lambda}$ so that $\mathcal L_\xi A \to \mathcal L_\xi A + \mathcal L_\xi \dd{\lambda} = \dd{(\lambda_0 + \mathcal L_\xi \lambda)}$, and we can always choose a gauge such that $\lambda_0 + \mathcal L_\xi \lambda = \text{constant}$, so $\mathcal L_\xi A = 0$ in this gauge. A further question arises when there are multiple interrelated fields with gauge symmetry, e.g., when there is a charged scalar field $\varphi$ which is coupled to the gauge field. For stationary background, the gauge field and the charged fields are stationary up to a gauge parameter, but these parameters may be different, and it   might be a concern whether  there exists a gauge which sets the Lie derivatives  of all the fields with respect to $\xi$ simultaneously to zero.\footnote{We thank the anonymous referee for raising this subtle point.} In fact, for any physical theory, the gauge invariance/covariance of the coupling ensures that a consistent gauge choice is possible such that the stationarity condition $\mathcal L_\xi \phi = 0$ holds for all matter fields. To illustrate this, we examine the case of a charged scalar field $\varphi$ coupled to a U(1) gauge field $A$. Consider the most basic gauge covariant quantity constructed out of $A$, $\varphi$ and its first derivative: the covariant derivative $\mathcal D \varphi = (\mathrm d - i q A) \varphi$. Under an infinitesimal gauge transformation $\varphi \to \varphi + i q \tilde \lambda \varphi $ with $q$ the charge of $\varphi$ and $\tilde \lambda$ the gauge parameter, hence $\mathcal D \varphi \to \mathcal D \varphi + i q \tilde \lambda \mathcal D \varphi$. Now, for a stationary charged scalar $\varphi$ and stationary gauge field $A$, the Lie derivative with respect to the Killing vector $\xi$ is zero up to a gauge factor: $\mathcal L_\xi A = \dd{\lambda_0}$ and $\mathcal L_\xi \varphi = i q \tilde \lambda \varphi$, for some $\lambda_0, \tilde \lambda$. In other words, the Killing flow is a pure gauge transformation. Any gauge covariant quantity must transform accordingly, so we have $\mathcal L_\xi \mathcal D \varphi = i q \tilde \lambda \mathcal D \varphi$, which imposes constraints on the relationship between $\lambda_0$ and $\tilde \lambda$:
    \begin{equation}
        \begin{aligned}
            \mathcal L_\xi \mathcal D \varphi & = \dd{(\mathcal L_\xi \varphi)} - i q (\varphi \mathcal L_\xi A + A \mathcal L_\xi \varphi) \\
            i q \tilde \lambda \mathcal D \varphi  & = i q (\tilde \lambda \dd{\varphi} + \varphi \dd{\tilde \lambda} - \varphi \dd{\lambda_0} - i q \tilde \lambda A \varphi)\\
            i q \tilde \lambda \mathcal D \varphi  & = i q \tilde \lambda \mathcal D \varphi + i q \varphi \dd{(\tilde \lambda - \lambda_0)}\\
            0 & = \dd{(\tilde \lambda - \lambda_0)}
        \end{aligned}
    \end{equation}
    for all $q, \varphi$. Hence, $\tilde \lambda$ differs from $\lambda_0$ by a constant, which corresponds to a global phase factor. We now carry out a gauge transformation $\varphi \to e^{i q \lambda} \varphi $, $A \to A + \dd{\lambda}$ and demand that it makes $\mathcal L_\xi A = 0 = \mathcal L_\xi \varphi$. We solve 
    \begin{equation}
        \mathcal L_\xi (A + \dd{\lambda}) = \dd{(\lambda_0 + \mathcal L_\xi \lambda)} = 0
    \end{equation}
    and 
    \begin{equation}
        \mathcal L_\xi (e^{i q \lambda} \varphi) = i q e^{i q \lambda} \varphi (\mathcal L_\xi \lambda + \tilde \lambda) = 0 
    \end{equation}
    to obtain $\lambda$ such that $\mathcal L_\xi \lambda = - \tilde \lambda$, which can consistently set $\mathcal L_\xi \varphi$ and $\mathcal L_\xi A$ to zero.

     The above discussion guarantees that (at least in a fixed gauge) $\bm \omega(\phi,\delta \phi, \mathcal{L}_\xi \phi) = 0$ on  a stationary background. Thus, the fundamental   identity becomes
    \begin{equation}\label{eq:q-theta-c}
            \dd{(\delta_\phi \mathbf Q_\xi - \xi \cdot  \bm \Theta(\phi, \delta \phi))} = \delta_\phi \mathbf C_\xi - \xi \cdot  \mathbf E(\phi) \delta \phi.
    \end{equation}
     The term $\xi \cdot \mathbf E(\phi)$ vanishes when we pull it back to the horizon, as $\xi$ is tangent to the horizon. As we will see shortly, this equation plays a central role in the derivation of the  first law for  non-stationary  black holes.

    \subsection{The Non-Stationary Comparison First Law}\label{ssec:comparison-1st-law}

        Previously in \cite{Wald:1993nt,Iyer:1994ys}, 
        a comparison version of the first law was derived for arbitrary diffeomorphism covariant theories of gravity using the Noether charge method. The entropy of stationary black holes was defined   solely in terms of the Noether charge $\mathbf Q_\xi$ integrated over a horizon cross-section. However, for dynamical black holes, the Noether charge has a major drawback --- it is subject to JKM ambiguities \cite{Jacobson:1993vj} away from the bifurcation surface (see Section \ref{sec:jkm-inv}). And, at an arbitrary cross-section of the horizon, the  Noether charge entropy does not satisfy a comparison  first law for dynamical black holes. Here, for dynamical black holes obtained by perturbing stationary black holes with a bifurcate Killing horizon, we show a different definition of entropy (the ``improved'' Noether charge)  does satisfy  a   comparison first law at   linear order in perturbation theory.   Similar results were previously  obtained in  \cite{Hollands:2024vbe}.

    We assume that the background field equations are satisfied, $ \mathbf E(\phi)=0,$ and we   require the perturbed fields to obey the linearised constraint equations, $\delta_\phi \mathbf C_\xi = 0$. Under these assumptions the fundamental variational identity \eqref{eq:q-theta-c} becomes 
    \begin{equation}
        \dd{(\delta_\phi \mathbf Q_\xi - \xi \cdot  \bm \Theta(\phi, \delta \phi))} = 0. \label{eq:d-Qimprov-0}
    \end{equation} 
We integrate this expression over a codimension-1 spatial hypersurface between the horizon $\mathcal H^+$ and spatial infinity $\mathcal S_{\infty}$ (a codimension-2 sphere at infinity). 
Because of Stokes' theorem, the boundary integral at   $\mathcal S_{\infty}$   is equal to the boundary integral at a   cross-section $\mathcal C$  of $\mathcal H^+$
    \begin{equation}
        \int_{\mathcal S_{\infty}}  (\delta_\phi \mathbf Q_\xi - \xi \cdot  \bm \Theta(\phi, \delta \phi))= \int_{\mathcal C} (\delta_\phi \mathbf Q_\xi - \xi \cdot  \bm \Theta(\phi, \delta \phi))\,. \label{eq:stokes}
    \end{equation}   
 For the boundary integral at infinity we make the identification
    \begin{equation}
    \label{massvarnew}
            \int_{\mathcal S_\infty} (\delta_\phi \mathbf Q_\xi - \xi \cdot  \bm \Theta(\phi, \delta \phi)) = \delta M - \Omega_{\mathcal H} \delta J.
    \end{equation}
    For stationary, axisymmetric  black holes, the horizon Killing vector field may be normalised at spatial infinity as  
    \begin{equation}
        \xi^a = (\partial_t)^a + \Omega_{\mathcal H} (\partial_\vartheta)^a\,,
    \end{equation}
    where $t$, $\vartheta$ are the temporal and angular coordinates, respectively. Then,  the variation of the canonical mass and angular momentum can be defined as 
    \begin{equation}
        \delta M = \int_{\mathcal S_\infty} (\delta_\phi \mathbf Q_{\partial_t} - \partial_t \cdot \bm \Theta(\phi, \delta \phi)), \qquad \delta J = - \int_{\mathcal S_\infty} \delta_\phi \mathbf Q_{\partial_\vartheta}\,,
    \end{equation}
    where $\partial_\vartheta \cdot \bm \Theta = 0$ on $\mathcal S_\infty$, because $\partial_\vartheta$ is parallel to $\mathcal S_\infty$. The mass $M$ is well defined if there exists a codimension-1 form $\mathbf B_\infty(\phi)$ at a timelike codimension-1 hypersurface whose radial coordinate $r$ tends to   infinity,   such that 
    \begin{equation}
        \bm \Theta(\phi, \delta \phi) \overset{r\to \infty}{=} \delta \mathbf B_\infty(\phi).\label{eq:consistency-infty}
    \end{equation}
 The definitions of the mass and angular momentum are then   
    \begin{align}
    \label{massadm}
     M &=  \int_{\mathcal S_\infty}  (\mathbf Q_{\partial_t}- \partial_t \cdot \mathbf B_\infty ) \,,\\
     J &=  - \int_{\mathcal S_\infty} \mathbf Q_{\partial_{\vartheta}}\,.
        \label{jadm}
 \end{align}
    In \cite{Iyer:1994ys}, the exactness in $\delta $ of $ \bm \Theta$ at   asymptotic infinity is shown for general relativity, assuming suitable fall-off conditions for the metric, and the   form $\mathbf B_\infty$   is explicitly constructed. Moreover, they recover the ADM definitions of the mass and angular momentum from \eqref{massadm} and \eqref{jadm}.

  Furthermore, for the boundary integral at the horizon, in the original works \cite{Wald:1993nt,Iyer:1994ys} this was evaluated at the bifurcation surface, where $\xi \cdot \bm \Theta =0$. Subsequently, they identified the Noether charge with the black hole entropy, which for arbitrary theories of gravity is given by the Wald entropy \eqref{iyerwald}, 
  \begin{equation}
  \label{idwald1}
      \int_{\mathcal B}  \delta_\phi \mathbf Q_\xi = \frac{\kappa}{2\pi} \delta S_{\text{Wald}}\,.
\end{equation}
 For stationary black holes the integral of the Noether charge is independent of the choice of horizon cross-section, so the choice for $\mathcal B$ is innocuous. This is because the difference between integrals of $\mathbf Q_\xi$ over different cross-sections is given by the integral of $\mathbf J_\xi$ over the horizon in between. But, because of the stationarity condition $\mathcal L_\xi \phi =0$ and the fact that the pullback of $\xi \cdot \mathbf L$ vanishes at the horizon, according to the definition of $\mathbf J_\xi$, \eqref{noethercurrent}, its pullback to $\mathcal H^+$ vanishes.  However,  for non-stationary perturbations   \eqref{idwald1} is really only true at the bifurcation surface, since $\mathcal L_\xi \phi   \neq 0 $ in the perturbed geometry. Further, for stationary perturbations it is not necessary to evaluate the boundary integral at $\mathcal B$. Indeed, Gao \cite{Gao:2003ys} (see also \cite{Compere:2006my}) showed that the boundary integral, including the $\xi \cdot \bm \Theta$ terms, gives the black hole entropy for arbitrary horizon cross-sections 
     \begin{equation}\label{gao}
        \int_{\mathcal C } (\delta_\phi \mathbf Q_\xi - \xi \cdot  \bm \Theta(\phi, \delta \phi)) = \frac{\kappa}{2 \pi} \delta S_\text{Wald}  \qquad \text{for} \qquad \delta (\mathcal L_\xi \phi ) = 0,
    \end{equation}
    Gao established this identity for general relativity (using results from \cite{Bardeen:1973gs}), that is, he obtained the variation of $S_{\text{BH}}$ from the left-hand side. But the identity   holds more generally for any diffeomorphism covariant theory of gravity, which follows   as a special case from our results.

 We generalise these derivations by considering non-stationary perturbations and  a spatial slice that extends from an arbitrary  horizon cross-section to spatial infinity.  Hence, we make the identification    
    \begin{equation}
 \label{vardynennoether}
        \int_{\mathcal C } (\delta_\phi \mathbf Q_\xi - \xi \cdot  \bm \Theta(\phi, \delta \phi)) = \frac{\kappa}{2 \pi} \delta S_\text{dyn}  \qquad \text{for} \qquad \delta (\mathcal L_\xi \phi ) \neq 0,
    \end{equation}
    where  $\kappa$ is the  surface gravity of the unperturbed Killing horizon. Hence, instead of only the Noether charge, the symplectic potential also contributes to the (variation of) the dynamical black hole entropy.
        \begin{figure}[t]
        \centering
        \tikzset{every picture/.style={line width=0.75pt}} 

\begin{tikzpicture}[x=0.75pt,y=0.75pt,yscale=-1,xscale=1]

\draw    (90,20) -- (230,160) ;
\draw    (90,160) -- (230,20) ;
\draw    (230,20) -- (300,90) ;
\draw    (230,160) -- (300,90) ;
\draw    (20,90) -- (90,160) ;
\draw    (20,90) -- (90,20) ;
\draw    (90,20) .. controls (91.67,18.33) and (93.33,18.33) .. (95,20) .. controls (96.67,21.67) and (98.33,21.67) .. (100,20) .. controls (101.67,18.33) and (103.33,18.33) .. (105,20) .. controls (106.67,21.67) and (108.33,21.67) .. (110,20) .. controls (111.67,18.33) and (113.33,18.33) .. (115,20) .. controls (116.67,21.67) and (118.33,21.67) .. (120,20) .. controls (121.67,18.33) and (123.33,18.33) .. (125,20) .. controls (126.67,21.67) and (128.33,21.67) .. (130,20) .. controls (131.67,18.33) and (133.33,18.33) .. (135,20) .. controls (136.67,21.67) and (138.33,21.67) .. (140,20) .. controls (141.67,18.33) and (143.33,18.33) .. (145,20) .. controls (146.67,21.67) and (148.33,21.67) .. (150,20) .. controls (151.67,18.33) and (153.33,18.33) .. (155,20) .. controls (156.67,21.67) and (158.33,21.67) .. (160,20) .. controls (161.67,18.33) and (163.33,18.33) .. (165,20) .. controls (166.67,21.67) and (168.33,21.67) .. (170,20) .. controls (171.67,18.33) and (173.33,18.33) .. (175,20) .. controls (176.67,21.67) and (178.33,21.67) .. (180,20) .. controls (181.67,18.33) and (183.33,18.33) .. (185,20) .. controls (186.67,21.67) and (188.33,21.67) .. (190,20) .. controls (191.67,18.33) and (193.33,18.33) .. (195,20) .. controls (196.67,21.67) and (198.33,21.67) .. (200,20) .. controls (201.67,18.33) and (203.33,18.33) .. (205,20) .. controls (206.67,21.67) and (208.33,21.67) .. (210,20) .. controls (211.67,18.33) and (213.33,18.33) .. (215,20) .. controls (216.67,21.67) and (218.33,21.67) .. (220,20) .. controls (221.67,18.33) and (223.33,18.33) .. (225,20) .. controls (226.67,21.67) and (228.33,21.67) .. (230,20) -- (230,20) ;
\draw    (90,160) .. controls (91.67,158.33) and (93.33,158.33) .. (95,160) .. controls (96.67,161.67) and (98.33,161.67) .. (100,160) .. controls (101.67,158.33) and (103.33,158.33) .. (105,160) .. controls (106.67,161.67) and (108.33,161.67) .. (110,160) .. controls (111.67,158.33) and (113.33,158.33) .. (115,160) .. controls (116.67,161.67) and (118.33,161.67) .. (120,160) .. controls (121.67,158.33) and (123.33,158.33) .. (125,160) .. controls (126.67,161.67) and (128.33,161.67) .. (130,160) .. controls (131.67,158.33) and (133.33,158.33) .. (135,160) .. controls (136.67,161.67) and (138.33,161.67) .. (140,160) .. controls (141.67,158.33) and (143.33,158.33) .. (145,160) .. controls (146.67,161.67) and (148.33,161.67) .. (150,160) .. controls (151.67,158.33) and (153.33,158.33) .. (155,160) .. controls (156.67,161.67) and (158.33,161.67) .. (160,160) .. controls (161.67,158.33) and (163.33,158.33) .. (165,160) .. controls (166.67,161.67) and (168.33,161.67) .. (170,160) .. controls (171.67,158.33) and (173.33,158.33) .. (175,160) .. controls (176.67,161.67) and (178.33,161.67) .. (180,160) .. controls (181.67,158.33) and (183.33,158.33) .. (185,160) .. controls (186.67,161.67) and (188.33,161.67) .. (190,160) .. controls (191.67,158.33) and (193.33,158.33) .. (195,160) .. controls (196.67,161.67) and (198.33,161.67) .. (200,160) .. controls (201.67,158.33) and (203.33,158.33) .. (205,160) .. controls (206.67,161.67) and (208.33,161.67) .. (210,160) .. controls (211.67,158.33) and (213.33,158.33) .. (215,160) .. controls (216.67,161.67) and (218.33,161.67) .. (220,160) .. controls (221.67,158.33) and (223.33,158.33) .. (225,160) .. controls (226.67,161.67) and (228.33,161.67) .. (230,160) -- (230,160) ;
\draw [color={rgb, 255:red, 74; green, 144; blue, 226 }  ,draw opacity=1 ]   (160,90) -- (300,90) ;
\draw [color={rgb, 255:red, 144; green, 19; blue, 254 }  ,draw opacity=1 ]   (190,60) .. controls (218.6,53.2) and (272.6,74.8) .. (300,90) ;
\draw [color={rgb, 255:red, 74; green, 144; blue, 226 }  ,draw opacity=1 ]   (160,90) ;
\draw [shift={(160,90)}, rotate = 0] [color={rgb, 255:red, 74; green, 144; blue, 226 }  ,draw opacity=1 ][fill={rgb, 255:red, 74; green, 144; blue, 226 }  ,fill opacity=1 ][line width=0.75]      (0, 0) circle [x radius= 2.01, y radius= 2.01]   ;
\draw [color={rgb, 255:red, 144; green, 19; blue, 254 }  ,draw opacity=1 ]   (190,60) ;
\draw [shift={(190,60)}, rotate = 0] [color={rgb, 255:red, 144; green, 19; blue, 254 }  ,draw opacity=1 ][fill={rgb, 255:red, 144; green, 19; blue, 254 }  ,fill opacity=1 ][line width=0.75]      (0, 0) circle [x radius= 2.01, y radius= 2.01]   ;
\draw    (300,90) ;
\draw [shift={(300,90)}, rotate = 0] [color={rgb, 255:red, 0; green, 0; blue, 0 }  ][fill={rgb, 255:red, 0; green, 0; blue, 0 }  ][line width=0.75]      (0, 0) circle [x radius= 2.01, y radius= 2.01]   ;

\draw (109,87.9) node  [color={rgb, 255:red, 74; green, 144; blue, 226 }  ,opacity=1 ]  {$\frac{\kappa}{2\pi} \delta S_{\text{Wald}}[\mathcal B]$};
\draw (160,45) node  [color={rgb, 255:red, 144; green, 19; blue, 254 }  ,opacity=1 ]  {$\frac{\kappa}{2\pi} \delta S_{\text{dyn}}[\mathcal C]$};
\draw (341.76,100.3) node    {$\delta M-\Omega _{\mathcal H} \delta J$};

\end{tikzpicture}
        \caption{Penrose diagram of an eternal asymptotically flat black hole. The   comparison version of the first law relates the variations of the black hole  mass and angular  momentum at spatial infinity   to the variation of the entropy at the event horizon. At the bifurcation surface $\mathcal B$ the entropy is given by the Wald entropy, and on an arbitrary cross-section $\mathcal C$ the horizon entropy is the dynamical black hole entropy for non-stationary variations.  }
        \label{fig:comparison-1st-law}
    \end{figure}
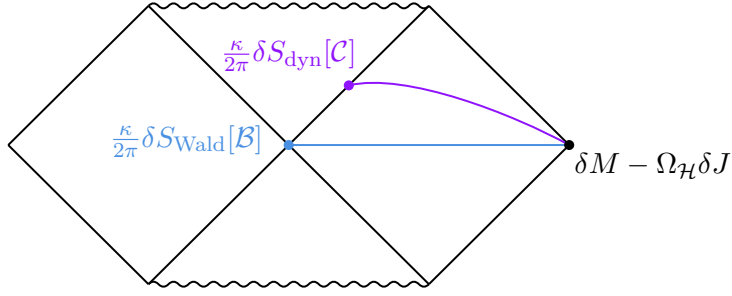    
\  Thus, by relating the two boundary integrals at  $\mathcal S_\infty$ and $\mathcal{C}$,  \eqref{massvarnew} and \eqref{vardynennoether} respectively, we      arrive at the non-stationary comparison first law for arbitrary horizon slices
    \begin{equation}
         \label{newfirstlaaa}
         \delta M - \Omega_{\mathcal H} \delta J =  \frac{\kappa}{2\pi}   \delta S_\text{dyn} .
    \end{equation}    
    When gauge fields are present, in general their minimal sector contributes additional terms in the first law associated to the charges of the black hole. For example, for Einstein-Maxwell gravity the first law should take the form $\delta M - \Omega_{\mathcal H} \delta J - \Phi_{\mathcal H} \delta Q = (\kappa/2\pi) \delta S_{\text{dyn}}$, where $\Phi_{\mathcal H} =- \xi^a A_a|_{\mathcal H}$ is the electric potential on the horizon, and $Q$ is the electric charge of the black hole. We do not have such charge terms in the first law in this paper, since we assume that the gauge field~$A_a$ is smooth on the entire horizon. Together with the stationary condition $\mathcal L_\xi A_a =0$ this implies that  the electric potential  vanishes on the horizon, since $ \Phi_{\mathcal H}$ is a constant in a stationary solution and $\xi^a$ vanishes on the bifurcation surface. If we relax the condition that $A_a$ is smooth everywhere on the horizon, then additional charge terms should be present in the black hole first law. For stationary perturbations this was shown in   \cite{Gao:2003ys} (notably, no additional conditions are imposed in \cite{Gao:2003ys} on the perturbation of the gauge field), and we plan to generalize this derivation to non-stationary perturbations in a forthcoming paper \cite{VisserYan}.  
    
     To serve as a well-posed definition for $S_\text{dyn}$, the variational formula \eqref{vardynennoether} is subject to two consistency conditions:
    \begin{itemize}
        \item[a)]  There exists a codimension-1 form $\mathbf B_{\mathcal H^+}(\phi)$ that satisfies 
        \begin{equation}
        \label{consistencya}
        \bm \Theta(\phi, \delta \phi) \fheq \delta \mathbf B_{\mathcal H^+}(\phi) \quad \text{and} \quad \mathbf B_{\mathcal H^+}(\phi) \fheq 0\,,
        \end{equation}
        where $\bm \Theta$ is the   pullback of the symplectic potential to the horizon, and  the second equality holds on the   background Killing horizon. This condition implies
        \begin{equation}
        \label{bphibla}
            \xi \cdot \bm \Theta(\phi, \delta \phi) \cceq \delta (\xi \cdot \mathbf B_{\mathcal H^+}(\phi) ) \cceq  \kappa \delta (\xi \cdot \mathbf B_{\mathcal H^+}/\kappa_3) \cceq \kappa \delta_\phi (\xi \cdot \mathbf B_{\mathcal H^+}/\kappa_3)
        \end{equation}
        simply because, when $\delta$ is not acting on $\mathbf B_{\mathcal H^+} $, it vanishes in the stationary background. 
        \item[b)] To extract the full variation $\delta$, instead of just the field-only variation $\delta_\phi$, from the first term, we also require
        \begin{equation}
            \delta_\phi \mathbf Q_\xi \cceq \kappa \delta (\mathbf Q_\xi / \kappa_3),        
        \end{equation}
        where, $\kappa_3$ is the surface gravity defined in \eqref{k3here}.
    \end{itemize}

\noindent     In Section \ref{sec:exactness}  we prove condition a) for general diffeomorphism covariant theories of gravity. There we will use the general structure of symplectic potential $\bm \Theta$ studied in \cite{Iyer:1994ys}, and we will work in GNC that allow us to use boost weight arguments. Later in Section \ref{sec:structural}, we prove condition~b) in the same setup as the proof for a). 

    An immediate corollary of the above two conditions is that $\delta_\phi S_\text{dyn} = \delta S_\text{dyn}$, i.e., the field-only variation is equivalent to full variation of $S_\text{dyn}$.   The reason for this is that a field variation does not act on $\kappa_3$, since the surface gravity variation only depends on the variation of the vector field $\delta \xi^a$, see \eqref{varkappa}, assuming  our gauge conditions for the perturbation hold. This implies $\delta_\phi \mathbf Q_\xi \cceq \kappa \delta_\phi (\mathbf Q_\xi / \kappa_3)$. Thus, combining this with \eqref{bphibla}, it follows    that we can write
    \begin{equation}
        \delta_\phi\left (  \int_{\mathcal C } \frac{2\pi}{\kappa_3}  (   \mathbf Q_\xi -  \xi \cdot  \mathbf B_{\mathcal H^+} )\right) =   \delta S_\text{dyn}\,.
    \end{equation}
    This means  that at first order  the variation of $S_\text{dyn}$ is independent of $\delta \xi$, which also follows immediately from   the defining relation \eqref{vardynennoether} for $\delta S_{\text{dyn}}$. Thus the variation of~the horizon Killing field, and hence also of the surface gravity, is absent in the first law. 
    
    Given that the  two conditions above are satisfied, we can define the dynamical entropy as the \emph{improved Noether charge} $\tilde{\mathbf Q}_\xi$ (see, e.g., \cite{Wald:1999wa,Harlow:2019yfa,Freidel:2020svx,Freidel:2020ayo, Shi:2020csw,Chandrasekaran:2020wwn,Freidel:2021cjp}) up to linear order in the perturbation away from stationarity,
    \begin{equation}
    \label{improvednoetherc}
        S_\text{dyn} =\int_{\mathcal C}  \frac{2 \pi}{\kappa_3} \tilde{\mathbf Q}_\xi =  \int_{\mathcal C} \frac{2 \pi}{\kappa_3}\left(\mathbf Q_\xi - \xi \cdot \mathbf B_{\mathcal H^+}\right).
    \end{equation}
At zeroth order in the perturbation, $\xi \cdot \mathbf B_{\mathcal H^+}(\phi) \cceq 0$, hence  $\tilde{\mathbf Q}_\xi \cceq \mathbf Q_\xi$ for Killing horizons. Moreover, at the bifurcation surface, $\xi^a =0$, hence we   recover the Iyer-Wald result \eqref{idwald1}. Therefore, at the bifurcation surface and on Killing horizons, $S_{\text{dyn}}$ reduces to the   Wald entropy.

    \subsection{The Physical Process First Law}
    \label{sec:physical2}
    In the derivation of the comparison version of the  first law above we treated the metric and matter field collectively, and defined  the dynamical entropy in terms of the  improved Noether charge for all fields $\phi$. In this section we consider the case where an external minimal matter source,  described by the stress-energy tensor $T_{ab}$, is switched on as a perturbation.  We assume that the external matter fields $\psi$ are decoupled from the already present matter fields $\varphi$,    and only provide a source for the  metric field via minimal coupling.\footnote{If a Maxwell field is present in the theory this implies the external matter is not allowed to be charged, like external charged scalar fields, which is the reason why we do not have charge terms in the first law.}  We show that the same dynamical entropy \eqref{improvednoetherc}  also satisfies
a physical process version of first law associated with this external stress-energy tensor.    
  This extends the proof of the physical process version of first law from the Raychaudhuri equation in Section  \ref{sec:ray}  to arbitrary diffeomorphism covariant theories. A similar proof was given  in \cite{Hollands:2024vbe}.

      \begin{figure}[H]
        \centering
        \tikzset{every picture/.style={line width=0.75pt}}      

\begin{tikzpicture}[x=0.75pt,y=0.75pt,yscale=-1,xscale=1]

\draw    (61.8,27.6) -- (233.8,203.6) ;
\draw    (61.8,203.6) -- (233.8,27.6) ;
\draw    (233.8,27.6) -- (319.8,115.6) ;
\draw    (233.8,203.6) -- (319.8,115.6) ;
\draw    (61.8,27.6) .. controls (63.47,25.93) and (65.13,25.93) .. (66.8,27.6) .. controls (68.47,29.27) and (70.13,29.27) .. (71.8,27.6) .. controls (73.47,25.93) and (75.13,25.93) .. (76.8,27.6) .. controls (78.47,29.27) and (80.13,29.27) .. (81.8,27.6) .. controls (83.47,25.93) and (85.13,25.93) .. (86.8,27.6) .. controls (88.47,29.27) and (90.13,29.27) .. (91.8,27.6) .. controls (93.47,25.93) and (95.13,25.93) .. (96.8,27.6) .. controls (98.47,29.27) and (100.13,29.27) .. (101.8,27.6) .. controls (103.47,25.93) and (105.13,25.93) .. (106.8,27.6) .. controls (108.47,29.27) and (110.13,29.27) .. (111.8,27.6) .. controls (113.47,25.93) and (115.13,25.93) .. (116.8,27.6) .. controls (118.47,29.27) and (120.13,29.27) .. (121.8,27.6) .. controls (123.47,25.93) and (125.13,25.93) .. (126.8,27.6) .. controls (128.47,29.27) and (130.13,29.27) .. (131.8,27.6) .. controls (133.47,25.93) and (135.13,25.93) .. (136.8,27.6) .. controls (138.47,29.27) and (140.13,29.27) .. (141.8,27.6) .. controls (143.47,25.93) and (145.13,25.93) .. (146.8,27.6) .. controls (148.47,29.27) and (150.13,29.27) .. (151.8,27.6) .. controls (153.47,25.93) and (155.13,25.93) .. (156.8,27.6) .. controls (158.47,29.27) and (160.13,29.27) .. (161.8,27.6) .. controls (163.47,25.93) and (165.13,25.93) .. (166.8,27.6) .. controls (168.47,29.27) and (170.13,29.27) .. (171.8,27.6) .. controls (173.47,25.93) and (175.13,25.93) .. (176.8,27.6) .. controls (178.47,29.27) and (180.13,29.27) .. (181.8,27.6) .. controls (183.47,25.93) and (185.13,25.93) .. (186.8,27.6) .. controls (188.47,29.27) and (190.13,29.27) .. (191.8,27.6) .. controls (193.47,25.93) and (195.13,25.93) .. (196.8,27.6) .. controls (198.47,29.27) and (200.13,29.27) .. (201.8,27.6) .. controls (203.47,25.93) and (205.13,25.93) .. (206.8,27.6) .. controls (208.47,29.27) and (210.13,29.27) .. (211.8,27.6) .. controls (213.47,25.93) and (215.13,25.93) .. (216.8,27.6) .. controls (218.47,29.27) and (220.13,29.27) .. (221.8,27.6) .. controls (223.47,25.93) and (225.13,25.93) .. (226.8,27.6) .. controls (228.47,29.27) and (230.13,29.27) .. (231.8,27.6) -- (233.8,27.6) -- (233.8,27.6) ;
\draw    (61.8,203.6) .. controls (63.47,201.93) and (65.13,201.93) .. (66.8,203.6) .. controls (68.47,205.27) and (70.13,205.27) .. (71.8,203.6) .. controls (73.47,201.93) and (75.13,201.93) .. (76.8,203.6) .. controls (78.47,205.27) and (80.13,205.27) .. (81.8,203.6) .. controls (83.47,201.93) and (85.13,201.93) .. (86.8,203.6) .. controls (88.47,205.27) and (90.13,205.27) .. (91.8,203.6) .. controls (93.47,201.93) and (95.13,201.93) .. (96.8,203.6) .. controls (98.47,205.27) and (100.13,205.27) .. (101.8,203.6) .. controls (103.47,201.93) and (105.13,201.93) .. (106.8,203.6) .. controls (108.47,205.27) and (110.13,205.27) .. (111.8,203.6) .. controls (113.47,201.93) and (115.13,201.93) .. (116.8,203.6) .. controls (118.47,205.27) and (120.13,205.27) .. (121.8,203.6) .. controls (123.47,201.93) and (125.13,201.93) .. (126.8,203.6) .. controls (128.47,205.27) and (130.13,205.27) .. (131.8,203.6) .. controls (133.47,201.93) and (135.13,201.93) .. (136.8,203.6) .. controls (138.47,205.27) and (140.13,205.27) .. (141.8,203.6) .. controls (143.47,201.93) and (145.13,201.93) .. (146.8,203.6) .. controls (148.47,205.27) and (150.13,205.27) .. (151.8,203.6) .. controls (153.47,201.93) and (155.13,201.93) .. (156.8,203.6) .. controls (158.47,205.27) and (160.13,205.27) .. (161.8,203.6) .. controls (163.47,201.93) and (165.13,201.93) .. (166.8,203.6) .. controls (168.47,205.27) and (170.13,205.27) .. (171.8,203.6) .. controls (173.47,201.93) and (175.13,201.93) .. (176.8,203.6) .. controls (178.47,205.27) and (180.13,205.27) .. (181.8,203.6) .. controls (183.47,201.93) and (185.13,201.93) .. (186.8,203.6) .. controls (188.47,205.27) and (190.13,205.27) .. (191.8,203.6) .. controls (193.47,201.93) and (195.13,201.93) .. (196.8,203.6) .. controls (198.47,205.27) and (200.13,205.27) .. (201.8,203.6) .. controls (203.47,201.93) and (205.13,201.93) .. (206.8,203.6) .. controls (208.47,205.27) and (210.13,205.27) .. (211.8,203.6) .. controls (213.47,201.93) and (215.13,201.93) .. (216.8,203.6) .. controls (218.47,205.27) and (220.13,205.27) .. (221.8,203.6) .. controls (223.47,201.93) and (225.13,201.93) .. (226.8,203.6) .. controls (228.47,205.27) and (230.13,205.27) .. (231.8,203.6) -- (233.8,203.6) -- (233.8,203.6) ;
\draw [color={rgb, 255:red, 74; green, 144; blue, 226 }  ,draw opacity=1 ]   (164.68,98.16) ;
\draw [shift={(164.68,98.16)}, rotate = 0] [color={rgb, 255:red, 74; green, 144; blue, 226 }  ,draw opacity=1 ][fill={rgb, 255:red, 74; green, 144; blue, 226 }  ,fill opacity=1 ][line width=0.75]      (0, 0) circle [x radius= 2.01, y radius= 2.01]   ;
\draw [color={rgb, 255:red, 144; green, 19; blue, 254 }  ,draw opacity=1 ]   (214.35,48.29) ;
\draw [shift={(214.35,48.29)}, rotate = 0] [color={rgb, 255:red, 144; green, 19; blue, 254 }  ,draw opacity=1 ][fill={rgb, 255:red, 144; green, 19; blue, 254 }  ,fill opacity=1 ][line width=0.75]      (0, 0) circle [x radius= 2.01, y radius= 2.01]   ;
\draw [color={rgb, 255:red, 139; green, 87; blue, 42 }  ,draw opacity=1 ][line width=1.5]    (178.13,67.71) -- (206.2,95.6) ;
\draw [shift={(176,65.6)}, rotate = 44.81] [color={rgb, 255:red, 139; green, 87; blue, 42 }  ,draw opacity=1 ][line width=1.5]    (8.53,-2.57) .. controls (5.42,-1.09) and (2.58,-0.23) .. (0,0) .. controls (2.58,0.23) and (5.42,1.09) .. (8.53,2.57)   ;
\draw  [draw opacity=0][fill={rgb, 255:red, 255; green, 255; blue, 255 }  ,fill opacity=1 ] (58.2,22.8) -- (107,22.8) -- (107,210) -- (58.2,210) -- cycle ;

\draw (146.61,79.4) node  [color={rgb, 255:red, 74; green, 144; blue, 226 }  ,opacity=1 ]  {$\delta S_{\text{dyn}
}( v _{1})$};
\draw (183.26,43.2) node  [color={rgb, 255:red, 144; green, 19; blue, 254 }  ,opacity=1 ]  {$\delta S_{\text{dyn}} ( v _{2})$};
\draw (220.17,103.38) node  [color={rgb, 255:red, 139; green, 87; blue, 42 }  ,opacity=1 ]  {$\delta T_{ab}$};

\end{tikzpicture}
        \caption{Physical process version of the first law. An external matter source, described by the stress-energy tensor variation $\delta T_{ab},$    crosses the horizon between two generic horizon slices $\mathcal C(v_1)$ and $\mathcal C(v_2)$.}
        \label{fig:physical-1st-law}
    \end{figure}
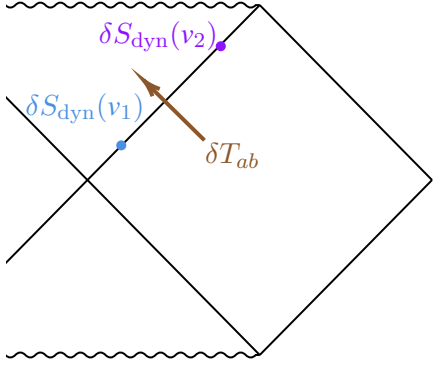

   In the presence of an external matter field, let us separate the total Lagrangian $\tilde{\mathbf L}$ into  that of the original system $\mathbf L(\phi)$ and that of the external matter $\mathbf L^\text{m}(\psi)$:
    \begin{equation}
        \tilde{\mathbf L}(\phi, \psi) = \mathbf L(\phi) + \mathbf L^{\text{m}}(g,\psi)\,.
    \end{equation}
    The variation of the original Lagrangian and external matter Lagrangian  can be expressed as 
    \begin{equation}
        \delta \mathbf L = \frac{1}{2} \bm E_{ab} \delta g^{ab} + \bm{\mathcal E}^A \delta \varphi_A + \dd{\bm \Theta(\phi,\delta \phi)}, \quad \delta \mathbf L^{\text{m}} = - \frac{1}{2} \bm T_{ab} \delta g^{ab} + \mathbf E^\text{m} \delta \psi + \dd{\bm \Theta^{\text{m}}(g,\psi, \delta \psi)}\,,
    \end{equation}
    where $E_{ab}, \mathcal E^A$ are the metric and matter field equations  of the original system, $T_{ab}$ is the stress-energy tensor of the minimally coupled external matter fields, and $E^\text{m}$ is the equation of motion for $\psi$. (The boldface symbols of the above equations of motion mean the product of that with the volume form $\bm \epsilon$.) Now, for the combined Lagrangian, the equation of motion for the metric takes the form 
    \begin{equation}
    \label{metriceomc}
        E_{ab} =  T_{ab},
    \end{equation}
    exactly as a generalisation of the Einstein equation, $\frac{1}{8\pi G}G_{ab} = T_{ab}$ in the case of general relativity coupled to minimal matter sources. 

    We repeat the steps in Section \ref{ssec:cps} for the original Lagrangian, keeping in mind that we have switched on external matter sources. Then, if  the equations of motion $\mathcal E^A = 0$ (which follows from our assumption that $\psi$ and $\varphi$ are decoupled) and $E_{ab}= T_{ab}$ are imposed, the fundamental identity \eqref{eq:q-theta-c} is modified as
    \begin{equation}
        \dd{(\delta_\phi \mathbf Q_{\xi} - \xi \cdot  \bm \Theta(\phi,\delta \phi))} = - \delta T^{ab} \xi_b \bm \epsilon_a - \frac{1}{2} \xi \cdot \bm T_{ab} \delta g^{ab} \,,\label{eq:grav-q-theta-c}
    \end{equation}
    where we have inserted the constraint form $\mathbf C_{\xi} = - E^{ab} \xi_b \bm \epsilon_a = - T^{ab} \xi_b \bm \epsilon_a$.

    Next, we integrate  \eqref{eq:grav-q-theta-c}  on an interval interpolating between two slices $\mathcal{C}(v_1)$ and $\mathcal{C}(v_2)$ on the horizon $\mathcal{H}^+$. The left-hand side then becomes  the difference in the first-order variation of the dynamical entropy between $\mathcal C(v_1)$ and $\mathcal C(v_2)$,
    \begin{equation}
        \left(\int_{\mathcal C(v_2)} - \int_{\mathcal C(v_1)}\right) (\delta_\phi \mathbf Q_\xi - \xi \cdot \bm \Theta(\phi, \delta \phi)) = \frac{\kappa}{2\pi}\left(\delta S_\text{dyn}(v_2) - \delta S_\text{dyn}(v_1)\right) = \frac{\kappa}{2\pi} \Delta \delta S_\text{dyn}\,,
    \end{equation}
    and integrating the right-hand side of \eqref{eq:grav-q-theta-c}  yields (note $\xi \cdot \bm T_{ab}$ vanishes as $\xi$ is tangent to the horizon)
    \begin{equation}
        - \int_{\mathcal C(v_1)}^{\mathcal C(v_2)} \delta T^{ab} \xi_b \bm \epsilon_a = \int_{\mathcal C(v_1)}^{\mathcal C(v_2)} \delta T_{ab} k^a \xi^b \bm \epsilon_{\mathcal H^+} = \int_{v_1}^{v_2} \dd{v} \int_{\mathcal{C}(v)} \dd A\, \delta T_{ab} \xi^a k^b.
    \end{equation}
    Here the symmetry of $T_{ab}$ is assumed, $\bm \epsilon_{\mathcal H^+} = - l \wedge \bm \epsilon_{\mathcal C}$ is the volume   on the horizon, and we used the following property for the pullback of $T^{ab} \bm \epsilon_a$ to the horizon: 
    \begin{equation}
       \delta T^{ab} \bm \epsilon_a = \delta T^{ab} \delta_a^c \bm \epsilon_c \fheq - \delta T^{ab} k_a l^c \bm \epsilon_c = - \delta T^{ab} k_a \bm \epsilon_{\mathcal H^+}, \label{eq:T-epsilon}
    \end{equation}
    and we identified $l^c \bm \epsilon_c = \bm \epsilon_{\mathcal H^+}$.

    Thus,   we obtain a   physical process first law between two arbitrary horizon slices
    \begin{equation}
        \Delta \delta S_{\text{dyn}}  = \frac{2 \pi}{\kappa}\int_{v_1}^{v_2} \dd{v} \int_{\mathcal{C}(v)} \dd A\, \delta T_{ab} \xi^a k^b, \label{eq:ppfl}
    \end{equation}
    which relates the difference in entropy between two horizon slices with the infalling null energy density of the external matter field across the horizon.

    \subsection{The Linearised Second Law}
    \label{sec:linearized2ndlaw}
    Here,   we derive a constant and non-decreasing  second law for dynamical black hole entropy at the linearised level from the comparison version and the physical process version of first laws, respectively. This was also observed in  \cite{Hollands:2024vbe}.

The comparison first law holds for any horizon cross-section, which implies that the dynamical entropy variation at one cross-section $\mathcal{C}(v_1)$ is equal to the dynamical entropy variation at another cross-section 
    \begin{equation}
            \delta S_\text{dyn}(v_1) = \delta S_\text{dyn}(v_2).
    \end{equation}
 This suggests that, locally, the dynamical entropy satisfies a constant linearised second law  
    \begin{equation}
        \partial_v \delta S_\text{dyn} = 0.
    \end{equation}
On the other hand, from  the physical process first law we can prove a stronger second law, namely that the entropy is non-decreasing on the horizon, assuming the null energy condition  $\delta T_{ab} k^a k^b \geq 0$ holds for the external matter source.  
This is just the local version of the physical process version of first law, \eqref{eq:ppfl}, between two infinitesimally close slices on the horizon:
    \begin{equation}\label{eq:grav-second-law}
        \partial_v \delta S_\text{dyn}   = \frac{2 \pi}{\kappa} \int_{\mathcal{C}(v)} \dd{A}\, \delta T_{ab}   \xi^a k^b \geq 0.
    \end{equation}
   For a typical field theory  the stress-energy tensor is quadratic in the perturbation. In order to have a non-trivial increase in the entropy at first order, we may tune the perturbation parameter for the external matter source  such that $\delta T_{ab} k^a k^b$ contributes to the first order perturbation of the entropy. That is,  we set $\delta \phi \sim \mathcal O(\epsilon)$ but $\delta \psi\sim \mathcal O(\epsilon^{1/2})$ for some small bookkeeping parameter~$\epsilon$ for the perturbation. 
   
   Thus, we conclude   that the linearised second law for dynamical black hole entropy, that we derived for general relativity   from the Raychaudhuri equation in Section \ref{sec:ray}, continues to hold for arbitrary theories of gravity.

 \subsection{Relation to Wall Entropy}
 \label{sec:relationwall}

 \label{sec:wallrelation}
    Here, we investigate the relationship between the dynamical entropy, defined as an improved Noether charge, and the Wall entropy.\footnote{We only consider the case  where the metric, scalar fields and vector fields are the dynamical fields. When matter fields with spin $s \geq 2$  are present, the Wall entropy may not be well defined unless certain  `integrability conditions'  are satisfied by the  matter fields. Once these conditions are imposed, there is a similar relation between the Wall entropy and $S_{\text{dyn}}$. This will be investigated in \cite{Yan2024}. }  This relationship was established before in \cite{Hollands:2024vbe} (see also~\cite{Kar:2024dqk}).    In \cite{Wall2015}, the Wall entropy is defined in terms of  the null-null component of the metric field equation  as  
    \begin{equation}
    \label{walldeff1}
        \partial_v^2 \delta S_{\text{Wall}} = - 2 \pi \int_{\mathcal{C}(v)} \dd{A} \delta E_{ab} k^a k^b.
    \end{equation}
    To obtain a similar expression for the dynamical entropy involving $\delta E_{ab}$, we contract the fundamental identity \eqref{eq:q-theta-c} with the null translation vector $k$ and pull it back to a horizon cross-section 
    \begin{equation}
        k \cdot \dd{(\delta_\phi \tilde{\mathbf Q}_\xi)} \cceq k \cdot \delta_\phi \mathbf C_\xi - k \cdot (\xi \cdot \mathbf E(\phi)) \delta \phi \cceq - \delta E^{ab} \xi_b (k \cdot \bm \epsilon_a) \cceq \delta E_{ab} k^a \xi^b \bm \epsilon_{\mathcal C} \,, 
    \end{equation}
    where the second equality follows from the fact that $k\cdot (\xi \cdot \mathbf E(\phi))$ vanishes because $\xi$ is proportional to $k$ on the horizon, and  $c_{ab}k^a \xi^b$ in the constraint form (see equation \eqref{eq:cab}) vanishes to first order in the perturbation for scalars or vectors. In the last equality, we used $\delta E^{ab} \bm \epsilon_a \fheq - \delta E^{ab} k_a \bm \epsilon_{\mathcal H^+}$ (using the same argument as \eqref{eq:T-epsilon}), and $k \cdot \bm \epsilon_{\mathcal H^+} \cceq \bm \epsilon_{\mathcal C}$. Rearrange the above equation, we arrive at
    \begin{equation}
        \mathcal L_k (\delta_\phi \tilde{\mathbf Q}_\xi) = \delta E_{ab} k^a \xi^b \bm \epsilon_{\mathcal C} + \dd{(k \cdot \delta_\phi \tilde{\mathbf Q}_\xi)},
    \end{equation}
    where we have  used  the Cartan-Killing equation \eqref{cartaneq}. Further, integrating the above identity on some compact horizon slice $\mathcal C(v)$ and interpreting $\mathcal L_k$  acting on   scalar quantities as $\partial_v$, we find a similar expression as \eqref{eq:grav-second-law} for the dynamical entropy 
    \begin{equation}
        \partial_v \delta S_\text{dyn} = \frac{2\pi}{\kappa} \int_{\mathcal{C}(v)} \dd{A} \delta E_{ab} \xi^a k^b\,.
    \end{equation}
     By substituting the expression for the   Killing field on the horizon $\xi^a \fheq \kappa v k^a$, this  reads 
    \begin{equation}
        \partial_v \delta S_\text{dyn} = 2\pi v \int_{\mathcal{C}(v)} \dd{A} \delta E_{ab} k^a k^b = - v \partial_v^2 \delta S_{\text{Wall}}\,,
    \end{equation}
  where   in the second equality we used  the definition of Wall entropy \eqref{walldeff1}. 
        As this should be valid at any affine time $v$, we have 
    \begin{equation}
        \partial_v \delta S_\text{dyn} = - v \partial_v^2 \delta S_{\text{Wall}} = - \partial_v \left( v \partial_v \delta S_{\text{Wall}}\right) + \partial_v \delta S_{\text{Wall}} = \partial_v \left((1 - v \partial_v) \delta S_{\text{Wall}}\right)\,.
    \end{equation}
    Thus we find that the relation between dynamical entropy   and Wall entropy  is (after pulling out the $\delta$)
    \begin{equation}
    \label{eq:walldynbla}
        S_\text{dyn} = (1 - v \partial_v) S_{\text{Wall}}\,,
    \end{equation}
    where the integration constant is zero because at the bifurcation surface $\mathcal{B}$ we have the following matching condition:
    \begin{equation}
        S_\text{dyn} \overset{\mathcal{B}}{=} S_{\text{Wall}} \overset{\mathcal{B}}{=} S_{\text{Wald}}
    \end{equation}
    from the properties of $S_\text{dyn}$ and $S_\text{Wall}$. We check this relation \eqref{eq:walldynbla} in Section \ref{sec:fofriemann} for any $f(\text{Riemann})$ theory by evaluating the improved Noether charge using GNC near the horizon.  

    Finally, by inserting the  relation \eqref{eq:walldynbla} for the dynamical entropy into the first law \eqref{newfirstlaaa}, we obtain the non-stationary comparison first law for general theories of gravity
 \begin{equation}
     \delta M - \Omega_{\mathcal H} \delta  J = T \delta \left (S_{\text{Wall}} - v \dv{v} S_{\text{Wall}} \right)\,.
 \end{equation}

    \subsection{Invariance under JKM Ambiguities to First Order}\label{sec:jkm-inv}
    Here, we give a brief review of the Jacobson-Kang-Myers (JKM) ambiguities \cite{Jacobson:1993vj} (see also~\cite{Iyer:1994ys}) of covariant phase space quantities, and we prove that the dynamical black hole entropy defined above is invariant under these ambiguities for first-order perturbations of a stationarity background (see also \cite{Hollands:2024vbe}). We also mention that the Wall entropy has recently been proven to be unambiguous to first order in the perturbation \cite{Wall:2024lbd}. 
    
    In the differential form language, ambiguities arise for quantities that are defined up to exact terms. In the covariant phase space formalism, there are three different types of ambiguities  present at different levels: for the Lagrangian form $\mathbf L$, the symplectic potential $\bm \Theta$ and the Noether charge $\mathbf Q$. The JKM ambiguities are
    \begin{align}
        \mathbf L(\phi) & \to \mathbf L(\phi) + \dd{\bm \mu(\phi)},\\
        \mathbf \Theta(\phi, \delta \phi) & \to \mathbf \Theta(\phi, \delta \phi) + \dd{\mathbf Y(\phi, \delta \phi)},\\
        \mathbf Q_\chi & \to \mathbf Q_\chi + \dd{\mathbf Z(\phi, \chi)}.
    \end{align}
    Here,   $\mathbf Y(\phi, \delta \phi)$ is required to be linear in $\delta \phi$  and $\mathbf Z(\phi, \chi)$ must be linear in the vector field $\chi^a$, in order to match $\mathbf \Theta$ and $\mathbf Q_\chi$, respectively.

    Under the above ambiguities, the quantities of our interest transform as 
    \begin{align}
        \mathbf \Theta(\phi, \delta \phi) &\to \mathbf \Theta(\phi, \delta \phi) + \delta \bm \mu(\phi) + \dd{\mathbf Y(\phi, \delta \phi)}\\
        \mathbf Q_\xi & \to \mathbf Q_\xi + \xi \cdot \bm \mu(\phi) + \mathbf Y(\phi, \mathcal L_\xi \phi) + \dd{\mathbf Z(\phi,\xi)}.
    \end{align}
    and the variation of $\mathbf Q_\xi$ transform as 
    \begin{equation}
        \delta \mathbf Q_\xi \to \delta \mathbf Q_\xi + \delta \xi \cdot \bm \mu(\phi) + \xi \cdot \delta \bm \mu(\phi) + \mathbf Y(\phi, \mathcal L_{\delta \xi} \phi) + \mathbf Y(\phi, \mathcal L_\xi \delta \phi) + \dd{(\delta \mathbf Z(\phi,\xi))}\,,
    \end{equation}
    where we have applied the product rule for $\delta $ to the variation of $\xi \cdot \bm \mu$ and $\bm Y$. Notice that in calculating $\delta \mathbf Y$, the only non-vanishing terms are those where $\delta$ is applied to the Lie derivative $\mathcal L_\xi \phi$, because $\mathcal L_\xi \phi = 0$ in the Killing background.

    We are interested in the ambiguities in the field   variation of the improved Noether charge, after which the dynamical entropy is defined, 
    \begin{equation}
        \delta_\phi \tilde {\mathbf{Q}}_\xi = \delta_\phi \mathbf Q_\xi - \xi \cdot \bm \Theta(\phi, \delta \phi)\,.
    \end{equation} Under the JKM ambiguities this transforms as 
    \begin{equation}
        \begin{split}
            \delta_\phi \tilde {\mathbf{Q}}_\xi & \to \delta_\phi \tilde {\mathbf{Q}}_\xi  + \mathbf Y(\phi, \mathcal L_\xi \delta \phi) - \xi \cdot \dd{\mathbf Y(\phi, \delta \phi)} + \dd{(\delta \mathbf Z(\phi, \xi) - \mathbf Z(\phi, \delta \xi))}\\
            & = \delta_\phi \tilde {\mathbf{Q}}_\xi + \dd{\left(\xi \cdot \mathbf Y(\phi, \delta \phi) + \delta_\phi \mathbf Z(\phi, \xi)\right)}\,,
        \end{split}
    \end{equation}
    where we have identified
    \begin{equation}
        \mathbf Y(\phi, \mathcal L_\xi \delta \phi) = \mathcal L_\xi \mathbf Y(\phi, \delta \phi) = \dd{(\xi \cdot \mathbf Y(\phi, \delta \phi))} + \xi \cdot \dd{\mathbf Y(\phi, \delta \phi)},
    \end{equation}
    using the  background stationarity condition $\mathcal L_\xi \phi = 0$, and we notice
    \begin{equation}
        \delta_\phi \mathbf Z(\phi, \xi) = \delta \mathbf Z(\phi, \xi) - \mathbf Z(\phi, \delta \xi). 
    \end{equation}
Thus, the  ambiguities that arise in $\delta_\phi \tilde {\mathbf{Q}}_\xi$ are exact differentials, which vanish when   integrated  over a compact horizon cross-section, due to Stokes' theorem. Hence, because of the definition in \eqref{vardynennoether},  $\delta S_{\text{dyn}} = (2\pi/ \kappa)\int_{\mathcal C} \delta_\phi \tilde {\mathbf{Q}}_\xi$, this implies the first-order variation of the dynamical black hole entropy   is JKM-invariant. In other words, the dynamical entropy itself is unambiguous to leading order for perturbations off stationary backgrounds. It is straightforward to apply this discussion to the case where   the gravitational and matter sectors are treated separately, as in Section \ref{sec:physical2}, because we can treat the gravitational part of the Lagrangian independently.

    \subsection{Exactness of the Symplectic Potential on the Horizon} \label{sec:exactness}
    Here, we systematically analyse the structure of $\bm \Theta(\phi, \delta \phi)$ using the powerful GNC near the future horizon and the associated boost weight argument, and we show that the symplectic potential in any diffeomorphism invariant theory would satisfy the consistency condition a) in Section \ref{ssec:comparison-1st-law} on a linearly perturbed Killing horizon. This is done by proving that, with a stationary background, $\bm \Theta(\phi, \delta \phi)$ is exact in $\delta$ when pulled back to the horizon, i.e., there exists some $\mathbf B_{\mathcal H^+}(\phi)$ on the horizon such that
    \begin{equation}
    \label{thetepulledbackk}
        \bm \Theta(\phi, \delta \phi) \fheq \delta \mathbf B_{\mathcal H^+}(\phi).
    \end{equation}
    We also demonstrate that the $\mathbf B_{\mathcal H^+}$ satisfying this relation vanishes on the background Killing horizon.  This relation \eqref{thetepulledbackk} was also established   in \cite{Hollands:2024vbe} using Killing field arguments, instead of boost weight arguments, for   diffeomorphism covariant Lagrangians that  depend only on the metric field. Below we  provide an independent, more general proof for arbitrary diffeomorphism covariant Lagrangians that depend on the metric and any (non-minimally coupled) bosonic matter field that is smooth on the horizon. 
    
    According to Iyer and Wald \cite{Iyer:1994ys}, the most general form of the symplectic potential $\bm \Theta(\phi, \delta \phi)$    is (up to certain JKM ambiguities)
    \begin{equation}
        \begin{aligned}
            \bm \Theta(\phi, \delta \phi) & = \bm \epsilon_a \Bigg( 2 X^{abcd} \nabla_d \delta g_{bc} + S^{abc} \delta g_{bc}\\
            & \qquad   + \sum_{k=0}^{n-1}T_{(k)}^{abcdef_1\cdots f_k} \delta \nabla_{(f_1\cdots f_k)} R_{bcde} + \sum_{k=0}^{m-1}U_{(k)}^{aAb_1\cdots b_k} \delta \nabla_{(b_1\cdots b_k)} \varphi_A\Bigg)
        \end{aligned}
    \end{equation}
    for a general diffeomorphism covariant theory with Lagrangian form
    \begin{equation}
        \mathbf L = L(g^{ab}, R_{abcd}, \nabla_{e_1} R_{abcd}, \cdots, \nabla_{(e_1 \cdots e_n)}R_{abcd}, \varphi_A, \nabla_{b_1} \varphi_A, \cdots, \nabla_{(b_1\cdots b_m)}\varphi_A) \bm \epsilon\,.
    \end{equation} 
    Here,
    \begin{equation}
        X^{abcd} = \frac{\delta L}{\delta R_{abcd}} = \pdv{L}{R_{abcd}} - \nabla_{e_1} \pdv{L}{(\nabla_{e_1}R_{abcd})} + \cdots + (-1)^n \nabla_{e_1 \cdots e_n} \pdv{L}{(\nabla_{(e_1 \cdots e_n)}R_{abcd})}
    \end{equation}
    which is the functional derivative of the Lagrangian $L$ with respect to the Riemann tensor $R_{abcd}$, and $S,T,U$ are some field-dependent tensors that can be derived from the Lagrangian using the methods elaborated in \cite{Iyer:1994ys}. For our discussion, we do not need the details of these tensors but only their general index structure.

    Consider the pullback of $\bm \Theta(\phi, \delta \phi)$ to the horizon $\mathcal H^+$. Expressing in GNC, we get
    \begin{equation}
        \begin{aligned}
            \bm \Theta(\phi, \delta \phi) & \fheq - l \wedge \bm \epsilon_{\mathcal C} \Bigg( 2 X^{u bcd} \nabla_d \delta g_{bc} + S^{u bc} \delta g_{bc}\\
            & \qquad \qquad   + \sum_{k=0}^{n-1}T_{(k)}^{u bcdef_1\cdots f_k} \delta \nabla_{(f_1\cdots f_k)} R_{bcde} + \sum_{k=0}^{m-1}U_{(k)}^{u Ab_1\cdots b_k} \delta \nabla_{(b_1\cdots b_k)} \varphi_A\Bigg)
        \end{aligned}
    \end{equation}
    where we have kept the only non-vanishing $(D-1)$-form $\bm \epsilon_u$ when pulling back to the horizon, and identified
    \begin{equation}
        \bm \epsilon_u = - l \wedge \bm \epsilon_{\mathcal C}.
    \end{equation}
    We analyse the structure term by term. First look at the $T$ terms. Depending on the boost weight $w$ of the combinations of indices $bcdef_1\cdots f_k$, we can write it as a sum of different weight combinations:
    \begin{equation}
        \begin{split}
            - l \wedge \bm \epsilon_{\mathcal C} \sum_w T_{(1-w)} \delta (\nabla_{\cdots}R)_{(w)} & \fheq - l \wedge \bm \epsilon_{\mathcal C} \sum_{w\geq 1} T_{(1-w)} \delta (\nabla_{\cdots}R)_{(w)}\\
            & \fheq \delta \left( - l \wedge \bm \epsilon_{\mathcal C} \sum_{w\geq 1} T_{(1-w)} (\nabla_{\cdots}R)_{(w)} \right).
        \end{split}
    \end{equation}
    Here, we have neglected the detailed indices combination as we only care about the boost weight $w$ they carry. The first equality holds as $T_{(1-w)}$'s with $w\leq 0$ are proportional to positive weight background quantities, which vanish. In the second equality we have ``pulled the $\delta$ to the front'' for the non-vanishing terms because, for a non-positive weight component $A_{(w\leq 0)}$ and any positive weight component $B_{(w>0)}$, we always have  
    \begin{equation}
        A_{(w\leq 0)} \delta B_{(w>0)} = \delta \left(A_{(w\leq 0)} B_{(w>0)}\right) - B_{(w>0)} \delta A_{(w\leq 0)} \fheq \delta \left(A_{(w\leq 0)} B_{(w>0)}\right)
    \end{equation}
    as background positive-weight tensor components vanish on the horizon. 
    
    The same argument also works for the last term involving $U$:
    \begin{equation}
        - l \wedge \bm \epsilon_{\mathcal C} \sum_{k=0}^{m-1}U_{(k)}^{u Ab_1\cdots b_k} \delta \nabla_{(b_1\cdots b_k)} \varphi_A \fheq \delta \left( - l \wedge \bm \epsilon_{\mathcal C} \sum_{w\geq 1} U_{(1-w)} (\nabla_{\cdots}\varphi)_{(w)} \right).
    \end{equation}
    Now we just need to show the first two terms in $\bm \Theta(\phi, \delta \phi)$ are exact. The easier one is the $S$-term:
    \begin{equation}
        S^{u ab} \delta g_{ab} \fheq S^{u ij} \delta \gamma_{ij} \fheq 0
    \end{equation}
    by our gauge conditions and boost weight analysis.

    To deal with the first term involving $X$ we first compute
    \begin{equation}
        \nabla_u \delta g_{v i} \fheq - \delta \omega_i, \quad \nabla_u \delta g_{ij} \fheq 2 \delta \bar K_{ij}, \quad \nabla_v \delta g_{ij} \fheq 2 \delta K_{ij}\,,
    \end{equation}
    where $\bar K_{ij}=\frac{1}{2}\partial_u \gamma_{ij}$ and $K_{ij}=\frac{1}{2}\partial_v \gamma_{ij}$ are the  extrinsic curvatures in the $u$- and $v$-directions, respectively,   
    and we identify the vanishing components of $X$ by a boost weight argument
    \begin{equation}
        X^{u v i u} \fheq X^{u i j u} \fheq X^{u ijk} \fheq 0. \label{eq:X-zero-cpts}
    \end{equation}
    Then, expanding the possible combinations of indices $b,c,d$ and keeping in mind the symmetries of the Riemann tensor, we obtain
    \begin{equation}
        - 2 l \wedge \bm \epsilon_{\mathcal C} X^{u bcd} \nabla_d \delta g_{bc} \fheq \delta \left(- 4 l \wedge \bm \epsilon_{\mathcal C} X^{u i j v} K_{ij}\right).
    \end{equation}
    The final result is
    \begin{equation}
        \bm \Theta(\phi, \delta \phi) \fheq \delta \left[ - l \wedge \bm \epsilon_{\mathcal C} \left( 4 X^{u i j v} K_{ij} + \sum_{w\geq 1} \Big( T_{(1-w)} (\nabla_{\cdots}R)_{(w)} + U_{(1-w)} (\nabla_{\cdots} \varphi)_{(w)} \Big)  \right) \right]
    \end{equation}
    Thus, we find  
    \begin{equation}
        \mathbf B_{\mathcal H^+}(\phi) = - l \wedge \bm \epsilon_{\mathcal C} \left( 4 X^{u i j v} K_{ij} + \sum_{w\geq 1} \Big( T_{(1-w)} (\nabla_{\cdots}R)_{(w)} + U_{(1-w)} (\nabla_{\cdots} \varphi)_{(w)} \Big)  \right).
    \end{equation}
    This closes the proof of exactness of $\bm \Theta(\phi, \delta \phi)$ in $\delta$ on the horizon, which was used in Section \ref{ssec:comparison-1st-law} to define the dynamical black hole entropy. The boost weight of $\mathbf B_{\mathcal H^+}$ is $+1$, because $X^{u i j v}$ has weight $0$ and $K_{ij}$ has   weight $+1$, hence it follows that $\mathbf B_{\mathcal H^+}$ vanishes on the background Killing horizon. 

    \subsection{Structural Analysis of the (Improved) Noether Charge}
    \label{sec:structural}
    In proving the exactness of $\bm \Theta(\phi, \delta \phi)$, we have analysed its structure using GNC. A similar procedure can be carried out for the Noether charge $\mathbf Q_\xi$ in any diffeomorphism covariant theory. We will do this in order to 
    \begin{enumerate}
        \item[1)] Prove the consistency condition b) for the Noether charge, which is required for the dynamical entropy to well defined (see Section \ref{ssec:comparison-1st-law});
        \item[2)] Analyse the general structure of the dynamical entropy, and highlight how it is different from the Wald entropy.
    \end{enumerate}

    \paragraph{Structure of Noether Charge} We first perform a structural analysis of the Noether charge in GNC. Iyer and Wald \cite{Iyer:1994ys} showed that the most general form of the Noether charge is given by 
    \begin{equation}
        \mathbf Q_\xi = \bm \epsilon_{ab} \left(- X^{abcd} \nabla_{[c} \xi_{d]} + W^{abc} \xi_c \right) + \mathbf Y(g,\varphi,\mathcal{L}_\xi g, \mathcal{L}_\xi \varphi) + \dd{\mathbf Z}
    \end{equation}
    where $W^{abc}$ is given by the theory, $\mathbf Y$ and $\mathbf Z$ are two different types of   JKM ambiguity~\cite{Jacobson:1993vj} that arise for the Noether charge. As shown in Section \ref{sec:jkm-inv}, the dynamical black hole entropy is invariant under JKM ambiguities up to an exact term (which integrates to zero on compact slices), therefore, in the following discussion, we will set $\mathbf Y = 0 $ and $\dd{\mathbf Z} = 0$ without loss of generality. These ambiguities will essentially cancel with the ambiguities appearing in $\xi \cdot   \mathbf B_{\mathcal H^+}$ in the improved Noether charge.

    When pulling back the Noether charge to a horizon slice, we obtain  in GNC, 
    \begin{equation}
        \mathbf Q_\xi \ceq - 2\bm \epsilon_{\mathcal C} \left(X^{u v c d}\nabla_{[c}\xi_{d]} + W^{u v c} \xi_c \right).
    \end{equation}
    Using equation \eqref{eq:X-zero-cpts}, equation \eqref{surfgrav1}, and calculating $\nabla_{[u}\xi_{i]} \fheq \frac{1}{2} \omega_i \xi_u$ in GNC, we have
 \begin{equation}
        \mathbf Q_\xi \ceq - 2\bm \epsilon_{\mathcal C} \left( 2 \kappa_3 X^{u v u v} - \xi_u  \tilde W^{u v u} \right). \label{eq:Q-xi}
    \end{equation}
    where we defined $\tilde W^{u v u} = W^{u v u} - X^{u v u i}\omega_i$.  This expression of $\mathbf Q_\xi$ contains both the background and the first-order contribution, so we are not immediately setting $\kappa_3$ and $\xi_u$ to their background values $\kappa$ and $- \kappa v$. 

    \paragraph{Consistency Condition for Noether Charge} Using the structure of the Noether charge in GNC on the horizon, we now prove the consistency condition b) in Section \ref{ssec:comparison-1st-law}.
   In GNC, we obtain from   \eqref{eq:Q-xi} the Noether charge associated to the varied Killing vector field $\delta \xi^a$: 
    \begin{equation}
    \label{corrrr}
        \mathbf Q_{\delta \xi} \ceq -4 (\delta \kappa_3) X^{uvuv} \bm \epsilon_{\mathcal C} + 2 (\delta \xi_u) \tilde W^{uvu} \bm \epsilon_{\mathcal C} = - 4 (\delta \kappa_3) X^{uvuv} \bm \epsilon_{\mathcal C} = \frac{\delta \kappa_3}{\kappa} \mathbf Q_\xi
    \end{equation}
    where we notice that the weight-1 quantity $\tilde W^{uvu} \fheq 0$ on the background, and in deriving the last equality, we have evaluated equation \eqref{eq:Q-xi} at zeroth order. Then, we prove the consistency condition as required
    \begin{equation}
        \kappa \delta \left(\mathbf Q_\xi/\kappa_3\right) = \delta \mathbf Q_\xi - \frac{\delta \kappa_3}{\kappa} \mathbf Q_\xi = \delta \mathbf Q_\xi - \mathbf Q_{\delta \xi} =\delta_\phi \mathbf Q_\xi\,, 
    \end{equation}
    where we used $\kappa_3 = \kappa$ for the Killing horizon background.  
    
    We also make explicit a corollary of \eqref{corrrr}, when combined with consistency condition a), given in \eqref{consistencya}. That is, the field-only variation $\delta_\phi$ is, in fact, equivalent to the full variation $\delta$ when it acts on $S_\text{dyn}$. We start the proof by acting with the full variation $\delta$ on the dynamical black hole entropy 
    \begin{equation}
        S_\text{dyn} = \int_{\mathcal C} \frac{2 \pi}{\kappa_3} (\mathbf Q_\xi - \xi \cdot \mathbf B_{\mathcal H^+}(\phi)),
    \end{equation}
    which yields
    \begin{equation}
        \begin{split}
            \delta S_\text{dyn} &= 2 \pi \int_{\mathcal C} \left(\delta \left(\mathbf Q_\xi/\kappa_3 \right) -  \xi \cdot \delta \mathbf B_{\mathcal H^+}/\kappa\right)\\
            &= 2 \pi \int_{\mathcal C} \left(\delta_\phi  \mathbf Q_\xi/\kappa  - \xi \cdot \delta_\phi \mathbf B_{\mathcal H^+}/\kappa\right)\\
            &= \delta_\phi \left(\int_{\mathcal C} \frac{2 \pi}{\kappa} (\mathbf Q_\xi - \xi \cdot \mathbf B_{\mathcal H^+}(\phi))\right) = \delta_\phi S_\text{dyn}\,,
        \end{split}
    \end{equation}
    where we have used the fact that $\mathbf B_{\mathcal H^+}(\phi) \fheq 0$ in the stationary background, $\mathbf B_{\mathcal H^+}(\phi)$ is independent of $\xi$ so $\delta \mathbf B_{\mathcal H^+} = \delta_\phi \mathbf B_{\mathcal H^+}$, and $\kappa_3 = \kappa$ on the unperturbed Killing horizon. Also, note that $\delta_\phi$ does not act on the surface gravities, since their variation depends on $\delta \xi^a$ only.

    \paragraph{Structure of Improved Noether Charge} To find the general structure for the improved Noether charge (hence the dynamical entropy), we calculate
    \begin{equation}
    \label{noethernewchgo}
        \delta_\phi \mathbf Q_\xi \ceq \delta_\phi \left( -4 \kappa_3 X^{u v u v} \bm \epsilon_{\mathcal C} - 2 \kappa_2 v \tilde W^{u v u} \bm \epsilon_{\mathcal C} \right)\,,
    \end{equation} 
    where we have inserted the background value $\xi_u = - \kappa_2 v$, as $\delta_\phi$ does not vary $\xi^a$ on the horizon. Moreover, under $\delta_\phi$ we may identify $\kappa_2 = \kappa_3 = \kappa$, because they remain background values.
    
    Combining \eqref{noethernewchgo} with the structural formula for $\xi \cdot  \bm \Theta(\phi, \delta \phi)$, which follows from the contraction $\xi \cdot \mathbf B_{\mathcal H^+}(\phi)$, we find the improved Noether charge at first order in perturbation theory  takes the form
    \begin{align}
            & \delta_\phi \tilde {\mathbf Q}_\xi \ceq - 4 \kappa \delta (X^{u v u v} \bm \epsilon_{\mathcal C})\\
            & \qquad \qquad - \kappa v\delta\left[\bm \epsilon_{\mathcal C} \left( 2 \tilde W^{u v u} + 4 X^{u i j v} K_{ij} + \sum_{w\geq 1} \Big( T_{(1-w)} (\nabla_{\cdots}R)_{(w)} + U_{(1-w)} (\nabla_{\cdots} \varphi)_{(w)} \Big)  \right) \right]. \nonumber
        \end{align}
    Hence, the general structure of  dynamical black hole entropy is
    \begin{equation}
        S_\text{dyn} = S_{\text{Wald}} + v S_{+}\,,
    \end{equation}
    where 
    \begin{equation}
        S_{\text{Wald}} = - 8 \pi \int_{\mathcal{C}(v)} X^{u v u v} \bm \epsilon_{\mathcal C}
    \end{equation}
    is the Wald  entropy, and 
    \begin{equation}
        S_+ = - 2 \pi \int_{\mathcal{C}(v)}\bm \epsilon_{\mathcal C} \left( 2 \tilde W^{u v u} + 4 X^{u i j v} K_{ij} + \sum_{w\geq 1} \Big( T_{(1-w)} (\nabla_{\cdots}R)_{(w)} + U_{(1-w)} (\nabla_{\cdots} \varphi)_{(w)} \Big) \right)
    \end{equation}
    is a collection of tensor components that have total boost weight $+1$. This implies directly that   1) when $S_{\text{dyn}}$ is evaluated on a Killing horizon, $S_+$ vanishes, and the dynamical entropy reduces to the Wald  entropy; 2) when $S_+$ is evaluated at the bifurcate surface, $v=0$, the dynamical entropy also becomes the Wald  entropy.

    \section{Examples of Dynamical Black Hole Entropy}\label{sec:examples}

   In this section we calculate the dynamical black hole entropy explicitly, using the improved Noether charge formula \eqref{improvednoetherc}, for three examples:  general relativity, $f(R)$ gravity, and $f(\text{Riemann})$ theory, respectively. For general relativity and $f(R)$ gravity, a purely covariant computation is carried out, whereas in the case of $f(\text{Riemann)}$ theories we employ Gaussian null coordinates and the associated boost weight analysis to simplify the calculation.

    \subsection{General Relativity}
   The Lagrangian form for general relativity with a cosmological constant $\Lambda$ is
    \begin{equation}
        \mathbf L = \frac{1}{16 \pi G} (R- 2 \Lambda)  \bm \epsilon,
    \end{equation}
  For this Lagrangian the  symplectic potential is \cite{Iyer:1994ys}
    \begin{equation}
        \bm \Theta(\phi, \delta \phi) = \frac{1}{16 \pi G} \bm \epsilon_{a} g^{ab} g^{cd} \left( \nabla_c \delta g_{bd} - \nabla_b \delta g_{cd} \right)
    \end{equation}
  and the Noether charge is given by 
    \begin{equation}
        \mathbf Q_\xi = - \frac{1}{16 \pi G} \bm \epsilon_{ab} \nabla^a \xi^b.
    \end{equation}
 We now decompose the symplectic potential and Noether charge on the horizon using the double null decomposition described in Section \ref{sec2.1}.  
 
    \paragraph{Symplectic Potential}
    The contraction of $\xi$ with $\bm \Theta(\phi, \delta \phi)$ evaluated at an arbitrary cross-section  $\mathcal{C} $ of the horizon $\mathcal{H}^+$ is  
    \begin{equation}
        \xi \cdot  \bm \Theta(\phi, \delta \phi) \ceq - \frac{\kappa v}{16 \pi G} \bm \epsilon_{\mathcal C} k^b g^{cd} \left( \nabla_c \delta g_{bd} - \nabla_b \delta g_{cd} \right)
    \end{equation}
   where we used that, on the horizon,
    \begin{equation}
        \bm \epsilon \fheq k \wedge l \wedge \bm \epsilon_{\mathcal C}
    \end{equation}
    with $\bm \epsilon_{\mathcal C}$   the codimension-2 area form. We calculate 
    \begin{equation}
        k^b g^{cd} \left( \nabla_c \delta g_{bd} - \nabla_b \delta g_{cd} \right) = k^b \delta \conn{c}{c}{b} - \frac{1}{2} k^b g^{cd} \nabla_b \delta g_{cd} = k^b \left( \gamma^c_a - k^c l_a - l^c k_a \right) \delta \conn{a}{b}{c} - \delta K = -2 \delta \theta_v \label{eq:gr-xi-theta}
    \end{equation}
    by using 
    \begin{equation}
        k^b k^c l_a \delta \conn{a}{b}{c} = \delta \left( l_a k^b \nabla_b k^a \right) \fheq 0, \quad k^b l^c k_a \delta \conn{a}{b}{c} = - \delta \left( l^a k^b \nabla_b k_a \right) \fheq 0
    \end{equation}
    and 
    \begin{equation}
        k^b \gamma^c_a \delta \conn{a}{b}{c} = - \frac{1}{2} k^b \gamma^{ac} \nabla_b \delta \gamma_{ac} \fheq -\delta \theta_v 
    \end{equation}
    where $\theta_v$ is the expansion along the $v$-direction, i.e., the trace of the $v$-extrinsic curvature $K_{ij}$. 
    Hence, we have 
    \begin{equation}
        \xi \cdot \bm \Theta(\phi, \delta \phi) \ceq \frac{1}{8 \pi G} \bm \epsilon_{\mathcal C}\, \kappa v\, \delta \theta_v.
    \end{equation}
We may now pull the $\delta$ to the front, as $\theta_v = 0$ for the stationary background, so we obtain
    \begin{equation}
        \xi \cdot \bm \Theta(\phi, \delta \phi) \ceq \delta \left(\frac{1}{8 \pi G} \bm \epsilon_{\mathcal C} \kappa v \theta_v \right),
    \end{equation}
    which verifies our proof that $\bm \Theta(\phi, \delta \phi)$ is exact in $\delta$ on the horizon.

    \paragraph{Noether Charge}
    We expand
    \begin{equation}
        \mathbf Q_\xi = - \frac{1}{16 \pi G} \bm \epsilon_{ab} \nabla^a \xi^b \ceq - \frac{1}{16 \pi G} \bm \epsilon_{\mathcal C} (k_a l_b - l_a k_b) \nabla^a \xi^b \ceq \frac{\kappa_3}{8 \pi G} \bm \epsilon_{\mathcal C}
    \end{equation}
    by using equation \eqref{surfgrav1} for the surface gravity $\kappa_3$.
    
    Now, the full variation on the horizon gives 
    \begin{equation}
        \delta \mathbf Q_\xi \ceq \frac{\kappa}{8 \pi G} \delta \bm \epsilon_{\mathcal C} + \frac{1}{8 \pi G} \bm \epsilon_{\mathcal C} \delta \kappa_3\,,
    \end{equation}
    where we have identified $\kappa_3 = \kappa$ on the  background Killing horizon.

    The $\delta \xi$ Noether charge, which we need to subtract from $\delta \mathbf Q_\xi$, reads 
    \begin{equation}
        \mathbf Q_{\delta \xi} \ceq - \frac{1}{16 \pi G} \bm \epsilon_{\mathcal C} (k_a l_b - l_a k_b) \nabla^a \delta \xi^b  \ceq \frac{1}{8 \pi G} \bm \epsilon_{\mathcal C} \delta \kappa_3\,,
    \end{equation}
    where we used   \eqref{varkappa}. So, we find 
    \begin{equation}
        \delta_\phi \mathbf{Q}_\xi = \delta \mathbf{Q}_\xi - \mathbf{Q}_{\delta \xi} \ceq \frac{\kappa}{8 \pi G} \delta \bm \epsilon_{\mathcal C}.
    \end{equation}

    \paragraph{Dynamical Black Hole Entropy}
    Finally, we obtain the entropy formula by evaluating \eqref{vardynennoether} at an arbitrary slice $\mathcal{C}(v)$  of the horizon $\mathcal{H}^+$
    \begin{equation}
        \delta S_\text{dyn} = \frac{2 \pi}{\kappa} \int_{\mathcal{C}(v)} \left(\delta_\phi \mathbf Q_\xi - \xi \cdot  \bm \Theta(\phi, \delta \phi) \right) = \frac{1}{4 G} \left( 1 - v \partial_v \right) \delta A = \delta \left( \frac{1}{4 G} \left( 1 - v \partial_v \right) A \right)
    \end{equation}
    where 
    \begin{equation}
        A = \int_{\mathcal{C}(v)} \bm \epsilon_{\mathcal C}
    \end{equation}
    is the area of $\mathcal{C}(v)$, and we have also identified
    \begin{equation}
        \partial_v \delta \bm \epsilon_{\mathcal C} = \delta (\theta_{v} \bm \epsilon_{\mathcal C}) = \bm \epsilon_{\mathcal C} \delta \theta_{v}.
    \end{equation}
    Here, the $\delta$ is pulled through as $\theta_{v} = 0$ on the background Killing horizon.

    Therefore, to first order in the perturbation, we obtain the entropy formula for dynamical black holes in general relativity 
    \begin{equation}
        S_\text{dyn} = \frac{1}{4 G} \left( 1 - v \partial_v \right) A\,,
    \end{equation}
      which coincides with the formula derived from the Raychaudhuri equation, and it agrees with the result  for the dynamical black hole entropy    in \cite{Hollands:2024vbe} (see also \cite{Rignon-Bret:2023fjq}).
    
    \subsection{\textit{f}(\textit{R}) Gravity}

   The Lagrangian form for $f(R)$ gravity is
    \begin{equation}
        \mathbf L = f(R) \bm \epsilon\,,
    \end{equation}
    where $f(R)= a_0 + a_1 R + a_2 R^2 + \cdots$ is a polynomial in $R$,  with coupling constants $a_0, a_1, a_2, \cdots $.
    
   The symplectic potential for this Lagrangian is
    \begin{equation}
        \bm \Theta(\phi, \delta \phi) = \bm \epsilon_d \left( f'(R) (g^{bd}g^{ac} - g^{cd} g^{ab}) \nabla_c \delta g_{ab} - \nabla_c(f'(R)) (g^{bd} g^{ac} - g^{cd} g^{ab}) \delta g_{ab} \right)
    \end{equation}
    where $f'(R) = \dv*{f}{R}$, and the Noether charge is
    \begin{equation}
        \mathbf Q_\xi = - \bm \epsilon_{ab} \left( f'(R) \nabla^{a} \xi^b + 2 \xi^a \nabla^b(f'(R)) \right).
    \end{equation}
    We carry out a similar procedure as in general relativity, by decomposing the volume form $\bm \epsilon$ into $\bm \epsilon_{\mathcal C}, k$ and $l$. 

    \paragraph{Symplectic Potential} The contraction of $\xi$ with $\bm \Theta(\phi, \delta \phi)$ on the horizon slice is
    \begin{equation}
        \begin{split}
            \xi \cdot  \bm \Theta(\phi, \delta \phi) & \ceq - \kappa v \bm \epsilon_{\mathcal C} \left( f'(R) k^a g^{bc} \left( \nabla_b \delta g_{ac} - \nabla_a \delta g_{bc} \right) - \nabla_c (f'(R)) (k^b g^{ac} - k^c g^{ab}) \delta g_{ab} \right)\\
            & \ceq 2 \kappa v \bm \epsilon_{\mathcal C} f'(R) \delta \theta_v\\
            & \ceq 2 \kappa v \partial_v \left( f'(R)  \delta \bm \epsilon_{\mathcal C} \right) \label{eq:sym-pot-f-R}
        \end{split}
    \end{equation}
    where we used \eqref{eq:gr-xi-theta}, the gauge condition $k^a \delta g_{ab} \fheq 0$, and 
    \begin{equation}
        k^c \nabla_c (f'(R)) = \partial_v (f'(R)) \fheq 0
    \end{equation}
    for a stationary background, and $\partial_v \delta \bm \epsilon_{\mathcal C} = \delta (\theta_{v} \bm \epsilon_{\mathcal C}) = \bm \epsilon_{\mathcal C} \delta \theta_{v}$ as above. 

    The above expression seems not to be exact in $\delta$. However, it is exact, because, as in GR, we may pull through the $\delta$ from the expression after the second equality:
    \begin{equation}
        \xi \cdot \bm \Theta(\phi, \delta \phi) \ceq \delta \left( 2 \kappa v \bm \epsilon_{\mathcal C} f'(R) \theta_v \right).
    \end{equation}
     In the following calculation we will still use the expression after the third equality of \eqref{eq:sym-pot-f-R}, which is not manifestly exact, as it is more convenient.

    \paragraph{Noether Charge}
    The Noether charge reads (without imposing the stationary condition of the background geometry)
    \begin{equation}
        \begin{split}
            \mathbf Q_\xi & \ceq - \bm \epsilon_{\mathcal C} (k_a l_b - l_a k_b) \left( f'(R) \nabla^{a} \xi^b + 2 \xi^a \nabla^b(f'(R)) \right)\\
            & \ceq 2 \kappa_3 f'(R) \bm \epsilon_{\mathcal C} - 2 l_a \xi^a \bm \epsilon_{\mathcal C} \partial_v (f'(R))\,,\\
        \end{split}
    \end{equation}
    where we have borrowed the results in general relativity. Hence, the variation reads 
    \begin{equation}
        \delta \mathbf Q_\xi \ceq 2 \delta \kappa_3 f'(R) \bm \epsilon_{\mathcal C} + 2 \kappa \delta \left( f'(R) \bm \epsilon_{\mathcal C} \right) - 2 (l_a \delta \xi^a) \bm \epsilon_{\mathcal C} \partial_v (f'(R)) - 2 \kappa v \delta \left( \bm \epsilon_{\mathcal C} \partial_v (f'(R)) \right)
    \end{equation}
    and the Noether charge for $\delta \xi$ reads 
    \begin{equation}
        \mathbf{Q}_{\delta \xi} \ceq 2 \delta \kappa_3 f'(R) \bm \epsilon_{\mathcal C} - 2 (l_a \delta \xi^a) \bm \epsilon_{\mathcal C} \partial_v (f'(R)).
    \end{equation}
    Then the field variation of the Noether charge is 
    \begin{equation}
        \delta_\phi \mathbf{Q}_\xi \ceq 2 \kappa \delta (f'(R)\bm \epsilon_{\mathcal C}) - 2 \kappa v \partial_v \left( \bm \epsilon_{\mathcal C} \delta (f'(R)) \right)\,,
    \end{equation}
    where we have used $\delta (\bm \epsilon_{\mathcal C} \partial_v (f'(R))) = \bm \epsilon_{\mathcal C} \partial_v (\delta (f'(R))) = \partial_v (\bm \epsilon_{\mathcal C} \delta (f'(R)))$.

    \paragraph{Dynamical Black Hole Entropy}

    Combining the previous results, we obtain 
    \begin{equation}
        \delta S_\text{dyn} = \frac{2 \pi}{\kappa} \int_{\mathcal{C}(v)} (\delta_\phi \mathbf Q_\xi - \xi \cdot  \bm \Theta(\phi, \delta \phi)) = 4 \pi \delta \left( \left( 1 - v \partial_v \right) \int_{\mathcal{C}(v)} f'(R) \bm \epsilon_{\mathcal C} \right),
    \end{equation}
    hence 
    \begin{equation}
    \label{dynamicalfR}
        S_\text{dyn} = \left( 1 - v \partial_v \right)  4 \pi\int_{\mathcal{C}(v)} f'(R) \bm \epsilon_{\mathcal C} =  (1- v \partial_v)S_{\text{JKM}}.
    \end{equation}
    At the bifurcation surface this coincides with the JKM entropy \cite{Jacobson:1995uq}, who showed that $S_{\text{JKM}}$ satisfies a second law for $f(R)$ gravity, similar to the area theorem for general relativity. On a Killing horizon, where $S_{\text{JKM}}$ is constant, the dynamical black hole entropy \eqref{dynamicalfR} coincides with the Wald entropy and JKM entropy.

   \subsection{\emph{f}(Riemann) Theories}
   \label{sec:fofriemann}
Next, we want    to calculate $\delta_\phi \mathbf Q_\xi - \xi \cdot  \bm \Theta(\phi, \delta \phi)$ for $f(\text{Riemann})$ theories. Here, the Lagrangian is a functional of the form $L = f(g^{ab}, R_{abcd})$, containing contractions of the inverse metric and the Riemann tensor. The symplectic potential  and the Noether charge are \cite{Azeyanagi:2009wf,Bueno:2016ypa} 
   \begin{equation}
        \bm \Theta(\phi, \delta \phi) = 2\bm \epsilon_d \left(X^{dabc} \nabla_c \delta g_{ab} - \left(\nabla_c X^{dabc}\right) \delta g_{ab}\right)
    \end{equation}
    and
    \begin{equation}
        \mathbf Q_{\xi} = - \bm \epsilon_{cd} \left(X^{cdab} \nabla_a \xi_b + 2 \xi_a \nabla_b X^{cdab}\right).
    \end{equation}
    Here, we denote
    \begin{equation}
        X^{abcd} = \pdv{L}{R_{abcd}}.
    \end{equation}
    Unlike general relativity and $f(R)$ gravity, for which we could compute the entropy formula covariantly, for $f(\text{Riemann})$ theories we will use GNC, which abide our gauge conditions in Section~\ref{ssec:gauge-cond},  in order to find a local geometric expression for the entropy $S_{\text{dyn}}$. A similar calculation was done in  \cite{Hollands:2024vbe} using Killing field arguments, instead of boost weight arguments, however they did not check that their expression (94) for  $S_{\text{dyn}}$ satisfies    the expected relation \eqref{eq:walldynbla} with the Wall entropy.
    
    \subsubsection*{Symplectic Potential}
    Our second gauge condition \eqref{fixednullnormal} implies 
    \begin{equation}
        \delta g_{v a} \fheq 0\, \quad \delta g_{u a} = 0\, .
    \end{equation}
    We calculate the non-zero components of $\nabla_c \delta g_{ab}$ in GNC
    \begin{equation}
        \nabla_u \delta g_{v i} \fheq - \delta \omega_i, \quad \nabla_u \delta g_{ij} \fheq 2 \delta \bar K_{ij}, \quad \nabla_v \delta g_{ij} \fheq 2 \delta K_{ij} \label{eq:del-dg-ngnc}\,,
    \end{equation}
    by using  
    \begin{equation}
        \nabla_c \delta g_{ab} = g_{bd} \delta \conn{d}{c}{a} + g_{ad} \delta \conn{d}{c}{b}\,.
    \end{equation}
    From the above expressions, we find
    \begin{align}
            \xi \cdot  \bm \Theta(\phi, \delta \phi) & \ceq  2 \kappa v \bm \epsilon_{u v} \left( X^{u abc} \nabla_c \delta g_{ab} - (\nabla_c X^{u abc}) \delta g_{ab} \right)\\
            & \ceq 2 \kappa v \bm \epsilon_{\mathcal C} \Big( - 2 X^{u v i u} \delta \omega_i + 2 X^{u i j u} \delta \bar K_{ij} + 2 X^{u i j v} \delta K_{ij} + X^{u ijk} \nabla_k \delta g_{ij} - (\nabla_c X^{u i j c}) \delta g_{ij} \Big)\,, \nonumber
       \end{align}
    where we have identified $\bm \epsilon_{u v} \fheq \bm \epsilon_{\mathcal C}$.

    We use a boost weight argument   to get 
    \begin{equation}
        X^{u v i u} \fheq X^{u i j u} \fheq X^{u ijk} \fheq \nabla_c X^{u ijc} \fheq K_{ij} \fheq 0 \label{eq:zero-cpts}\,,
    \end{equation}
    as these are proportional to positive weight affine GNC components on the background.

    So, we have 
    \begin{equation}
        \xi \cdot  \bm \Theta(\phi, \delta \phi) \ceq \delta ( 4 \kappa v \bm \epsilon_{\mathcal C}  X^{u i j v} K_{ij})\,,
    \end{equation}
    where we have pulled    the $\delta$ to the front in the term $ 4 \kappa v \bm \epsilon_{\mathcal C} X^{u ij v} \delta K_{ij}$, as 
    \begin{equation}
        4 \kappa v \bm \epsilon_{\mathcal C} X^{u ij v} \delta K_{ij} \fheq \delta \left(  4 \kappa v \bm \epsilon_{\mathcal C} X^{u ij v} K_{ij} \right) - \underbrace{K_{ij}}_{\fheq 0} \delta ( 4 \kappa v \bm \epsilon_{\mathcal C}  X^{u i j v}).
    \end{equation}
    Also, this suggests that $\xi \cdot \bm \Theta(\phi, \delta \phi)$ is exact in $\delta$, i.e., $\xi \cdot \bm \Theta(\phi, \delta \phi) \fheq \delta (\xi \cdot \mathbf B_{\mathcal H^+})$, as discussed in Section \ref{sec:exactness}, where
    \begin{equation}
        \mathbf B_{\mathcal H^+} \fheq - 4 (l \wedge \bm \epsilon_{\mathcal C}) X^{uijv} K_{ij}.
    \end{equation}
    In general, on the horizon, whenever we have a product of  a non-positive weight component on the background and the variation of a positive weight component, i.e., $A_{(w\leq 0)} \delta B_{(w>0)}$, we may pull the $\delta$   through 
    \begin{equation}
        A_{(w\leq 0)} \delta B_{(w>0)} \fheq \delta \left( A_{(w\leq 0)} B_{(w>0)} \right). \label{eq:factor-delta}
    \end{equation}
    This will be used implicitly throughout our following calculations.

    \subsubsection*{Noether Charge}
    First we  calculate 
    \begin{equation}
        \xi_a \fheq - \xi^v (\dd{u})_a, \quad \nabla_{[u} \xi_{v]} \fheq \kappa_3, \quad \nabla_{[u} \xi_{i]} \fheq -\frac{1}{2} \omega_i \xi^v, \quad \nabla_{[v} \xi_{i]} \fheq \nabla_{[i} \xi_{j]} \fheq 0 \label{eq:del-xi-ngnc}\,,
    \end{equation}
    so that  
    \begin{equation}
        \begin{split}
            \mathbf Q_\xi & \ceq - 2 \bm \epsilon_{u v} \left( X^{u v a b}\nabla_a \xi_b - 2 \xi^v \nabla_b X^{u v u b} \right)\\
            & \ceq - 2 \bm \epsilon_{\mathcal C} \left( 2 \kappa_3 X^{u v u v} - \xi^v \omega_i X^{u v u i} - 2 \xi^v \nabla_v X^{u v u v} - 2 \xi^v \nabla_i X^{u v u i} \right)\,,
        \end{split}
    \end{equation}
    where we have expanded the summation over dummy index $b$. Further, we compute 
    \begin{equation}
        \nabla_v X^{u v u v} = \partial_v X^{u v u v} - \omega_i X^{u v u i}
    \end{equation} 
    and 
    \begin{equation}
        \nabla_i X^{u v u i} = D_i X^{u v u i} + K_{ij} X^{u i j v} + \bar K_{ij} X^{u i u j} + \frac{1}{2} \omega_i X^{u v u i} + K X^{u v u v}
    \end{equation}
    using Table \ref{tb:covder}. Finally, we obtain
    \begin{equation}
        \begin{split}
            \mathbf Q_\xi & \ceq 4 \bm \epsilon_{\mathcal C} \left( \xi^v \left( (\partial_v + K) X^{uvuv} + K_{ij} X^{uijv} + \bar K_{ij} X^{uiuj} + D_i X^{uvui} \right) - \kappa_3 X^{uvuv} \right)\\
            & \ceq 4 \xi^v \left( \partial_v (X^{uvuv}\bm \epsilon_{\mathcal C}) + \bm \epsilon_{\mathcal C} (K_{ij} X^{uijv} + \bar K_{ij} X^{uiuj} + D_i X^{uvui}) \right) - 4 \kappa_3 \bm \epsilon_{\mathcal C} X^{uvuv}
        \end{split}
    \end{equation}
    by identifying
    \begin{equation}
        \partial_v \bm \epsilon_{\mathcal C} = K \bm \epsilon_{\mathcal C}.
    \end{equation}
    And we can compute 
    \begin{equation}
        \mathbf{Q}_{\delta \xi} \ceq 4 (\delta \xi^v) \left( \partial_v (X^{uvuv}\bm \epsilon_{\mathcal C}) + \bm \epsilon_{\mathcal C} (K_{ij} X^{uijv} + \bar K_{ij} X^{uiuj} + D_i X^{uvui}) \right) - 4 (\delta \kappa_3) \bm \epsilon_{\mathcal C} X^{uvuv}
    \end{equation}
    so that the field variation of the Noether charge is 
    \begin{equation}
        \delta_\phi \mathbf{Q} \ceq 4 \kappa v \delta \left( \partial_v (X^{uvuv}\bm \epsilon_{\mathcal C}) + \bm \epsilon_{\mathcal C} (K_{ij} X^{uijv} + \bar K_{ij} X^{uiuj} + D_i X^{uvui}) \right) - 4 \kappa \delta \left( \bm \epsilon_{\mathcal C} X^{uvuv} \right),
    \end{equation}
    where we inserted the background quantities $\xi^v = \kappa v$ and $\kappa_3 = \kappa$.

    \subsubsection*{Dynamical Black Hole Entropy}
    Combining the above calculations, we finally have 
    \begin{equation}
        \begin{split}
            \delta_\phi \mathbf Q_\xi - \xi \cdot  \bm \Theta(\phi, \delta \phi) & \ceq  4 \kappa \delta \left( - (1 - v \partial_v) (X^{uvuv}\bm \epsilon_{\mathcal C}) + v \bm \epsilon_{\mathcal C} \left( \bar K_{ij} X^{uiuj} + D_i X^{uvui} \right) \right).
        \end{split}
    \end{equation}
    Here, we notice  there is a codimension-2 total derivative term $D_i X^{u v u i}$. This is related to the \emph{entropy current} discussed in \cite{Bhattacharyya:2021jhr}. It is important when one discusses the local behaviour of entropy density-current on any horizon slice. In this paper, we investigate the entropy with respect to the whole horizon slice, and we assume the slice is compact. Hence, the total derivative will be integrated out and will not contribute to the black hole entropy. 
    
    The dynamical black hole entropy is obtained through 
    \begin{equation}
        \begin{split}
            \delta S_\text{dyn} & = \frac{2 \pi}{\kappa} \int_{\mathcal{C}(v)} \left( \delta \mathbf Q_\xi - \xi \cdot  \bm \Theta(\phi, \delta \phi) \right)\\
            & = - 8 \pi \int_{\mathcal{C}(v)} \dd[D-2]{x} \left( (1 - v \partial_v) \delta \left(\sqrt{\gamma}\, X^{u v u v}\right) - v \delta \left(\sqrt{\gamma}\, \bar K_{ij} X^{u i u j}\right) \right).
        \end{split}
    \end{equation}
    We further unpack the notations
    \begin{equation}
        \delta \left( \sqrt{\gamma}\, \bar K_{ij} X^{u i u j} \right) = \sqrt{\gamma}\, \bar K_{ij} \delta X^{u i u j}
    \end{equation}
    and
    \begin{equation}
        \delta X^{u i u j} = 4 \pdv{X^{u i u j}}{R_{v k v l}} \delta R_{v k v l}.
    \end{equation}
    Using the boost weight analysis, the only non-vanishing term after applying the chain rule to $\delta X^{u i u j}$ is proportional to $\delta R_{v k v l}$. This is   because $\delta X^{u i u j}$ is of weight 2, and after the chain rule the non-vanishing terms should be products of a weight 0 or lower background term and a weight 2 or higher variation. In this case there is only one such term available, as written above. A more detailed analysis is provided in \cite{Wall:2024lbd}.

    The Riemann component involved reads 
    \begin{equation}
        \delta R_{v k v l} \fheq g_{v u} \delta \left( \partial_v \conn{u}{l}{k} - \partial_l \conn{u}{v}{k} + \conn{a}{l}{k} \conn{u}{v}{a} - \conn{a}{v}{k} \conn{u}{l}{a} \right) = - \partial_v \delta K_{kl}
    \end{equation}
    and we calculate  
    \begin{equation}
        \begin{split}
            v\sqrt{\gamma}\, \bar K_{ij} \delta X^{uiuj} & \fheq - 4 v \sqrt{\gamma}\, \bar K_{ij} \pdv{X^{uiuj}}{R_{vkvl}} \partial_v \delta K_{kl}\\
            & \fheq -4 v \partial_v \left( \sqrt{\gamma}\, \bar K_{ij} \pdv{X^{uiuj}}{R_{vkvl}} \delta K_{kl} \right) + 4 \sqrt{\gamma} \left( v \partial_v  \bar K_{ij} \right) \pdv{X^{uiuj}}{R_{vkvl}} \delta K_{kl}\\
            & \fheq 4 (1 - v \partial_v) \left( \sqrt{\gamma}\, \bar K_{ij} \pdv{X^{uiuj}}{R_{vkvl}} \delta K_{kl} \right)
        \end{split}
    \end{equation}
    by using the boost weight argument and applying the Killing equation \eqref{eq:lie-deriv} to $\bar K_{ij}$, i.e., 
    \begin{equation}
        (v \partial_v - 1) \bar K_{ij} \fheq 0.
    \end{equation}
    Then, pulling the $\delta$'s through and recovering the expression of $X^{abcd}$, we find
    \begin{equation}
        \delta S_\text{dyn} = - 8 \pi \delta \left(\left( 1 - v \partial_v \right) \int_{\mathcal{C}(v)} \dd[D-2]{x} \sqrt{\gamma}\, \left( \pdv{L}{R_{u v u v}} - 4 \pdv{L}{R_{u i u j}}{R_{v k v l}} \bar K_{ij} K_{kl} \right)\right).
    \end{equation}
    So, to first order in the perturbation,  the dynamical entropy of the black hole is 
    \begin{equation}
        S_\text{dyn} = - 8 \pi \left( 1 - v \partial_v \right) \int_{\mathcal{C}(v)} \dd[D-2]{x} \sqrt{\gamma}\, \left( \pdv{L}{R_{u v u v}} - 4 \pdv{L}{R_{u i u j}}{R_{v k v l}} \bar K_{ij} K_{kl} \right),
    \end{equation}
    which is valid at any slice of the future horizon. 
    In a  covariant notation, the dynamical black hole entropy for $f(\text{Riemann})$ theory is
    \begin{equation}
        S_\text{dyn} = - 8 \pi (1 - v \partial_v) \int_{\mathcal C(v)} \bm \epsilon_{\mathcal C} k_a l_b k_c l_d \left(\pdv{L}{R_{abcd}} - 4 \pdv{L}{R_{aecf}}{R_{bgdh}} \bar K_{ef} K_{gh} \right).
    \end{equation}
  Thus, we confirm this is related to the Wall entropy \eqref{wallentropy} for $f(\text{Riemann})$  theories   by 
    \begin{equation}
        S_\text{dyn} = \left(1 - v\partial_v\right) S_{\text{Wall}}\,,
    \end{equation}
     as anticipated in Section \ref{sec:relationwall}.

    \section*{Acknowledgement}
       We are grateful to  Venkatesa Chandrasekaran, Ted Jacobson, Prahar Mitra, Antoine Rignon-Bret, Ronak Soni, Antony Speranza, Simone Speziale, Andrew Svesko, Erik Verlinde, Robert Wald, Aron Wall and Hongbao Zhang for extensive discussions.  This work was supported in part by AFOSR grant FA9550-19-1-0260 “Tensor Networks and Holographic Spacetime”, STFC grant ST/P000681/1 “Particles, Fields and Extended Objects”, and an Isaac Newton Trust Early Career grant.  MRV is supported by SNF Postdoc Mobility grant P500PT-206877 “Semi-classical thermodynamics of black holes and the information paradox”. ZY is supported by an Internal Graduate Studentship of Trinity College, Cambridge. ZY is also grateful for the hospitality of KITP and UCSB, where this work was completed. This research was supported in part by grant NSF PHY-2309135 to  KITP.

    \appendix
     \section{The Tale of Three Surface Gravities}\label{app:surf-grav}

In this Appendix we derive some relations for the  three   surface gravities, and their variations, that are defined on an arbitrary null hypersurface $\mathcal N$ through
    \begin{equation}
    \label{k1}
     \nabla_a (\xi_b \xi^b) \nheq - 2 \kappa_1 \xi_a,
    \end{equation}
    \begin{equation}
           \label{k2}
     \xi^b \nabla_b \xi^a \nheq \kappa_2 \xi^a,
    \end{equation}
    \begin{equation}
       \label{k3}
     (\nabla^a \xi^b )  ( \nabla_{[a} \xi_{b]})  \nheq -2 \kappa_3^2\,,
    \end{equation}
    where $\xi_a$ is the normal to $\mathcal N.$ We will not assume that $\xi^a$ is a Killing field in this Appendix, hence $\mathcal N$ is not necessarily a Killing horizon. \\
    
\noindent 1.    The first relation,  \eqref{k3relation1},  that we want to derive is \cite{Jacobson:1993pf,Belin:2022xmt}
    \begin{equation}
       \label{k3relation}
        \kappa_3 \nheq \frac{1}{2}(\kappa_1 + \kappa_2)\,.
        \end{equation}
        \textit{Proof:} Since $\xi^a$ is  orthogonal to the null hypersurface, by Frobenius's theorem it satisfies the irrotationality condition on $\mathcal N$
        \begin{equation}
           \xi_{[a}  \nabla_{b} \xi_{c]}\nheq 0\,.
        \end{equation}
        This is equivalent to
        \begin{equation}
              \xi_c \nabla_{[a} \xi_{b]} \nheq -  \xi_{[a} \nabla_{b]} \xi_c +   \xi_{[a} \nabla_{|c|} \xi_{b]}\,.
        \end{equation}
Contracting both sides with $\nabla^a \xi^b$ yields
\begin{equation}
\label{intermediatek3}
    \xi_c( \nabla^a \xi^b )(\nabla_{[a}\xi_{b]}) \nheq - \xi_a (\nabla^{[a} \xi^{b]})( \nabla_b \xi_c) + \xi_a (\nabla^{[a}\xi^{b]})(\nabla_c \xi_b)\,.
\end{equation}
Note  if $\xi^a$ were a Killing field,   the two terms on the rhs would be equal to each other because of Killing's equation $\nabla_{b}\xi_c = - \nabla_c \xi_b$. However, for a generic null normal they differ from each other. Expanding the rhs gives
\begin{align}
     - \xi_a (\nabla^{[a} \xi^{b]})( \nabla_b \xi_c) + \xi_a (\nabla^{[a}\xi^{b]})(\nabla_c \xi_b)&= - \frac{1}{2} ( \xi_a\nabla^a \xi^b) (\nabla_{b}\xi_c) + \frac{1}{2}(\xi_a \nabla^b \xi^a) (\nabla_b \xi_c)  \nonumber  \\
    &\quad + \frac{1}{2}(\xi_a \nabla^a \xi^b )(\nabla_c \xi_b) - \frac{1}{2}(\xi_a \nabla^b \xi^a) (\nabla_c \xi^b ) \nonumber  \\
     &\nheq - \frac{1}{2}\kappa_2 \xi^b \nabla_b \xi_c - \frac{1}{2} \kappa_1 \xi^b \nabla_b \xi_c + \frac{1}{2}\kappa_2 \xi^b \nabla_c \xi_b + \frac{1}{2}\kappa_1 \xi^b \nabla_c \xi^b\nonumber \\
     &\nheq  - \frac{1}{2} \xi_c (\kappa_2^2 + 2 \kappa_1 \kappa_2 + \kappa_1^2)\,.
\end{align}
By equating this to the lhs of \eqref{intermediatek3}, which is equal to $- 2 \xi_c \kappa_3^2$, we obtain the simple relation~\eqref{k3relation} between the  surface gravities. $\Box$ \\

 \noindent 2. Next, we  prove the set of relations    \eqref{surfgrav1},
\begin{equation}
\label{newkappas}
         \kappa_1 \nheq  l^a \nabla_a (  k_b\xi^b ), \quad   \kappa_2 \nheq- k^a \nabla_a ( l_b  \xi^b)\, ,  \quad  \kappa_3  \nheq  l_{[a} k_{b]} \nabla^a \xi^b\,,
    \end{equation} 
 where $k^a$ is the  affinely parameterised null normal to $\mathcal N$, and $l^a$ is the auxiliary  null vector fields, satisfying 
\begin{equation}
    k_a k^a \nheq 0, \quad k^b \nabla_b k^a \nheq 0 , \quad l_a l^a \nheq 0, \quad k_a l^a \nheq-1, \quad  k_a\xi^a \nheq 0\,, \quad l^b \nabla_b k^a = k^b \nabla_b l^a.
\end{equation}
\textit{Proof:} We start by decomposing the vector field $\xi^a$   in terms of the null vector fields $k^a$ and $l^a$, i.e.
\begin{equation}
\label{decxiapp}
    \xi^a =  -(\xi^b l_b ) k^a - (\xi^b k_b) l^a\,. 
\end{equation}
The vector field $\xi^a$ has no components along the codimension-2 spatial directions, since it has to be tangent to the null geodesic generators of $\mathcal N$. Moreover, we have chosen $k^a$ and $l^a$ off the null surface such that the form  \eqref{decxiapp} remains true away from $\mathcal N$ (similar to the construction of $k^a$ and $l^a$ in Section \ref{sec:geometricsetup}).
We first derive the expression for $\kappa_1$. Contracting definition \eqref{k1} with $l^a$ and inserting the decomposition \eqref{decxiapp} for $\xi^a$ once yields
\begin{equation}
    -2\kappa_1 \xi_a l^a \nheq l^a \nabla_a( -(\xi_c l^c  k_b + \xi_ck^c l_b)\xi^b) = - 2 l^a \nabla_ a(\xi_c l^c \xi_b k^b) \nheq  - 2 \xi_c l^c l^a  \nabla_ a( \xi_b k^b)\,.
\end{equation}
In the final equality we used $\xi_b k^b \nheq 0$. Dividing both sides by $\xi^c l_c$ leads to the desired expression for $\kappa_1 $ in \eqref{newkappas}.

Next, in order to derive the   expression for $\kappa_2$, we contract definition \eqref{k2} with $l_a$ and insert the decomposition \eqref{decxiapp} twice
\begin{equation}
    \kappa_2 \xi^a l_a \nheq l_a (\xi^c l_c k^b + \xi^ck_c l^b ) \nabla_b (\xi^c l_c k^a + \xi^c k_c l^a) \nheq l_a  \xi^c l_c k^b  \nabla_b (\xi^c l_c k^a  ) \nheq - \xi^c l_c k^b \nabla_b (\xi^c l_c)\,.
\end{equation}
In the second equality we used $\xi^c k_c \nheq 0 $ and $k^b \nabla_b (\xi^c k_c) \nheq 0.$ And, to obtain   the third equality, we employed $k^b \nabla_b k^a \nheq 0$ and $l_a k^a \nheq -1.$ Thus, we find the expression for $\kappa_2$ in \eqref{newkappas}.    

Finally, the expression for $\kappa_3$ is a consequence of    relation \eqref{k3relation} between the surface gravities  
  \begin{equation}
      \kappa_3 \nheq  \frac{1}{2}(l^a \nabla_a (k_b \xi^b) - k^a \nabla_a (l_b \xi^b)) = \frac{1}{2}(l_ak_b   - l_bk_a  )\nabla^a \xi^b\,,
    \end{equation}
    where the second equality follows from the assumption that $l^a$ and $k^a$ commute with each other, i.e. $\mathcal L_k l^a =  k^b \nabla_b l^a- l^b \nabla_b k^a =0.$ $\Box$ \\

    \noindent 3. 
As an aside, we mention that the surface gravities also satisfy the relations
\begin{equation}
    \kappa_1 \nheq \beta_a \xi_b \nabla^a \xi^b, \quad \kappa_2 \nheq - \beta_a \xi_b \nabla^b \xi^a, \quad  \kappa_3 \nheq \beta_{[a} \xi_{b]} \nabla^a \xi^b\,,
\end{equation}
  where $\beta^a$ is an auxiliary null vector field that satisfies $\beta_a \xi^a=-1  $ on $\mathcal N$, see \eqref{beta}. We will not use these expressions in the main text, however, since our gauge conditions keep $k^a$ and $l^a$ fixed (and not $\beta^a$ and/or $\xi^a$). \\

  \noindent 4. Finally,   we derive the variational relations \eqref{varkappa} for the surface gravities
\begin{equation}
\label{varkappa2}
        \delta \kappa_1 \nheq   l^a \nabla_a (k_b \delta \xi^b), \quad \delta \kappa_2 \nheq- k^a \nabla_a(l_b \delta \xi^b)\, , \quad \delta \kappa_3 \nheq  l_{[a} k_{b]} \nabla^a \delta \xi^b\,.
    \end{equation}
\textit{Proof:} In taking the variation of \eqref{newkappas} we assume that our gauge conditions in Section \ref{ssec:gauge-cond} hold on the null surface $\mathcal N$ (instead of $\mathcal H^+$)  or away from $\mathcal N$. In particular, \eqref{fixedkandl} states that $\delta k^a =0$, $\delta l^a =0$, and together with \eqref{fixednullnormal} we have $\delta l_a =0$ everywhere, but   $\delta k_a =0$ only holds on the null hypersurface (since $k^a \delta g_{ab}\nheq 0$). We immediately see that the variation of $\kappa_2$ depends only on the variation of the normal $\xi^a$, since the variation $\delta$ commutes with the partial derivative~$\partial_a$. (Note the covariant derivatives in the expressions for $\kappa_1$ and $\kappa_2$, \eqref{newkappas},  can be replaced with partial derivatives.)

To calculate the variation of $\kappa_1$ we have to do a bit more work. Since $\delta k_a$ does not vanish away from~$\mathcal N,$  it is not immediately clear that the derivative   $l^a \nabla_a (\xi^b \delta k_b)$  vanishes on $\mathcal N.$ However, we now show that it does, by first writing it as $l^ak^c \xi^b  \nabla_a   \delta g_{bc}$, which follows from $\delta k^c =0$ and $\xi^b \delta g_{bc}=k^c \delta g_{bc}=0$ on $\mathcal N$. Since $\xi^a$ is proportional to $k^a$ on $\mathcal N$, we will consider the following quantity
\begin{equation}
          l_c k^a k^b \delta \conn{c}{a}{b}   = \frac{1}{2} l^c k^a k^b (2 \nabla_{(a} \delta g_{b)c} - \nabla_c \delta g_{ab}) \,.
\end{equation}
The lhs vanishes because of the gauge condition \eqref{affinegaugecond}, $\delta(k^a \nabla_a k^b) =0,$ and the first term on the rhs is zero since
\begin{equation}
    k^a k^b \nabla_a \delta g_{bc} = k^a \nabla_a (k^b \delta g_{bc}) - k^a \nabla_a k^b (\delta g_{bc}) \nheq 0\,.
\end{equation}
Hence, it follows that
\begin{equation}
    l^c k^a k^b \nabla_c \delta g_{ab} \nheq 0,
\end{equation}
which implies that  $l^a \nabla_a (\xi^b \delta k_b)$  vanishes on $\mathcal N.$ Therefore, also the variation of $\kappa_1$ only depends on the variation of $\xi^a$, and we obtain the first relation in \eqref{varkappa2}. 

To obtain the variation of $\kappa_3$, we vary \eqref{k3relation},
\begin{equation}\delta \kappa_3 = \frac{1}{2}(\delta \kappa_1 + \delta \kappa_2)\,.
\end{equation} By inserting the expressions for $\delta \kappa_1$ and $\delta \kappa_2$ in \eqref{varkappa2} and using the fact that  $l^a$ and $k^a$ commute, we arrive at the desired expression. $\Box$

    \section{Geometric Quantities in   Gaussian Null Coordinates}
\label{appGNC}
In this Appendix we calculate the Christoffel connections and certain components of the covariant derivative in  affine Gaussian null coordinates $(v,u,x^i)$ and   Killing (non-affine) Gaussian null coordinates ($\tau,\rho,x^i$). We recall these coordinates  are related by the transformation \eqref{coordtransf},
\begin{equation}
  v = \frac{1}{\kappa} \exp(\kappa \tau), \quad u = \rho \exp (- \kappa \tau)  .
\end{equation}

    \subsection{Affine GNC}

   \label{app:gnc-q}

 The metric components in   Gaussian null coordinates ($u,v, x^i$),  based upon the affine parameterisation of the null geodesics of the horizon, are given by \eqref{gnc},
    \begin{equation}
        g_{u u} = g_{u i} = 0, \quad g_{u v} = -1, \quad g_{v v} = - u^2 \alpha, \quad g_{v i} = - u \omega_i, \quad g_{ij} = \gamma_{ij},
    \end{equation}
    and the inverse metric components are 
    \begin{equation}
        g^{u u} = u^2 (\alpha + \omega^2), \quad g^{u v} = -1, \quad g^{u i} = u \omega^i, \quad g^{v v} = g^{v i} = 0, \quad g^{ij} = \gamma^{ij},
    \end{equation}
    where $\gamma^{ij}$ is the inverse of $\gamma_{ij}$, $\omega^i = \gamma^{ij} \omega_j$ and $\omega^2 = \gamma_{ij} \omega^i \omega^j$. The quantities $\alpha, \omega_i$ and $\gamma_{ij} $ generically depend on all coordinates $(v,u,x^i)$, but for stationary metrics with a boost Killing field $\xi  = \kappa (v \partial_v - u \partial_u)$ they  depend only on the product $\kappa uv$ and $x^i$.

 The Christoffel connections on $\mathcal H^+$ in GNC can be computed to be 
    \begin{align}
        \conn{u}{u}{u} & = \conn{v}{u}{u} = \conn{v}{u}{v} = \conn{v}{u}{i} = \conn{i}{u}{u} = 0\\
        \conn{u}{u}{v} & = \frac{1}{2} \partial_u(u^2 \alpha) - \frac{1}{2} u \omega^i \partial_u \left( u \omega_i \right) \fheq 0\\
        \conn{u}{u}{i} & = \frac{1}{2} \partial_u \left( u \omega_i \right) + u \omega^j \bar K_{ij} \fheq \frac{1}{2} \omega_i\\
        \conn{u}{v}{v} & = \frac{1}{2} u^2 (\alpha + \omega^2) \partial_u (u^2 \alpha) + \frac{1}{2} u^2 \partial_v \alpha + \frac{1}{2} u^2 \omega^i \left( u \partial_i \alpha - 2 \partial_v \omega_i \right) \fheq 0\\
        \conn{u}{v}{i} & = \frac{1}{2} u^2 (\alpha + \omega^2) \partial_u (u \omega_i) + \frac{1}{2} u^2 \partial_i \alpha + u \omega^j \left( K_{ij} + u \partial_{[j} \omega_{i]} \right)\fheq 0\\
        \conn{u}{i}{j} & = K_{ij} - u^2(\alpha + \omega^2) \bar K_{ij} + u D_{(i}\omega_{j)} \fheq K_{ij}\\
        \conn{v}{v}{v} & = - \frac{1}{2} \partial_u (u^2 \alpha) \fheq 0\\
        \conn{v}{v}{i} & = - \frac{1}{2} \partial_u (u \omega_i) \fheq - \frac{1}{2} \omega_i\\
        \conn{v}{i}{j} & = \bar K_{ij}\\
        \conn{i}{u}{v} & = - \frac{1}{2} \gamma^{ij} \partial_u (u \omega_j) \fheq - \frac{1}{2} \omega^i\\
        \conn{i}{u}{j} & = \bar K\indices{^i_j}\\
        \conn{i}{v}{v} & = \frac{1}{2} u \gamma^{ij} \left( \omega_j \partial_u (u^2 \alpha) + \partial_j \alpha - 2 \partial_v \omega_j \right) \fheq 0\\
        \conn{i}{v}{j} & = K\indices{^i_j} + u \gamma^{ik} \partial_{[k}\omega_{j]} + \frac{1}{2} u \omega^i \partial_u (u \omega_j) \fheq K\indices{^i_j}\\
        \conn{i}{j}{k} & = \Gamma[\gamma]^i_{jk} - u \omega^i \bar K_{jk} \fheq \Gamma[\gamma]^i_{jk}
    \end{align}
    where the extrinsic curvatures in the $v$- and $u$-directions are defined, respectively, as
    \begin{equation}
        K_{ij} = \frac{1}{2} \partial_v \gamma_{ij}, \quad \bar K_{ij} = \frac{1}{2} \partial_u \gamma_{ij}\,,
    \end{equation}
    and $D_i$ is the intrinsic covariant derivative of the codimension-2 surface.

    Finally, on the horizon,   the components of the $v$- and $i$-covariant derivatives of a tensor $T$  are shown in  Table \ref{tb:covder}.

    \begin{table}[H]
        \begin{tabular}{|c|c|c|}
            \hline
            $\nabla$ & $T^{\cdots a\cdots }$ & $T_{\cdots a \cdots}$ \\
            \hline
            & & \\
            $v$ & $\begin{aligned}
                \nabla_v T^{\cdots v\cdots} & = \partial_v T^{\cdots v\cdots} - \frac{1}{2} \omega_i T^{\cdots i\cdots}\\
                \nabla_v T^{\cdots u\cdots} & = \partial_v T^{\cdots u\cdots} \\
                \nabla_v T^{\cdots i\cdots} & = \partial_v T^{\cdots i\cdots} - \frac{1}{2} \omega^i T^{\cdots u\cdots} + K\indices{^i _{j}}T^{\cdots j\cdots}
            \end{aligned}$ & $\begin{aligned}
                \nabla_v T_{\cdots u\cdots} & = \partial_v T_{\cdots u\cdots} + \frac{1}{2} \omega^i T_{\cdots i\cdots} \\
                \nabla_v T_{\cdots v\cdots} & = \partial_v T_{\cdots v\cdots}\\
                \nabla_v T_{\cdots i\cdots} & = \partial_v T_{\cdots i\cdots} + \frac{1}{2} \omega_i T_{\cdots v\cdots} - K\indices{_i^j}T_{\cdots j\cdots}
            \end{aligned}$\\
            & & \\
            \hline
            & & \\
            $i$ & $\begin{aligned}
                \nabla_i T^{\cdots v\cdots} & = D_i T^{\cdots v\cdots} + \bar K _{ij}T^{\cdots j\cdots} - \frac{1}{2} \omega_i T^{\cdots v\cdots}\\
                \nabla_i T^{\cdots u\cdots} & = D_i T^{\cdots u\cdots} + K _{ij}T^{\cdots j\cdots} + \frac{1}{2} \omega_i T^{\cdots u\cdots}\\
                \nabla_i T^{\cdots j\cdots} & = D_i T^{\cdots j\cdots} + K\indices{_i^j} T^{\cdots v\cdots} + \bar K\indices{_i^j} T^{\cdots u\cdots}
            \end{aligned}$ & $\begin{aligned}
                \nabla_i T_{\cdots u\cdots} & = D_i T_{\cdots u\cdots} - \bar K\indices{_i^j}T_{\cdots j\cdots} - \frac{1}{2} \omega_i T_{\cdots u\cdots}\\
                \nabla_i T_{\cdots v\cdots} & = D_i T_{\cdots v\cdots} - K\indices{_i^j}T_{\cdots j\cdots} + \frac{1}{2} \omega_i T_{\cdots v\cdots}\\
                \nabla_i T_{\cdots j\cdots} & = D_i T_{\cdots j\cdots} - \bar K\indices{_{ij}}T_{\cdots v\cdots} - K\indices{_{ij}} T_{\cdots u\cdots}
            \end{aligned}$ \\
            & & \\
            \hline
        \end{tabular}
        \caption{Components of tensor covariant derivatives on the horizon in affine GNC. } \label{tb:covder}
    \end{table}

     \subsection{Killing GNC}\label{app:ngnc-q}

   In non-affine Gaussian null coordinates ($\tau , \rho, x^i$),    based upon the Killing parameterisation of the null geodesics of the horizon,,   the metric components are 
    \begin{equation}
        g_{\rho \rho} = g_{\rho i} = 0, \quad g_{\rho \tau} = -1, \quad g_{\tau \tau} = - \rho {\tilde \alpha}, \quad g_{\tau i} = - \rho \omega_i, \quad g_{ij} = \gamma_{ij},
    \end{equation}
    and the inverse metric components are 
    \begin{equation}
        g^{\rho \rho} = \rho {\tilde \alpha} + \rho^2 \omega^2, \quad g^{\rho \tau} = -1, \quad g^{\rho i} = \rho \omega^i, \quad g^{\tau \tau} = g^{\tau i} = 0, \quad g^{ij} = \gamma^{ij}
    \end{equation}
    where $\gamma^{ij}$ is the inverse of $\gamma_{ij}$, $\omega^i = \gamma^{ij} \omega_j$ and $\omega^2 = \gamma_{ij} \omega^i \omega^j$. The quantities $\omega_i$ and $\gamma_{ij}$ are the same as for affine GNC, but $\tilde \alpha$ is different from the metric function $\alpha$ in affine GNC. $\tilde \alpha$ is related to $\alpha$ by $\tilde \alpha = - 2 \kappa + \rho \alpha$ where $\kappa$ is the background surface gravity. All metric quantities in Killing GNC only depend on the coordinates  $(\rho,x^i)$ since $\tau$ is a Killing parameter.

 The Christoffel connections in Killing GNC on the Killing horizon are
    \begin{align}
        \conn{\rho}{\rho}{\rho} & = \conn{\tau}{\rho}{\rho} = \conn{\tau}{\rho}{\tau} = \conn{\tau}{\rho}{i} = \conn{i}{\rho}{\rho} = 0\\
        \conn{\rho}{\rho}{\tau} & = \frac{1}{2} \partial_\rho(\rho {\tilde \alpha}) - \frac{1}{2} \rho \omega^i \partial_\rho \left( \rho \omega_i \right) \heq \frac{1}{2} {\tilde \alpha} = - \kappa\\
        \conn{\rho}{\rho}{i} & = \frac{1}{2} \partial_\rho \left( \rho \omega_i \right) + \rho \omega^j \bar {\mathcal K}_{ij} \heq \frac{1}{2} \omega_i\\
        \conn{\rho}{\tau}{\tau} & = \frac{1}{2} \rho ({\tilde \alpha} + \rho \omega^2) \partial_\rho (\rho {\tilde \alpha}) + \frac{1}{2} \rho \partial_\tau {\tilde \alpha} + \frac{1}{2} \rho^2 \omega^i \left( \partial_i {\tilde \alpha} - 2 \partial_\tau \omega_i \right) \heq 0\\
        \conn{\rho}{\tau}{i} & = \frac{1}{2} \rho ({\tilde \alpha} + \rho \omega^2) \partial_\rho (\rho \omega_i) + \frac{1}{2} \rho \partial_i {\tilde \alpha} + \rho \omega^j \left( {\mathcal K}_{ij} + \rho \partial_{[j} \omega_{i]} \right)\heq 0\\
        \conn{\rho}{i}{j} & = {\mathcal K}_{ij} - \rho({\tilde \alpha} + \rho \omega^2) \bar {\mathcal K}_{ij} + \rho D_{(i}\omega_{j)} \heq {\mathcal K}_{ij}\\
        \conn{\tau}{\tau}{\tau} & = - \frac{1}{2} \partial_\rho (\rho {\tilde \alpha}) \heq - \frac{1}{2} {\tilde \alpha} = \kappa\\
        \conn{\tau}{\tau}{i} & = - \frac{1}{2} \partial_\rho (\rho \omega_i) \heq - \frac{1}{2} \omega_i\\
        \conn{\tau}{i}{j} & = \bar {\mathcal K}_{ij}\\
        \conn{i}{\rho}{\tau} & = - \frac{1}{2} \gamma^{ij} \partial_\rho (\rho \omega_j) \heq - \frac{1}{2} \omega^i\\
        \conn{i}{\rho}{j} & = \bar {\mathcal K}\indices{^i_j}\\
        \conn{i}{\tau}{\tau} & = \frac{1}{2} \rho \gamma^{ij} \left( \omega_j \partial_\rho (\rho {\tilde \alpha}) + \partial_j {\tilde \alpha} - 2 \partial_\tau \omega_j \right) \heq 0\\
        \conn{i}{\tau}{j} & = {\mathcal K}\indices{^i_j} + \rho \gamma^{ik} \partial_{[k}\omega_{j]} + \frac{1}{2} \rho \omega^i \partial_\rho (\rho \omega_j) \heq {\mathcal K}\indices{^i_j}\\
        \conn{i}{j}{k} & = \Gamma[\gamma]^i_{jk} - \rho \omega^i \bar {\mathcal K}_{jk} \heq \Gamma[\gamma]^i_{jk}
    \end{align}
    where the extrinsic curvatures in the $\tau$- and $\rho$-directions are, respectively, given by 
    \begin{equation}
        {\mathcal K}_{ij} = \frac{1}{2} \partial_\tau \gamma_{ij}, \quad \bar  {\mathcal K}_{ij} = \frac{1}{2} \partial_\rho \gamma_{ij}\,. 
    \end{equation}

\noindent In   Table \ref{tb:covder2} we summarise the components of the     $\tau$- and $i$-covariant derivatives of a tensor $T$ on the horizon in Killing GNC.

    \begin{table}[H]
        \begin{tabular}{|c|c|c|}
            \hline
            $\nabla$ & $T^{\cdots a\cdots }$ & $T_{\cdots a \cdots}$ \\
            \hline
            & & \\
            $\tau$ & $\begin{aligned}
                \nabla_\tau T^{\cdots \tau\cdots} & = \partial_\tau T^{\cdots \tau\cdots} + \kappa T^{\cdots \tau \cdots } - \frac{1}{2} \omega_i T^{\cdots i\cdots}\\
                \nabla_\tau T^{\cdots \rho\cdots} & = \partial_\tau T^{\cdots \rho\cdots} - \kappa T^{\cdots \rho \cdots} \\
                \nabla_\tau T^{\cdots i\cdots} & = \partial_\tau T^{\cdots i\cdots} - \frac{1}{2} \omega^i T^{\cdots \rho\cdots} + {\mathcal K}\indices{^i _{j}}T^{\cdots j\cdots}
            \end{aligned}$ & $\begin{aligned}
                \nabla_\tau T_{\cdots \rho\cdots} & = \partial_\tau T_{\cdots \rho\cdots} + \kappa T_{\cdots \rho \cdots} + \frac{1}{2} \omega^i T_{\cdots i\cdots} \\
                \nabla_\tau T_{\cdots \tau\cdots} & = \partial_\tau T_{\cdots \tau\cdots} - \kappa T_{\cdots \tau \cdots}\\
                \nabla_\tau T_{\cdots i\cdots} & = \partial_\tau T_{\cdots i\cdots} + \frac{1}{2} \omega_i T_{\cdots \tau\cdots} - {\mathcal K}\indices{_i^j}T_{\cdots j\cdots}
            \end{aligned}$\\
            & & \\
            \hline
            & & \\
            $i$ & $\begin{aligned}
                \nabla_i T^{\cdots \tau\cdots} & = D_i T^{\cdots \tau\cdots} + \bar {\mathcal K} _{ij}T^{\cdots j\cdots} - \frac{1}{2} \omega_i T^{\cdots \tau\cdots}\\
                \nabla_i T^{\cdots \rho\cdots} & = D_i T^{\cdots \rho\cdots} + {\mathcal K} _{ij}T^{\cdots j\cdots} + \frac{1}{2} \omega_i T^{\cdots \rho\cdots}\\
                \nabla_i T^{\cdots j\cdots} & = D_i T^{\cdots j\cdots} + {\mathcal K}\indices{_i^j} T^{\cdots \tau\cdots} + \bar {\mathcal K}\indices{_i^j} T^{\cdots \rho\cdots}
            \end{aligned}$ & $\begin{aligned}
                \nabla_i T_{\cdots \rho\cdots} & = D_i T_{\cdots \rho\cdots} - \bar {\mathcal K}\indices{_i^j}T_{\cdots j\cdots} - \frac{1}{2} \omega_i T_{\cdots \rho\cdots}\\
                \nabla_i T_{\cdots \tau\cdots} & = D_i T_{\cdots \tau\cdots} - {\mathcal K}\indices{_i^j}T_{\cdots j\cdots} + \frac{1}{2} \omega_i T_{\cdots \tau\cdots}\\
                \nabla_i T_{\cdots j\cdots} & = D_i T_{\cdots j\cdots} - \bar {\mathcal K}\indices{_{ij}}T_{\cdots \tau\cdots} - {\mathcal K}\indices{_{ij}} T_{\cdots \rho\cdots}
            \end{aligned}$ \\
            & & \\
            \hline
        \end{tabular}
        \caption{Components of tensor covariant derivatives on the horizon in Killing GNC.} \label{tb:covder2}
    \end{table}

    \bibliographystyle{JHEP}
    \bibliography{bibliography}
\end{document}